\documentclass[twocolumn,pra,superscriptaddress,amssymb,amsmath]{revtex4}
\usepackage{graphicx}
\usepackage{epstopdf}
\usepackage{hyperref}

\hypersetup{%
  pdfpagemode=None, 
  pdfstartpage=1,
  pdfstartview=FitH,
  pdftoolbar=true,
  colorlinks = true,
  linkcolor=blue,
  citecolor=blue,
  bookmarksopen=false
}

\def\openone{\leavevmode\hbox{\small1\kern-3.3pt\normalsize1}}

\newcommand{\Op}[1]{\boldsymbol{\mathsf{\hat{#1}}}}
\newcommand{\gOp}[1]{\boldsymbol{\hat{#1}}}

\newcommand{\Fkt}[1]{\,\mathsf {#1}}

\ifx\Tr\renewcommand{\Tr}{\Fkt{Tr}}
\else\newcommand{\Tr}{\Fkt{Tr}}
\fi

\renewcommand{\dagger}{+}

\begin{document}

\title{Controlling open quantum systems: Tools, achievements, and limitations}

\author{Christiane P. Koch}
\email{christiane.koch@uni-kassel.de}
\affiliation{Theoretische Physik,  Universit\"at Kassel,
  Heinrich-Plett-Str. 40, 34132 Kassel, Germany}

\date{\today}

\begin{abstract}
The advent of quantum devices, which exploit the two essential
elements of quantum physics, coherence and entanglement, has 
sparked renewed interest in the control of open quantum systems. 
Successful implementations face the challenge to preserve the
relevant  nonclassical features at the level of device operation. A
major obstacle is decoherence which is caused by interaction with the
environment. 
Optimal control theory is a tool that can be used to identify control
strategies in the presence of decoherence. 
We review here recent advances in optimal control methodology that allow
for tackling typical tasks in device operation for open quantum
systems and discuss examples of relaxation-optimized dynamics. 
Optimal control theory is also a useful tool to exploit the
environment for control. We discuss  examples and point out 
possible future extensions. 
\end{abstract}


\maketitle

\section{Introduction}

Control refers to the ability to steer a dynamical system from an
initial to a final state with a desired accuracy; optimal control does
so with minimum expenditure of effort and resources. 
A famous example is the Apollo space mission where optimal control was
used to safely land the spacecraft on the moon. The essence of this
control task can be stripped down to a textbook example where students
calculate the change in acceleration, that is, the rate of burning
fuel, required to reach the moon's surface with zero velocity. This
example highlights the central role of optimal control in any type of
engineering, its importance being rivaled only by feedback, a subject
not covered in this review. 

In \textit{quantum} optimal control~\cite{GlaserEPJD15}, 
Newton's equations governing the motion of
the spacecraft are replaced by the quantum mechanical laws of motion,
of course. In contrast, 
the control, corresponding for example to a radio-frequency
(RF) amplitude or the electric field of a laser, is assumed to be
classical.  Quantum optimal control represents one variant of quantum
control~\cite{DAlessandroBook} and is closely related to coherent
control~\cite{ShapiroBook}. The latter requires exploitation of
quantum coherence, i.e., matter wave interference. In contrast, quantum control could also refer 
to inducing a desired dynamics, for example by amplitude modulations that avoid driving certain 
transitions,  without matter wave interference. 

Despite of its prominence  in mathematics and
engineering~\cite{PontryaginBook,KrotovBook}, optimal control
was introduced to NMR spectroscopy~\cite{Conolly1986,Mao1986}
and to the realm of matter wave dynamics~\cite{TannorJCP85,PeircePRA88,RonnieDancing89} only in the 1980s. In the latter case, the idea was to
calculate, via numerical optimization, laser fields that would steer a
photoinduced chemical reaction in the desired way~\cite{TannorJCP85,PeircePRA88,RonnieDancing89,Tannor92,GrossJCP92,SomloiCP93}. 
It was triggered by
the advent of femtosecond lasers and pulse shaping capabilities that
opened up seemingly endless possibilities to create intricate laser pulse
trains. While a controlled breaking of chemical bonds was indeed
demonstrated soon after~\cite{AssionSci98}, the pulses were obtained
by closed-loop optimizations in the experiments~\cite{JudsonPRL92}
rather than from theoretical calculations. In experimental closed-loop optimization, a shaped 
pulse is applied to the sample, and the outcome is measured. Based on the outcome, the pulse 
shape is modified, typically by a genetic algorithm. However, even for a
chemical reaction as simple as breaking the bond in a  diatomic alkali molecule,
the calculated optimized laser field cannot directly be used in the 
experiment~\cite{BorisJPC04}. The reason for this is two-fold: The way
how the optimized laser fields are obtained is rather different in
theory and experiment. Whereas the field in calculations is shaped as a
function of time~\cite{SomloiCP93,ZhuJCP98}, experiments employ
spectral shaping~\cite{WeinerRSI00}. As a result, calculated
pulses  are often incompatible with experimental pulse shaping
capabilities. Second,  
the theoretical modeling is simply not accurate enough. This results in pulses which are optimal for the 'wrong' dynamics and which can therefore not 
directly be applied in the experiments. 

These obstacles are not present, or at least much less severe, 
in other fields of application~\cite{GlaserEPJD15}. 
Once the timescale of the relevant dynamics is nanoseconds or slower, 
pulse shaping in the experiment is also done in time
domain~\cite{WrightRSI04}. While device response might still be an  
issue~\cite{MotzoiPRA11,JaegerPRA13}, the overall 
approaches in theory and experiment are similar in spirit.
Moreover, Hamiltonians and relaxation
parameters may be known much more accurately than is currently the
case in photoinduced chemical reactions. A prominent example is 
NMR where the development of optimal control in theory and experiment 
went hand in hand, yielding beautiful results, for example on
arbitrary excitation profiles~\cite{KobzarJMR05}, or robust broadband 
excitation~\cite{SkinnerJMR06,NimbalkarJMR13}. 

Given these observations, quantum information processing (QIP)
and related technologies offer themselves as an obvious playground for
quantum optimal control: In these applications, typically
the quantum system to be controlled is well characterized, and
timescales are sufficiently slow to use electronics for pulse shaping.
Not surprisingly, quantum optimal control
has attracted much interest in these fields over the last decade. This
included the adaptation of optimal control tools, for example to gate
optimization~\cite{PalaoPRL02,TeschPRL02,PalaoPRA03}, 
creation of entanglement~\cite{PlatzerPRL10,MuellerPRA11,CanevaNJP12},
or measurement~\cite{EggerPRA14}. Gate optimizations were carried out
for almost all QIP platforms, notably comprising ions~\cite{PoulsenPRA10}, 
atoms~\cite{TreutleinPRA06,GoerzJPB11,GoerzPRA14}, 
nitrogen vacancy (NV) centers in diamond~\cite{ScheuerNJP14,WaldherrNat14},
and superconducting qubits~\cite{RebentrostPRL09,GoerzNJP14}. 
Other QIP tasks, such as state
preparation~\cite{GrondPRA09,WangPRA10,MachnesPRA11,DoriaPRL11,RojanPRA14}, 
transport~\cite{DeChiaraPRA08,CanevaPRL09,FuerstNJP14}, 
and storage~\cite{GorshkovPRA08}, have also been the subject of optimal
control studies. These tasks are not only relevant for quantum
computing and communication but also for related applications that
exploit coherence and entanglement, for example quantum sensing or 
quantum simulation. Protocols derived with optimal control have by now 
reached a maturity that allows them to be tested in experiments.
Examples include the crossing of a phase transition studied with
trapped, cold atoms~\cite{RosiPRA13,vanFrank15}; the improvement of
the imaging capabilities of a single NV center~\cite{HaeberlePRL13};
and the creation of spin entanglement~\cite{DoldeNatComm14}, 
quantum error correction~\cite{WaldherrNat14} and matter wave
interferometry~\cite{vanFrankNatComm14,vanFrank15}.    

All of these examples share a generic feature that is typical for quantum
engineering: Control over the system, which inevitably also brings
about noise, needs to be balanced with sufficient isolation of the
desired quantum features. This sets the theme for controlling open
quantum systems. Traditionally, a quantum system is defined to be open
when it interacts with its
environment~\cite{WeissBook,BreuerBook}. This interaction results in
loss of energy and phase information. It can be modeled
phenomenologically within the semigroup approach or microscopically,
by embedding the system in a bath. Besides coupling to a bath, the
dynamics of a quantum system becomes effectively dissipative also when
the system is subject to measurements or noisy controls. 

Dissipative
processes pose a challenge to quantum control. At the same time,
desired dissipation may act as an enabler for control. We will 
review control strategies in both cases and then explain how optimal
control theory can be used to adapt them to more complex quantum
systems.
 
This topical review is organized as follows: 
Section~\ref{sec:OQS} briefly recalls the basic concepts in the theory
of open quantum systems, introducing the distinction between Markovian
and non-Markovian dynamics in Sec.~\ref{subsec:MarknonMark} and
addressing the issue of gauging success of control for open quantum
systems in Sec.~\ref{subsec:success}. 
The problem of analyzing controllability of open quantum systems, an
important prerequisiste to synthesizing control fields, is
introduced in Sec.~\ref{subsec:controllability}. 
Progress in the control of open
quantum systems is reviewed in Sections~\ref{subsec:strategies}
and~\ref{sec:OC-OQS} with Sec.~\ref{subsec:strategies} dedicated to control
strategies that were constructed with analytical methods and
Sec.~\ref{sec:OC-OQS} covering numerical optimal control. 
In Sec.~\ref{subsec:OCT}, the numerical methodology is explained in
detail for a simple example, followed by a discussion of the 
modifications required to adapt it to more advanced control
targets. The remainder of Sec.~\ref{sec:OC-OQS} reviews applications
of numerical optimal control to open quantum systems, starting with
examples for fighting or avoiding decoherence in
Sec.~\ref{subsec:fighting}. Control strategies that rely on the
presence of the environment are discussed in
Secs.~\ref{subsec:cooling}
and~\ref{subsec:exploiting}. Section~\ref{sec:concl} concludes.

\section{Open quantum systems}
\label{sec:OQS}
The state of an open quantum system is described by the density
operator $\gOp\rho_S$ which is an element of Liouville space. 
Any theory that aims at the control of an open quantum system is faced
with two basic prerequisites -- the ability to calculate the system's
dynamics, $\gOp\rho_S(t)$, and the ability to quantify success of
control.  

\subsection{Markovian vs non-Markovian dynamics} 
\label{subsec:MarknonMark}
Formally, the time evolution of any open quantum system can be
described by a dynamical map, $\gOp\rho_S(t)=\mathcal
D_{t,0}\left(\gOp\rho_S(0)\right)$ which is 
completely positive and trace preserving (CPTP)~\cite{BreuerBook}. The
dynamical map is divisible if it can be written as the composition of
two CPTP maps $\mathcal D_{t,0} =\mathcal D_{t,t^\prime}\mathcal
D_{t^\prime,0}$ $\forall t^\prime \le t$. If the dynamical map is
divisible, the  open system's time evolution is memoryless and called
Markovian. Various scenarios can lead to Markovian dynamics, weak
coupling between system and environment together with a decay of
environmental correlations much faster than the timescales of the
system dynamics being the most common case~\cite{BreuerBook}. 
However, open systems often exhibit pronounced memory
effects, in particular in condensed matter experiments, which reflect
characteristic features of the environment. The dynamics are then
called non-Markovian. Memory effects are caused by 
structured spectral densities, nonlocal correlations between
environmental degrees of freedom and correlations in the initial
system-environment state~\cite{BreuerReview,RivasRPP14}. 

In the Markovian case, the dynamics can be described by a master
equation in Lindblad form~\cite{BreuerBook}. In general, it  
needs to be solved numerically to determine $\gOp\rho(t)$. 
This can be done with arbitrarily high precision~\cite{KosloffARPC94}.
However, the computational effort may quickly become challenging due
to the exponential scaling of the size of Hilbert and Liouville
space. To date, room for improvement seems to be 
limited~\cite{LucasPRA14,SwekePRA15}. 

The situation is worse for non-Markovian dynamics, where a unified
framework such as the master 
equation in Lindblad form does not exist. A variety of methods has
been developed~\cite{deVegaRMP16}, each with different assumptions and
hence a different range of applicability. They include time-local
non-Markovian master  equations~\cite{BreuerBook}, stochastic
unravellings~\cite{PiiloPRL08,KochPRL08,DiosiPRL14}, and an auxiliary
density matrix approach~\cite{MeierJCP99}. A common feature of these
methods is their ability to correctly describe thermalization of the
system. 
Slightly different in philosophy are methods which attempt to solve the
dynamics of both system and environment~\cite{KochPRL03,HughesJCP09}. 
Key is to account only for that part of the environment that is
relevant for the system's dynamics, i.e., for the 'effective modes', which 
can be spins or harmonic oscillators. The underlying idea is that of
quantum simulation on a classical computer~\cite{GeorgescuRMP14},
where the true environment is replaced by a surrogate one that
generates the same dynamics. 
If one is interested in short times, the number of modes in the
surrogate Hamiltonian can be truncated with a prespecified
error~\cite{GualdiPRA13}. 
Longer propagation times than those computationally affordable with
exact dynamics of system and environment become possible by separating the
environment into two baths, one that is responsible for the
memory effects and that is modeled by effective modes as
explained above, and a second one that by itself would lead to
Markovian dynamics only. The secondary bath can be accounted for in
terms of a Markovian master equation in Lindblad
form~\cite{RebentrostPRL09,ReichSciRep15} or via a stochastic
unravelling using a single secondary bath mode~\cite{KatzJCP08}.   
A more comprehensive overview over methods to tackle non-Markovian
dynamics is found in Ref.~\cite{deVegaRMP16}. 

Understanding the influence of memory effects requires the ability 
to quantify them. An obvious way to define a measure of
non-Markovianity is to quantify deviation from
divisibility~\cite{WolfPRL08,RivasPRL10}. Interestingly,
this corresponds to an  increase of correlations if the
system is bi- or multipartite~\cite{RivasPRL10,DharPRA15}. 
Other measures to
capture  memory effects focus on specific features 
such as recoherence and information backflow. These can be
characterized in terms of distinguishability between quantum
states~\cite{BreuerPRL09,LainePRA10,VasilePRA11}, 
re-expansion of the volume of accessible states~\cite{LorenzoPRA13},
or the capacity to reliably transmit quantum 
information~\cite{BylickaSciRep14}. A comprehensive overview over the
different measures is found
in Ref.~\cite{BreuerReview}, and an illustrative comparison for a toy
model is presented in Ref.~\cite{AddisPRA14}.

The proposed non-Markovianity measures can be classified in a
hierarchy by generalizing the notion of
divisibility~\cite{ChenPRA15,Bae16}. While this is gratifying from a
theoretical perspective, it is still an open question
how non-Markovianity can be measured in 
a condensed phase experiment. Although some
of the measures have been evaluated in
experiment~\cite{LuiNatPhys11,GessnerNatPhys14}, dissipation in these 
examples was either engineered or artificial, in the sense that
different degrees of freedom within one particle were considered to
play the role of system and bath. A true condensed phase setting is
more challenging due to limited control and thus limited access to
measurable quantities.

Current interest in non-Markovian dynamics is fueled by the revival of
genuine quantum properties such as quantum coherence and correlations
that non-Markovianity entails. It has sparked the hope to exploit
non-Markovianity as a resource. Quantum control in particular, which
relies on properties such as coherence, should  
be more powerful in the non-Markovian compared to the Markovian
regime. 

\subsection{Measuring success of control in open quantum systems} 
\label{subsec:success}
When the goal is to control an open quantum system, the ability to
gauge success of control is even more important than that to measure
the degree of non-Markovianity. Any suitable figure of merit needs to
fulfill two conditions: (i) It should take its optimum value if and
only if the control target is reached. (ii) It needs to be computable. 
An obvious control target is to drive a given initial state to a
desired target state. The corresponding figure of merit is the
projection onto the target state. For open quantum systems, 
this is given in terms of the Hilbert Schmidt product of the state of
the system at the final time and the target state,
\begin{equation}
  \label{eq:Fstate2state}
  F_T = \mathrm{Tr}\left\{ 
    \mathcal D_{T,0}\left(\gOp\rho_{initial}\right)\gOp\rho_{target}\right\}
\end{equation}
This figure of merit has been used for example in control studies of
cooling where the timescales of the dissipative process and the
coherent system dynamics are
comparable~\cite{BartanaJCP97,SchmidtPRL11}. Typically, the target for
cooling is a pure state. Sometimes the timescale of dissipation is much
slower than that of the coherent dynamics. This situation is 
encountered when using femtosecond lasers for laser
cooling~\cite{ViteauSci08}. In order to avoid the very long
propagation times for repeated cooling cycles consisting of coherent
excitation and spontaneous emission as well as the 
expensive description in Liouville space, one can expand
$\gOp\rho_{initial}$ in a basis of Hilbert space vectors 
and tailor the dynamics of the Hilbert space vectors
such that the target will be
reached irrespective of the initial state~\cite{ReichNJP13}. 
The construction of the proper
figure of merit in that case will be discussed below in
Section~\ref{subsec:cooling}. 

An important control target in the context of quantum information
processing is the implementation of unitary operations, or quantum
gates. This corresponds to simultaneous state-to-state transitions for
all states in the logical
basis~\cite{PalaoPRL02,TeschPRL02,PalaoPRA03}. 
A straightforward way to express this control target in Liouville
space is given by~\cite{KallushPRA06,ToSHJPB11}
\begin{equation}
  \label{eq:Fgate}
  F_T = \frac{1}{d}\sum_{j=1}^{d^2} \mathrm{Tr}\left\{
    \Op O\gOp\rho_j\Op O^+ \mathcal D_{T,0}\left(\gOp\rho_j\right)
    \right\}\,,
\end{equation}
where $\Op O$ denotes the desired target operation, defined on the logical subspace of dimension $d$.  The set of 
$\gOp\rho_j$ forms a suitable orthonormal basis of the $d^2$-dimensional 
Liouville (sub)space or, more simply, all $d^2$ matrices for which one entry is equal to one and all other zero. 
The Hilbert Schmidt product in Eq.~\eqref{eq:Fgate}
checks how well  the actual evolution, $\mathcal
D_{T,0}\left(\gOp\rho_j\right)$, matches the desired one, 
$\Op O\gOp\rho_j\Op O^+$. The scaling of Eq.~\eqref{eq:Fgate}
with system size $d$ restricts its use to examples with very few
qubits. 

If the target is the implementation of a unitary operation,
and not an arbitrary dynamical map, much less resources are required
to gauge success of control. This observation is at the basis of all
current proposals to estimate the average gate
fidelity~\cite{BenderskyPRL08,daSilvaPRL11,FlammiaPRL11,MagesanPRL11,ReichPRL13}
which forego the full knowledge of $\mathcal D_{T,0}$ that is obtained 
in quantum process tomography in favor of efficiency. 
One way to understand the reduction in effort is to start from
Eq.~\eqref{eq:Fgate} and ask how many states are required in the sum
to have a well-defined figure of merit, i.e., a figure of merit that
takes its optimum value if and only if the target operation is
realized. Surprisingly, the answer to this question is 
three, independent of system
size~\cite{ReichPRA13,GoerzNJP14}:
\begin{equation}
  \label{eq:Fgate3}
  F_T = \frac{1}{3}\sum_{j=1}^3 \mathrm{Tr}\left\{
    \Op O\gOp\rho_j\Op O^+ \mathcal D_{T,0}\left(\gOp\rho_j\right)
    \right\}\,.
\end{equation}
One state in Eq.~\eqref{eq:Fgate3}
measures the departure from unitarity or, more precisely,
from unitality in the logical subspace, and two more states are
necessary to distinguish any two unitaries~\footnote{%
The problem is slightly more involved if one seeks an estimate of the
error instead of a simple yes-no answer. It is addressed in
Refs.~\cite{ReichPRL13,GualdiPRA14,ReichJPA14}. 
For control applications, however,
Eq.~\eqref{eq:Fgate3} and variants are sufficient\cite{GoerzNJP14}. 
}. The latter requires two
states because one needs to determine the basis in which the unitary
is diagonal and then check whether the phases on the diagonal are
identical. Only two states are required because it is possible to
construct one  density operator that 'fixes' the complete Hilbert
space basis, 
$\gOp\rho_2=\sum_{i=1}^d\lambda_i|\varphi_i\rangle\langle\varphi_i|$,
using one-dimensional orthogonal projectors, 
$|\varphi_i\rangle\langle\varphi_i|$, 
with non-degenerate eigenvalues, $\lambda_i\neq\lambda_j$. A
variant of Eq.~\eqref{eq:Fgate3} is obtained by replacing the two 
states for the basis and phases by $d$ 
states $\rho_j=|\varphi_i\rangle\langle\varphi_i|$. This is still a
reduction compared to the $d^2$ states in Eq.~\eqref{eq:Fgate} and was
found to lead to faster convergence in control calculations than 
Eq.~\eqref{eq:Fgate3}~\cite{GoerzNJP14}.
Evaluation of both Eq.~\eqref{eq:Fgate3} and its $d+1$ state variant
are much more efficient than that of
Eq.~\eqref{eq:Fgate}~\cite{GoerzNJP14}. 

Both Eqs.~\eqref{eq:Fgate} and~\eqref{eq:Fgate3} target implementation
of a specific unitary $\Op O$. For difficult control problems, where
a numerical search can easily get stuck, it is desirable to formulate
the control 
target in the most flexible way. For example, instead of implementing a
specific two-qubit gate such as CNOT, it may be sufficient to realize a
gate that is locally equivalent to CNOT, i.e., that differs from CNOT
by local operations. The corresponding figure of merit is based on the
so-called local invariants~\cite{MuellerPRA11,ReichJCP12}. Similarly,
one can formulate a figure of merit for targeting an arbitrary perfect
entangler~\cite{WattsPRA15,GoerzPRA15}. Since these figures of merit
are based on the local invariants which in turn are calculated 
from the unitary evolution, extension to non-unitary dynamics requires
to first determine the unitary part of the overall evolution. This is
possible, using the same mathematical concepts that have
led to Eq.~\eqref{eq:Fgate3}~\cite{ReichPhD}.

\section{Control of open quantum systems}
\label{sec:COQS}

The theory of controlling open quantum system can be divided into two
main questions---analysis of controllability and synthesis of
controls (called motion planning in the classical automatic control
community): 
When the goal is to realize a certain desired dynamics, it is
worthwhile to check first whether performance of the task is possible
at all, before starting to search for controls that lead to the target.
Section~\ref{subsec:controllability} briefly reviews the current state
of the art in controllability analysis of open quantum systems,
whereas Sec.~\ref{subsec:strategies} summarizes strategies for control
synthesis that are based on certain properties of the system's
interaction with its environment, and optimal control theory as a tool for
control synthesis will be presented in Sec.~\ref{sec:OC-OQS}.

\subsection{Controllability of open quantum systems}
\label{subsec:controllability}
Controllability analysis provides the mathematical tools for answering
the question whether the target is reachable~\cite{JurdjevicBook}. In
particular, a complete framework exists  
for finite-dimensional systems undergoing unitary dynamics: 
Separating the Hamiltonian into drift and control terms, 
\begin{equation}
  \label{eq:Hgeneric}
\Op H = \Op H_0 + \sum_j u_j(t) \Op H_j\,,  
\end{equation}
the system is controllable provided the Lie algebra spanned by 
the nested commutators of $\Op H_0$ and $\Op H_j$ is full
rank~\cite{Jurdjevic72,HuangJMP83,DAlessandroBook}.  
The elements of the Lie algrebra can be interpreted as
tangential vectors, i.e., as the directions, 
of the unitary evolution (which is an element of a
Lie group). If evolution into all directions can be generated,
the system is controllable. Controllability can also be viewed in
terms of connectivity between Hilbert space basis
states~\cite{PolackPRA09}; it then corresponds to 
presence of the respective matrix elements. 

For open quantum systems, the evolution is not unitary anymore, and 
analysis of the Hamiltonian alone is not sufficient to decide
controllability: The dissipative part of the evolution may prevent or
enable certain states to be reached. For example, even if the full
rank condition for the Lie algebra of the Hamiltonian is fulfilled,
fast decoherence will inhibit transitions between pure states by
turning any pure state into a mixed one. Merely the presence of
non-vanishing matrix elements in the Hamiltonian is thus not
sufficient to decide controllability. 
Their magnitude  matters as 
well, in particular of those in the drift Hamiltonian $\Op H_0$ that
cannot be tuned by external controls $u_j$. To date, rigorous controllability
analysis does not take such a 
dependence on operator norms, or competing time scales
into account. 

As a result, one needs to turn to numerical search for
most open quantum systems, even for an example as simple as the central spin model~\cite{ArenzNJP14}. This is rather unsatisfactory since a numerical search cannot provide rigorous answers to controllability, in particular lack thereof
(in the sense of reachability of
the target with a predetermined error $\epsilon$), due to its local
character.  An extension of controllability analysis to the
needs of open quantum systems would address this issue but remains an
open problem to date. It would be 
particularly relevant for open quantum systems with almost unitary
dynamics, which are often encountered e.g. in quantum
technology applications.  

As an example where the dissipative part of the overall evolution
enables certain states to be reached, consider cooling which turns
mixed states into pure ones. Obviously, controllability analysis based
on the Lie rank condition for the Hamiltonian alone cannot provide any
information on time evolutions which change the purity of the system's
state, $\mathcal P =\Tr\left[\gOp\rho^2_S(t)\right]$. 

By and large, controllability of open quantum systems still remains
uncharted territory to date~\cite{GlaserEPJD15}. 
This refers in particular to dynamic
controllability where the analysis accounts for available dynamical
resources such as coupling to external fields and environmental
degrees of freedom, or measurements, in contrast to kinematic
controllability. The latter implies existence of a dynamical map that
transforms any initial state into the target state. While this
existence can be proven for finite-dimensional
systems~\cite{WuJPhysA07}, it is of limited relevance for practical
applications since, in general, one cannot derive any dynamical
information from the proof.

To tackle controllability of open quantum systems,  two routes can be
followed: Either one starts from a complete description of 
the system and its environment, or one considers the reduced
description of the system alone. In the first case, the tools of
controllability analysis for unitary dynamics can be employed. This
way it was possible to show, for example, that even for completely
controllable system-environment dynamics, cooling is
possible only if the environment contains a sufficiently large
virtual subsystem which
is in a state with the desired purity~\cite{TicozziSciRep14}. 
While exact solution of the combined system-environment dynamics is
extremely challenging (and impossible in many cases), controllability
analysis from this perspective is expected to significantly advance our
understanding already for simple models and for generic questions, as
in the example of Ref.~\cite{TicozziSciRep14}.
Moreover, it appears to be promising for two reasons 
-- it does not rely on \textit{a priori} assumptions on the reduced
dynamics, and most likely it will benefit from recent progress in the
controllability of infinite-dimensional
systems~\cite{ChambrionAnnHP09,BeauchardCMP10,BoussaidIEEETrans13,BoscainCMP15}.

Rigorous controllability analyses for reduced dynamics have been
limited to date to  
the Markovian
case~\cite{AltafiniJMP03,LorenzaPRA01,ToSHRMP09,MearaIEEE12,AltafiniAC12,Dive16}. In 
particular, the sets of reachable states were
characterized~\cite{AltafiniJMP03}, and 
the Lindblad operators were shown to form a Lie
wedge~\cite{ToSHRMP09}. While the Lie wedge provides a sufficient, but not necessary condition for controllability, necessary but not sufficient criteria can be identified by considering isotropy of the generator of the dissipative motion~\cite{Dive16}. 
Numerically, non-Markovian dynamics were
shown to lead to $SU(N)$ controllability 
(in the sense of reachability of
the target with a predetermined error $\epsilon$)
for a system whose
Hamiltonian  allows for realization of $SO(N)$
operations only~\cite{ReichSciRep15}.  However, no rigorous analysis of 
controllability for non-Markovian dynamics has been performed to
date; and it is not yet clear whether and under which conditions
non-Markovian effects can improve controllability of an open quantum
system. 

\subsection{Control strategies for open quantum systems}
\label{subsec:strategies}

Control strategies that are obtained by analytical methods can be
roughly divided into two classes -- those  that exploit
symmetries in the system-bath interaction and those that make
assumptions on the timescale of this interaction. 
In the first case, protection from decoherence is achieved by
keeping the system's state in a so-called decoherence-free
subspace~\cite{LidarPRL98,LidarReviewDFS03}:
If the system-bath interaction contains a symmetry,
for example qubits couple indistinguishably to their environment,
it is possible to construct a set of system states that are invariant
under the system-bath interaction. These states form a
decoherence-free subspace in the system's total Hilbert
space~\cite{LidarPRL98,LidarReviewDFS03}. 
Stimulated Raman Adiabatic Passage (STIRAP) and
electromagnetically induced transparency represent
examples for decoherence-free subspaces~\cite{LidarReviewDFS03}. 
In the context of quantum information, two or more physical qubits,
carried for example by atoms, can be used to encode one logical qubit that is
decoherence-free~\cite{LMDuanPRL97}. Decoherence-free subspaces have been
demonstrated, for example, in liquid-state nuclear magnetic
resonance~\cite{FortunatoNJP02} and with trapped ions~\cite{MonzPRL09}.  
Numerical methods can be employed to identify (approximate)
decoherence-free subspaces~\cite{WangJacobsPRA13} and to find an
external control that drives the
system dynamics into a decoherence-free subspace~\cite{YiPRA09}. 

More generally, in physics the presence of a symmetry implies
existence of a conserved quantity. In the context of decoherence, 
a symmetry in the system-bath interaction leads to a quantum number
which is preserved under this interaction. The
eigenstates belonging to the preserved quantum number define a 
noiseless subsystem, i.e., a logical
subsystem that is intrinsically protected from
noise~\cite{LorenzaPRL00,LorenzaSci01}.
The main limitation of this set of approaches is imposed by the
existence of a suitable symmetry which is not necessarily
available. 

In the second case of control strategies that are built on assumptions
on the timescale of the system-bath interaction, a
trivial strategy is obtained for slow decoherence: One simply needs to
perform the desired operation on a time scale much faster than that of
decoherence. But this may not always be possible, and slow decoherence
also allows for eliminating the effects of decoherence based on
average Hamiltonian theory~\cite{AHT} or, in more intuitive terms,
spin echo techniques from nuclear magnetic resonance and
generalizations thereof. This set of strategies is often referred to
as dynamical
decoupling~\cite{LorenzaPRL99,LorenzaJMO04,SouzaPTRSA12,SuterRMP15}. It 
relies on  many quasi-instantaneous actions of control
fields that do not allow the system to interact with its environment, 
creating an effective dynamics of the system alone. 

Extensive work
over the last two decades has allowed to account for e.g. pulse
imperfections~\cite{KhodjastehPRL05} and extend the technique beyond
spin dynamics for which it was originally developed, for example to
ions~\cite{BiercukNature09}, and superconducting
circuits~\cite{BylanderNatPhys11}.
The main limitation of the dynamical decoupling approach is the finite
duration of any  control field which cannot always be made
sufficiently short. By now dynamical decoupling has grown into a field
of its own which has been covered by several
reviews~\cite{LorenzaJMO04,SouzaPTRSA12,SuterRMP15}, and the
reader is referred to these for a more in-depth analysis. 

The time-dependent perspective used in dynamical decoupling can be
translated into a frequency-space picture using the generalized
transfer filter function approach~\cite{GreenPRL12,Paz-SilvaPRL14}. 
This allows to efficiently predict fidelities, at least for weak
noise~\cite{SoareNatPhys14}, and provides a connection to control
strategies that decouple the system from its environment by
engineering a spectral separation~\cite{ClausenPRL10}. In particular,
it was shown that the system-bath interaction can be cancelled, at least to second 
order, by choosing the time-dependence of the control field such
that its spectrum becomes orthogonal to the bath or noise
spectrum~\cite{ClausenPRL10}. Such an approach requires weak
coupling and negligible initial system-bath correlations.
Moreover, the bath spectrum needs to be
known. Another example of engineering a spectral separation is given by the protection of so-called edge states in topological insulators via band Liouvillians~\cite{ViyuelaPRB12,RivasPRB13}. 

In addition to clarifying the relation between time-domain based and frequency-domain based control strategies, the 
transfer filter function perspective also 
provides a practical tool for characterizing the noise
spectral density~\cite{CywinskiPRB08}. Its
applicability to a variety of physical platforms has already been
demonstrated~\cite{CywinskiPRB08,KotlerNature11,KotlerPRL13}. Noise
spectroscopy provides an excellent starting point for deriving
microscopic models for the system-environment interaction and 
thus a more thorough understanding of noise at the quantum
mechanical level. It also allows for tailoring control synthesis to
specific  spectral features of the noise, either, if possible, by
using the transfer filter function approach
directly~\cite{Paz-SilvaPRL14}, or by exploiting knowledge of these
features using optimal control theory, 
in Sec.~\ref{sec:OC-OQS} below.

\section{Optimal control of open quantum systems} 
\label{sec:OC-OQS}

Quantum optimal control theory refers to a set of methods that synthesize
external control fields from knowledge of the control target,
including constraints, and the time evolution of the quantum
system~\cite{GlaserEPJD15}. It is based on the calculus of variations,
i.e., on formulating the control
target as a functional of external controls that realize the
desired dynamics. Knowledge of the system's time evolution enters
implicitly via evaluation of the target functional and, possibly, its
derivatives. 

For a few exceptional cases, for example one or two
spins (or qubits)~\cite{KhanejaPNAS03,KhanejaJMR03,%
LapertPRL10,LapertPRA13,MukherjeePRA13,MukherjeeNJP15}, a harmonic
oscillator~\cite{RezekEPL09,HoffmannEPL11}, 
or a sequence of $\Lambda$-systems subject to
decay~\cite{SklarzPRA04,HaidongPRA12},  
the external controls can be determined using geometric techniques
based on Pontryagin's maximum principle~\cite{PontryaginBook}. 
Typically, however, the control problem cannot be solved in closed
form, and one needs to resort to numerical optimization. 

Most often, controlling open quantum systems is a difficult control
task such
that efficiency of the optimization algorithm is important. Therefore,
mainly algorithms based on the target functional's gradient have 
been employed for open quantum systems to date. They will
be reviewed in Sec.~\ref{subsec:OCT}. The remainder of this
section is dedicated to control strategies that were developed using
numerical optimization, starting with strategies avoiding decoherence
in Sec.~\ref{subsec:fighting}, followed by strategies exploiting
presence of the environment in Sec.~\ref{subsec:cooling}
and~\ref{subsec:exploiting}. 

\subsection{Optimal control theory applied to open quantum systems} 
\label{subsec:OCT}

Conceptually the simplest control problem is represented by
state-to-state transfer: Given a known initial state,
$\gOp\rho_{ini}$, at time $t=0$,
find the external field that drives this state at final time $t=T$
into the target state, $\gOp\rho_{target}$, with prespecified error,
$\epsilon$.  The corresponding target functional is found in
Eq.~\eqref{eq:Fstate2state} where dependence on the set of external
controls $\{u_j\}$ is implicit in the time evolution, $\mathcal
D_{T,0}=\mathcal D_{T,0}[\{u_j\}]$. 

This control problem was first
stated in the context of laser cooling molecular
vibrations~\cite{BartanaJCP93,BartanaJCP97,TannorJPCA99}. The time evolution
was modeled by a Markovian master equation in Lindblad form, and later
extended to a non-Markovian example~\cite{SchmidtPRL11,SchmidtPS12}. 
For the sake
of conceptual clarity, we present here the algorithm for a Markovian
master equation, 
\begin{equation}
  \label{eq:LvN}
  \frac{d\Op\rho}{dt} = \mathcal{L}(\Op\rho) = 
  -i\left[\Op H\left[u(t)\right],\Op\rho\right]
  +\mathcal{L}_D(\Op\rho)\,,
\end{equation}
with the Hamiltonian of the form~\eqref{eq:Hgeneric}. For simplicity,
we assume a single control $u_1(t)=u(t)$ in Eq.~\eqref{eq:Hgeneric}. In the
example of laser cooling molecular vibrations, $u(t)$ would be the
electric field of a short laser pulse, and $\Op H_1$ in Eq.~\eqref{eq:Hgeneric}
the transition dipole moment of the molecule.
The dissipative part of Eq.~\eqref{eq:LvN} is given by~\cite{BreuerBook}
\begin{equation}
  \label{eq:Lindblad}
  \mathcal L_D(\Op\rho) = \sum_k \gamma_k \left(
    \Op{A}_k \Op{\rho} \Op{A}_k^\dagger
    - \frac{1}{2} \left\{\Op{A}_k^\dagger \Op{A}_k, \Op{\rho} \right\}
  \right) \,,
\end{equation}
where the Lindblad operators $\Op A_k$
model the various dissipative channels. For
example, $\Op{A}_k = |0\rangle\langle k|$ for spontaneous decay from a
level $k$ to the ground state with rate $\gamma_k$, inversely
proportional to the level's lifetime. 

Optimization algorithms are 
obtained by seeking an extremum of $F_T$,
cf. Eq.~\eqref{eq:Fstate2state}, with respect to the control,
$u(t)$. This can be done by direct evaluation of the extremum
condition~\cite{OhtsukiJCP99,KhanejaJMR05} 
or by building in monotonic convergence \textit{a
  priori} using Krotov's method~\cite{BartanaJCP97,ReichJCP12}. The
resulting set of coupled equations are, surprisingly, rather similar. 
The main difference is in the update of the control which can be
performed concurrently~\cite{KhanejaJMR05}, i.e., for all times $t$ at
once, or sequentially in
time~\cite{BartanaJCP97,OhtsukiJCP99,ReichJCP12}.  Guaranteed
monotonic convergence is only obtained with a sequential update of the
control. In this case, the equation for determining the control reads
\begin{eqnarray}
  \label{eq:update}
  \Delta u(t) &=& u^{(i+1)}(t)-u^{(i)}(t) \\ \nonumber
  &=&
  \frac{S(t)}{\lambda} \mathfrak{Im}\left\{
    \Tr\left[\Op\sigma^{(i)}(t)\,
      \frac{\partial \mathcal{L}\left(\gOp\rho\right)}{\partial u}
      \Big|_{\rho^{(i+1)},u^{(i+1)}}\right]
      \right\}\,.
\end{eqnarray}
$S(t)$ denotes a shape function that can be used to switch the control
on and off smoothly~\cite{SundermannJCP99} 
or to impose an initial or final ramp~\cite{ReichSciRep15}; and
$\lambda$ is a parameter of the algorithm 
whose choice determines the step size in the
change of the control. Its optimal value can be estimated in an
automated way, similarly to a line search in quasi-Newton
methods~\cite{EitanPRA11}.  $\gOp\sigma(t)$ denotes the so-called co-state or adjoint state, and 
the derivative with respect to the control is given by the commutator
\[
 \frac{\partial \mathcal{L}\left(\gOp\rho\right)}{\partial u}
      \Big|_{\rho^{(i+1)},u^{(i+1)}} =
    -i \left[ \frac{\partial \Op H[u(t)] }
      {\partial u}\Big|_{u^{(i+1)}},\gOp\rho^{(i+1)}\right]\,.
\]
For the common case of linear coupling to the control, as in
Eq.~\eqref{eq:Hgeneric}, the explicit dependence on $u^{(i+1)}$
vanishes and the commutator simply becomes $\left[\Op
  H_1,\gOp\rho^{(i+1)}\right]$. Nonetheless, the right-hand side of
Eq.~\eqref{eq:update} depends on $u^{(i+1)}$ via
$\gOp\rho^{(i+1)}(t)$, i.e., it is an implicit equation. Solution of
the implicit equation can usually be avoided by a low order
approximation in the iterative algorithm, 
employing two shifted time discretizations to represent the time
dependence of states and control, $\gOp\rho(t_i)$ and $u(t_i+\Delta
t/2)$~\cite{PalaoPRA03}.

Equation~\eqref{eq:update} also depends on the state of system at time
$t$, $\gOp\rho(t)$, and the co-state, $\gOp\sigma(t)$. These are obtained by solving 
\begin{subequations}\label{eq:forward}
\begin{eqnarray}
  \label{eq:LvNforward}
  \frac{d\Op\rho^{(i)}}{dt} = 
  -i\left[\Op H[u^{(i)}(t)],\Op\rho^{(i)}\right]
  +\mathcal{L}_D(\Op\rho^{(i)})
\end{eqnarray}
with initial condition
\begin{equation}
  \label{eq:rho_ini}
  \gOp\rho^{(i)}(t=0)=\gOp\rho_{ini}\,,
\end{equation}
\end{subequations}
and 
\begin{subequations}\label{eq:backward}
\begin{eqnarray}
  \label{eq:LvNbackward}
  \frac{d\Op\sigma^{(i)}}{dt} = 
  -i\left[\Op H\left[u^{(i)}(t)\right],\Op\sigma^{(i)}\right]
  -\mathcal{L}_D(\Op\sigma^{(i)})\,,  
\end{eqnarray}
which is solved backward in time. The 'initial' condition is derived
from the target functional at final time $F_T$,
\begin{equation}
  \label{eq:sigma_ini}
  \gOp\sigma^{(i)}(t=T) =   \gOp\rho^{(i)}_{target}\,.
\end{equation}
\end{subequations}
This coupled set of equations needs to be solved iteratively, 
starting with some guess for the control, $u^{(i=0)}(t)$, where the
index $(i)$ denotes iteration.

The algorithm represented by Eqs.\eqref{eq:update}, \eqref{eq:forward}
and \eqref{eq:backward} is straightforwardly extended from targeting a
single state to targeting a unitary
operation~\cite{KallushPRA06,ToSHJPB11,GoerzNJP14}: A unitary
operation $\Op U$ can be viewed as several simultaneous state-to-state
transfers~\cite{TeschPRL02,PalaoPRA03} which are all driven by the
same control. Consequently, Eq.~\eqref{eq:update} becomes
\begin{equation}
  \label{eq:updateN}
  \Delta u(t) =
  \frac{S(t)}{\lambda} \sum_{j=1}^M \mathfrak{Im}\left\{
    \Tr\left[\gOp\sigma^{(i)}_j(t)\,
      \frac{\partial \mathcal{L}\left(\gOp\rho_j\right)}{\partial u}
      \Big|_{\rho_j^{(i+1)},u^{(i+1)}}\right]
      \right\}\,,
\end{equation}
and  Eqs.~\eqref{eq:forward} and \eqref{eq:backward} need to be solved
for $M$ states $\gOp\rho_j(t)$ and $M$ co-states $\gOp\sigma^{(i)}_j(t)$
simultaneously. As explained in Sec.~\ref{subsec:success}, it was
first believed that the sum in Eq.~\eqref{eq:updateN} has to run over
$M=d^2$ states where $d$ is the dimension of the space on which the
desired operation is defined~\cite{KallushPRA06,ToSHJPB11}. The
initial conditions $\gOp\rho_j(t=0)$ are then given by orthogonal
basis states spanning this space.   
Recently it was shown that $M$ can be reduced all the way down to 3 in
which case the states discussed in Sec.~\ref{subsec:success} need to
be taken as initial conditions $\gOp\rho_j(t=0)$~\cite{GoerzNJP14}.
The initial conditions for the co-states are always given in terms of
the desired target operation, $\gOp\sigma_j(t=T)\sim \Op U 
\gOp\rho_j(t=0)\Op U^+$, up to a suitable normalization~\cite{GoerzNJP14}.  
Moreover, weights may be introduced in $\gOp\sigma_j(t=T)$ to speed up
convergence by attaching different importance
to different basis states $\gOp\rho_j(t=0)$~\cite{GoerzNJP14}. 

Extension of the optimization algorithm to more flexible control
targets, such as an arbitrary perfect
entangler~\cite{WattsPRA15,GoerzPRA15} instead of a specific 
unitary, requires two steps. First, a modified target functional
results in a modification of Eq.~\eqref{eq:sigma_ini}: The right-hand
side of Eq.~\eqref{eq:sigma_ini} will be replaced by the derivative of
the new target functional with respect to the states, $\gOp\rho_j$,
evaluated at time $t=T$. For a target functional based on the local
invariants, this requires, in particular, to determine the unitary part
of the overall actual evolution, as explained in
Sec.~\ref{subsec:success}.
Second, dependence of the functional on the states $\gOp\rho_j$
may be non-convex. In this case, Eq.~\eqref{eq:updateN} needs to be
amended by a second term in its right-hand side~\cite{ReichJCP12}
which depends additionally on the change in the states,
$\gOp\rho_j^{(i+1)}-\gOp\rho_j^{(i)}$. While the additional
computational effort for evaluating the control update is negligible,
storage of all $\gOp\rho_j^{(i)}$ is necessary and may become a
limiting factor when scaling up the system size. Such extensions of
the optimization algorithm to control targets other than a specific
state or unitary have been applied to coherent
dynamics~\cite{MuellerPRA11,ReichJCP12,WattsPRA15,GoerzPRA15}. For 
open quantum systems, they are still under exploration.

Typically, more than one solution exists to a quantum control problem.
When using the basic optimization algorithm presented above, it will
then depend on the initial guess $u^{(0)}(t)$ which solution $u^\star(t)$ the
algorithm identifies. Two strategies can be employed in order to
fine-tune the iterative search and guide it toward a solution with
certain desired properties -- careful selection of the initial
guess by scanning or preoptimization~\cite{GoerzEPJQT15} or use of 
additional constraints~\cite{PalaoPRA13,ReichJMO14}. In the first
case, the initial guess needs to be parametrized in a suitable form,
for example in terms of amplitudes and phases of Fourier components, or
amplitudes, widths and positions of Gaussians. These parameters can
easily be pre-optimized within a prespecified range employing a standard
gradient-free optimization method~\cite{GoerzEPJQT15}. When the result
is used as initial guess in the optimization algorithm presented
above, the ensuing fine-tuning will usually not lead to  drastic
changes in the field, keeping it close to the parametrized
form~\cite{GoerzEPJQT15}. 

A more stringent way to enforce certain desired properties of the
control solution is obtained by employing additional
constraints~\cite{PalaoPRA13,ReichJMO14}. This comes at the expense of a
modified optimization algorithm. An explicit description for deriving
the modified algorithms is available when using Krotov's
method~\cite{ReichJCP12}. It allows for formulating constraints as
a functional of the control, with the only condition that the
functional be positive (or negative)
semidefinite~\cite{ReichJMO14}. This 
requirement is necessary to ensure monotonic convergence. As an
example, consider a spectral constraint on the
control~\cite{ReichJMO14},
\begin{equation}
  \label{eq:Jspectral}
  J_{spec}[u(t)] = \frac{1}{2\pi} \int_0^T \int_0^T
  \Delta u(t) K(t-t^\prime) \Delta u(t^\prime) dt^\prime dt\,,
\end{equation}
where $K(t-t^\prime)$ is the Fourier transform of a spectral filter
$\tilde K(\omega)\ge 0$. As a result, the update equation for the
control becomes a Fredholm equation of the second
kind~\cite{ReichJMO14}. A judicious choice of the shape function
$S(t)$ allows for analytically solving the Fredholm equation in
frequency domain such that the additional numerical effort for
including the spectral constraint consists merely in two additional
Fourier transforms~\cite{GoetzPRA16}. More than one constraint may be
employed at a time, with different weights allowing to emphasize
importance of one compared to another~\cite{PalaoPRA13,GoetzPRA16}. 
While additional constraints provide information that guides the
optimization algorithm, they also restrict the space of admissible
solutions~\cite{MooreJCP12,RivielloPRA15}. Therefore care needs to be
taken to balance their benefits and their disadvantages.

\subsection{Fighting and avoiding decoherence}
\label{subsec:fighting}

While it is probably the dream of every control engineer to discover
unthought of control schemes, a more realistic scenario starts from 
known a control strategy, as those described in
Sec.~\ref{subsec:strategies}, 
and extends it to a wider range of conditions, 
new types of systems, or new types of dissipative
processes using the optimization techniques described
above.

One of the  most popular control strategy in the area of
quantum technologies currently is dynamical
decoupling~\cite{SouzaPTRSA12,SuterRMP15}. While already powerful in
itself, dynamical decoupling can be made 
more robust by numerical optimization that targets specific noise
features that were previously unaccounted for, using, for example, 
the gradient-ascent technique~\cite{GormanPRA12} or genetic
algorithms~\cite{QuirozPRA13}. In these examples, optimization did not
compromise feasibility of the pulse sequences. Moreover, the length of
dynamical decoupling sequences can be minimized~\cite{Paz-SilvaSciRep13}. 
Dynamical decoupling and
numerically optimized pulses can also complement each other, as
recently demonstrated for entanglement generation and distribution in
NV centers in diamonds~\cite{DoldeNatComm14}. 

A second successful control strategy described in
Sec.~\ref{subsec:strategies} is based on decoherence-free subspaces and
noiseless subsystems. These are somewhat less often used than
dynamical decoupling, mainly due to the difficulty of identifying
them for more complex systems. While direct identification of
decoherence-free subspaces is hampered by presence of numerous traps
in the search space~\cite{WangJacobsPRA13}, quantum optimal control
may be used to dynamically identify them. Indeed, optimization of an
open system's dynamics for 
targets that rely on quantum coherence is intrinsically biased toward
those subspaces in Hilbert (or Liouville) space that are least
affected by decoherence~\cite{KatzPRL07}. For example, transfer of 
coherence and polarization between coupled heteronuclear
spins was improved by cross-correlated
relaxation optimized pulse (CROP)
sequences and  relaxation optimized pulse elements
(ROPE)~\cite{KhanejaPNAS03,KhanejaJMR03,StefanatosPRA04,GershenzonJMR08}. 
Optimization can take both longitudinal and transveral
relaxation into account~\cite{StefanatosPRA04}. The
underlying mechanism  was revealed to consist in  tuning 
cross-correlated to autocorrelated relaxation
rates~\cite{KhanejaPNAS03}. Counterintuitively, maximum polarization
transfer between coupled spins was achieved with sequences that are
longer than conventional ones~\cite{KhanejaJMR03}, highlighting the
importance to include the dissipative dynamics in the
optimization. 

Finally, if all regions in Hilbert space are similarly affected by
decoherence, 
an obvious control strategy consists in beating decoherence by the
fastest possible operation. This strategy is faced, however, with the so-called
quantum speed limit, i.e., a fundamental bound on the shortest
operation time~\cite{GiovannettiPRA03,LevitinPRL09}. For a two-level
system, it can be estimated analytically~\cite{HegerfeldtPRL13}. For
more complex systems, optimal control theory can be used as a tool to
both identify the quantum speed limit and determine controls that
drive the system at the quantum speed limit~\cite{CanevaPRL09}. 
For example, the shortest possible duration of entangling quantum
gates was determined for cold, trapped atoms~\cite{GoerzJPB11} and for
superconducting qubits~\cite{GoerzNJP14,GoerzPhD}. In quantum dots, 
phonon-assisted decoherence was
minimized~\cite{HohenesterPRL04,HohenesterJPB07}. 
Interestingly, presence of the environment may improve 
the quantum speed limit~\cite{DeffnerPRL13,CimmarustiPRL15}. 
This has not yet been explored systematically but could be done, using 
quantum optimal control. 

\subsection{Cooling and quantum reservoir engineering}
\label{subsec:cooling}

The example of cooling~\cite{BartanaJCP97,TannorJPCA99,%
SchmidtPRL11,SchmidtPS12,%
MachnesPRL12,ReichNJP13,RahmaniPRA13,MukherjeePRA13,MukherjeeNJP15}
was already taken as 
reference to introduce optimal control of open quantum systems in
Sec.~\ref{subsec:OCT}. Since cooling changes the purity of the state, 
it relies on the presence of the environment. When using the algorithm
outlined in Sec.~\ref{subsec:OCT}, the control that drives the cooling
process is determined for each initial state
separately. Alternatively, one may seek a control that will lead to
cooling irrespective of the initial state~\cite{ReichNJP13}. Such an
approach is particularly useful if the timescales of the coherent and
the dissipative dynamics is different, as in the case of optical
pumping~\cite{ViteauSci08} which consists of many repeated cycles of
excitation and spontaneous emission. The theoretical framework lends
itself to generalization to quantum reservoir engineering which is why
it is outlined in more detail in the following.

The idea is to start from an orthogonal
basis for the space on which the initial states are defined and ensure
transitions favorable to cooling for every basis
state~\cite{ReichNJP13}. In the case of optical pumping, these are
transitions into states from which spontaneous emission preferrably
occurs to the cooling target. As is usual in quantum optimal control,
this control task is stated in terms of a 
yield functional. Additionally, the control should not
excite any population which has already been accumulated in the
cooling target state. In other words, the cooling target should remain
a steady state of the evolution. This requirement is translated into a
second term in the optimization functional, besides the
yield. Moreover, it is often not possible to cool arbitrary initial
states, due to limitations on control bandwidth. Instead, one can seek
to cool states which are defined on a certain subspace. This results
in an additional term in the optimization functional that suppresses
leakage from this subspace in order to keep the cooling cycle 
closed. Finally, one needs to guarantee that for all basis states
cooling occurs with the same efficiency; otherwise the cooling might
get stuck. This can be achieved either
by imposing symmetric excitation of all basis states or by having all
basis states form an 'assembly line', i.e., enforce one specific
excitation pathway for all states. All requirements need to be met
simultanously and, consequently, the optimization
functional consists of four terms -- one for the yield, one of the
steady state, one to suppress leakage and one to ensure symmetric
excitation or enforce a specific excitation pathway for all
states~\cite{ReichNJP13}. 

Application of this optimization framework showed that laser cooling
of molecular vibrations is possible even in cases where the molecular
structure favors heating rather than cooling~\cite{ReichNJP13}. It
also answers the question about the minimal requirement on the
molecular structure to realize, with shaped pulses, cooling instead of
heating, assuming no constraints on the control -- existence of one 
state which undergoes spontaneous emission with moderate probability
into the cooling target. 

If the molecular structure is favorable to cooling vibrations, an
optimized laser pulse 
results in a substantially smaller number of cooling cycles than an
unshaped pulse~\cite{ReichNJP13}. A similar speed up of the cooling
due to pulse sequences 
obtained from quantum optimal control theory has also been reported
for an optomechanical resonator~\cite{MachnesPRL12} and for trapped,
quasi-condensed cold atoms~\cite{RahmaniPRA13}. 

Laser cooling can be viewed as a particular example of quantum
reservoir engineering~\cite{PoyatosPRL96,PielawaPRL07,DiehlNatPhys08},
where a desired state becomes the 'ground state' of a
driven dissipative system. It holds the promise of a particularly
robust control strategy. However, applications of quantum reservoir
engineering have been limited to quantum optics to date. In a condensed phase
settings, both desired and undesired dissipative channels come into
play, and non-Markovian effects may occur. Thus, quantum reservoir
engineering in the condensed phase represents a 
challenging control problem. 

As with any control problem, two questions
need to be tackled -- that of controllability and that of control
synthesis, i.e., what states are attainable and how can the
necessary driving be realized. The first question has been answered
for generic models, such as a two- and a four-level system and a
harmonic oscillator, that undergo Markovian
dynamics~\cite{SchirmerPRA10}. The obtained understanding of controllability
can be exploited to construct dissipative channels that allow for the
robust generation of long-distance entanglement~\cite{Motzoi16}. 
The question which states are attainable in the presence of additional
undesired dissipative channels, a generic feature of any condensed
phase setting, has not yet been tackled to date. 
A possible influence of non-Markovian effects on the reachable states
has also not yet been addressed. 

Control synthesis may be achieved in several ways. First, incoherent
control by the environment, for example, via certain population
distributions in the environmental modes, may be used to control the
system~\cite{PechenPRA06}. However, this contradicts the assumption
that the environment by definition is uncontrollable. Alternatively, 
measurements effectively lead to dissipative dynamics and may thus be
used to generate desired dissipation~\cite{PechenPRA06b,BurgarthNatComm14,Arenz16}. They may be
augmented by suitably tailored coherent excitation for more effective
control~\cite{ShuangJCP07}. Quantum reservoir engineering may also be
formulated as an optimization problem where the target is a certain
desired steady state. An optimization algorithm  is obtained by
generalizing the theoretical framework for laser
cooling~\cite{ReichNJP13} outlined above. However, the search space is
even larger 
than for a standard quantum control problem; and the efficient 
numerical implementation is an open challenge. Meeting this challenge
would allow for exploring quantum reservoir engineering in the
presence of undesired dissipation typical for condensed phase settings
and in the non-Markovian regime.

\subsection{Exploiting non-Markovianity for quantum control}
\label{subsec:exploiting}

It was shown already several years ago that non-Markovian evolution
may ease control~\cite{RebentrostPRL09,FloetherNJP12}, and 
cooperative effects of dissipation and driving were
reported~\cite{SchmidtPRL11}. However, a more thorough understanding
of the nature of non-Markovianity was required to understand its
interplay with quantum control. As described in
Sec.~\ref{subsec:MarknonMark}, non-Markovianity has in the meantime
been characterized in terms of information flow from the environment to
the system~\cite{BreuerPRL09}, increase of correlations in a bipartite
system~\cite{RivasPRL10}, or re-expansion of the volume of
accessible states~\cite{LorenzoPRA13}, among others. 
The important point in the
context of quantum control is that each of these measures holds a
promise for better control:
Correlations between system and environment 
may improve fidelities of single qubit gates~\cite{RebentrostPRL09},
cooperative effects of control and dissipation may allow for entropy
export and thus cooling~\cite{SchmidtPRL11,MukherjeeNJP15}; and harnessing
non-Markovianity may enhance the efficiency of quantum 
information processing and communication~\cite{LaineSciRep14,BylickaSciRep14}. 

A first example  of exploiting non-Markovianity for quantum
control  was reported in Ref.~\cite{ReichSciRep15}, 
showing that, for an anharmonic ladder system, the  environment may be
utilized to extend possible operations from $SO(N)$ to
$SU(N)$. Presence of at least one two-level defect in the environment 
that is sufficiently isolated and sufficiently
strongly coupled to the system was identified as prerequisite for the
observed controllability enhancement.   Such conditions are found in
current experiments with 
  superconducting circuits and other systems which are immersed in a
  small, 'natural' spin bath, for example color centers in diamond.


The limited number of optimal control studies of open quantum systems
with non-Markovian
dynamics~\cite{RebentrostPRL09,FloetherNJP12,SchmidtPRL11,ReichSciRep15,MukherjeeNJP15} 
testifies to the fact that control of these systems remains largely
uncharted territory. 
The full potential of the specific features of non-Markovian dynamics
for quantum control remains yet to be explored. 
Open questions include, for example, 
how the build-up of memory influences control; whether specific
features of the spectral density can be exploited for control, and if
so, how.  The tools for performing these studies, both in terms of
simulating non-Markovian dynamics~\cite{deVegaRMP16} and carrying out
optimal control calculations, cf. Sec.~\ref{subsec:OCT}, have been
developed and are there to be used.

\section{Conclusions} 
\label{sec:concl}
The present review has been focused on control of open quantum systems
as they are encountered in the field of quantum technologies. It will
be concluded by briefly mentioning examples from other fields of
current interest: Quantum optimal control for open systems has been
employed in the context of quantum thermodynamics, in order to
determine the optimal efficiency of a noisy heat
engine~\cite{StefanatosPRE14}; biological chromophore
complexes, in order to maximize exciton
transfer~\cite{BrueggemannJPPA07,HoyerNJP14}; 
molecules immersed in dissipative media, in order to maximally align
them with respect to a laboratory
axis~\cite{PelzerJCP08,ZhdanovPRA11,VieillardPRA13};    
molecular junctions, in order to control the current, shot noise and
Fano factors~\cite{KleinekathoeferEPJB10}; as well as chemical reaction
dynamics~\cite{SugnyJPPA07,ChenelJPCA12,ChenelJCP15}, including 
charge transfer in molecules~\cite{TremblayPRA08}, and 
surface
photochemistry~\cite{MishimaJCP09,AsplundPRL11,TremblayPRB12}. 
The numerous applications attest to the maturity as well as 
versatility of the quantum control toolbox~\cite{GlaserEPJD15}.

At this stage, three rules for controlling open quantum systems may
tentatively be formulated: 
\begin{enumerate}
\item If the desired operation shall keep pure states pure,
  Markovian dynamics are unwanted. The effect of the
environment in this case 
is detrimental. A suitable strategy is then to perform any
desired operation as fast as possible. Quantum optimal control theory
is a viable tool to determine both the shortest operation time and the
control that drives the desired dynamics. The actual dissipative
processes may be neglected during the optimization for computational
simplicity. Explicit account of the dissipative processes comes at a
significantly larger numerical cost but allows for identifying
subspaces which are less affected by or even immune to decoherence. 
\item If the desired operation changes the purity of the system's
  state, presence of the environment is necessary for realizing the
  control target. In this case, Markovian dynamics may be 
  desired: The control target is reachable if it is a fixed
  state of the Liouvillian.  External control fields may be used to
  ensure that this is the case. The corresponding control strategy is
  referred to as quantum reservoir engineering. 
  If the Liouvillian has several fixed
  points, external fields may also be used to drive the dynamics to
  the desired one. The role of non-Markovian effects in this type of
  desired dissipation has not been explored to date. 
\item Non-Markovian dynamics in general may have both beneficial and
  detrimental effects on controlled quantum dynamics. Improved
  controllability is a first example of a benefit. It requires
  presence of a few strongly coupled and sufficiently
  isolated environmental modes which can effectively 
  act as ancillas~\footnote{An
    important   difference to true ancilla systems is controllability:
    Environmental modes are by definition uncontrollable and can only
    be accessed via their coupling to the system.}, and use of quantum
  optimal control for properly exploiting these modes. 
  An improved quantum speed limit is a second
  example.
\end{enumerate}

While the number of examples for successful control of open quantum
systems is growing, our current understanding of controllability
and the most promising control strategies for open quantum systems 
is still rather limited. In
particular, a thorough understanding of the role of non-Markovian
effects is lacking to date, and it is currently still unknown which
features of non-Markovianity 
can be exploited for quantum control. 

Investigation of a larger range of
models, with both small and large baths, consisting of harmonic modes and
spins, and  a systematic analysis of non-Markovianity
may elucidate this question. 
Such an improved understanding would not only be crucial for advancing
quantum technologies but would also be beneficial for adjacent fields
such as condensed matter physics or chemical reaction dynamics.

\begin{acknowledgments}
  I would like to thank Ugo Boscain, Steffen
  Glaser, Dominique Sugny, and David Tannor for their comments on the manuscript.
  Recent advances in the control of open quantum systems in my own group
  were made possible by the work of 
  Michael Goerz, Giulia Gualdi and Daniel Reich whom I would like to
  thank  for their dedication and perserverance. 
\end{acknowledgments}


\begin{thebibliography}{210}
\expandafter\ifx\csname natexlab\endcsname\relax\def\natexlab#1{#1}\fi
\expandafter\ifx\csname bibnamefont\endcsname\relax
  \def\bibnamefont#1{#1}\fi
\expandafter\ifx\csname bibfnamefont\endcsname\relax
  \def\bibfnamefont#1{#1}\fi
\expandafter\ifx\csname citenamefont\endcsname\relax
  \def\citenamefont#1{#1}\fi
\expandafter\ifx\csname url\endcsname\relax
  \def\url#1{\texttt{#1}}\fi
\expandafter\ifx\csname urlprefix\endcsname\relax\def\urlprefix{URL }\fi
\providecommand{\bibinfo}[2]{#2}
\providecommand{\eprint}[2][]{\url{#2}}

\bibitem[{\citenamefont{Glaser et~al.}(2015)\citenamefont{Glaser, Boscain,
  Calarco, Koch, K\"ockenberger, Kosloff, Kuprov, Luy, Schirmer,
  Schulte-Herbr\"uggen et~al.}}]{GlaserEPJD15}
\bibinfo{author}{\bibfnamefont{S.~J.} \bibnamefont{Glaser}},
  \bibinfo{author}{\bibfnamefont{U.}~\bibnamefont{Boscain}},
  \bibinfo{author}{\bibfnamefont{T.}~\bibnamefont{Calarco}},
  \bibinfo{author}{\bibfnamefont{C.~P.} \bibnamefont{Koch}},
  \bibinfo{author}{\bibfnamefont{W.}~\bibnamefont{K\"ockenberger}},
  \bibinfo{author}{\bibfnamefont{R.}~\bibnamefont{Kosloff}},
  \bibinfo{author}{\bibfnamefont{I.}~\bibnamefont{Kuprov}},
  \bibinfo{author}{\bibfnamefont{B.}~\bibnamefont{Luy}},
  \bibinfo{author}{\bibfnamefont{S.}~\bibnamefont{Schirmer}},
  \bibinfo{author}{\bibfnamefont{T.}~\bibnamefont{Schulte-Herbr\"uggen}},
  \bibnamefont{et~al.}, \bibinfo{journal}{Eur. Phys. J. D}
  \textbf{\bibinfo{volume}{69}} (\bibinfo{year}{2015}).

\bibitem[{\citenamefont{D'Alessandro}(2007)}]{DAlessandroBook}
\bibinfo{author}{\bibfnamefont{D.}~\bibnamefont{D'Alessandro}},
  \emph{\bibinfo{title}{Introduction to Quantum Control and Dynamics}}
  (\bibinfo{publisher}{Chapman \& Hall/CRC}, \bibinfo{year}{2007}).

\bibitem[{\citenamefont{Shapiro and Brumer}(2012)}]{ShapiroBook}
\bibinfo{author}{\bibfnamefont{M.}~\bibnamefont{Shapiro}} \bibnamefont{and}
  \bibinfo{author}{\bibfnamefont{P.}~\bibnamefont{Brumer}},
  \emph{\bibinfo{title}{Quantum Control of Molecular Processes}}
  (\bibinfo{publisher}{Wiley Interscience}, \bibinfo{year}{2012}),
  \bibinfo{edition}{2nd} ed.

\bibitem[{\citenamefont{Pontryagin et~al.}(1962)\citenamefont{Pontryagin,
  Boltyanskii, Gamkrelidze, and Mishchenko}}]{PontryaginBook}
\bibinfo{author}{\bibfnamefont{L.~S.} \bibnamefont{Pontryagin}},
  \bibinfo{author}{\bibfnamefont{V.~G.} \bibnamefont{Boltyanskii}},
  \bibinfo{author}{\bibfnamefont{R.~V.} \bibnamefont{Gamkrelidze}},
  \bibnamefont{and} \bibinfo{author}{\bibfnamefont{E.~F.}
  \bibnamefont{Mishchenko}}, \emph{\bibinfo{title}{The Mathematical Theory of
  Optimal Processes}} (\bibinfo{publisher}{Wiley}, \bibinfo{address}{New York},
  \bibinfo{year}{1962}).

\bibitem[{\citenamefont{Krotov}(1996)}]{KrotovBook}
\bibinfo{author}{\bibfnamefont{V.~F.} \bibnamefont{Krotov}},
  \emph{\bibinfo{title}{Global Methods in Optimal Control}}
  (\bibinfo{publisher}{Marcel Dekker, New York}, \bibinfo{year}{1996}).

\bibitem[{\citenamefont{Conolly et~al.}(1986)\citenamefont{Conolly, Nishimura,
  and Macovski}}]{Conolly1986}
\bibinfo{author}{\bibfnamefont{S.}~\bibnamefont{Conolly}},
  \bibinfo{author}{\bibfnamefont{D.}~\bibnamefont{Nishimura}},
  \bibnamefont{and} \bibinfo{author}{\bibfnamefont{A.}~\bibnamefont{Macovski}},
  \bibinfo{journal}{IEEE Trans. Med. Imaging} \textbf{\bibinfo{volume}{MI-5}},
  \bibinfo{pages}{106} (\bibinfo{year}{1986}).

\bibitem[{\citenamefont{Mao et~al.}(1986)\citenamefont{Mao, Mareci, Scott, and
  Andrew}}]{Mao1986}
\bibinfo{author}{\bibfnamefont{J.}~\bibnamefont{Mao}},
  \bibinfo{author}{\bibfnamefont{T.~H.} \bibnamefont{Mareci}},
  \bibinfo{author}{\bibfnamefont{K.~N.} \bibnamefont{Scott}}, \bibnamefont{and}
  \bibinfo{author}{\bibfnamefont{E.}~\bibnamefont{Andrew}},
  \bibinfo{journal}{J. Magn. Reson.} \textbf{\bibinfo{volume}{70}},
  \bibinfo{pages}{310} (\bibinfo{year}{1986}).

\bibitem[{\citenamefont{Tannor and Rice}(1985)}]{TannorJCP85}
\bibinfo{author}{\bibfnamefont{D.}~\bibnamefont{Tannor}} \bibnamefont{and}
  \bibinfo{author}{\bibfnamefont{S.}~\bibnamefont{Rice}}, \bibinfo{journal}{J.
  Chem. Phys.} \textbf{\bibinfo{volume}{83}}, \bibinfo{pages}{5013}
  (\bibinfo{year}{1985}).

\bibitem[{\citenamefont{Peirce et~al.}(1988)\citenamefont{Peirce, Dahleh, and
  Rabitz}}]{PeircePRA88}
\bibinfo{author}{\bibfnamefont{A.~P.} \bibnamefont{Peirce}},
  \bibinfo{author}{\bibfnamefont{M.~A.} \bibnamefont{Dahleh}},
  \bibnamefont{and} \bibinfo{author}{\bibfnamefont{H.}~\bibnamefont{Rabitz}},
  \bibinfo{journal}{Phys. Rev. A} \textbf{\bibinfo{volume}{37}},
  \bibinfo{pages}{4950} (\bibinfo{year}{1988}).

\bibitem[{\citenamefont{Kosloff et~al.}(1989)\citenamefont{Kosloff, Rice,
  Gaspard, Tersigni, and Tannor}}]{RonnieDancing89}
\bibinfo{author}{\bibfnamefont{R.}~\bibnamefont{Kosloff}},
  \bibinfo{author}{\bibfnamefont{S.}~\bibnamefont{Rice}},
  \bibinfo{author}{\bibfnamefont{P.}~\bibnamefont{Gaspard}},
  \bibinfo{author}{\bibfnamefont{S.}~\bibnamefont{Tersigni}}, \bibnamefont{and}
  \bibinfo{author}{\bibfnamefont{D.}~\bibnamefont{Tannor}},
  \bibinfo{journal}{Chem. Phys.} \textbf{\bibinfo{volume}{139}},
  \bibinfo{pages}{201} (\bibinfo{year}{1989}).

\bibitem[{\citenamefont{Tannor et~al.}(1992)\citenamefont{Tannor, Kazakov, and
  Orlov}}]{Tannor92}
\bibinfo{author}{\bibfnamefont{D.}~\bibnamefont{Tannor}},
  \bibinfo{author}{\bibfnamefont{V.}~\bibnamefont{Kazakov}}, \bibnamefont{and}
  \bibinfo{author}{\bibfnamefont{V.}~\bibnamefont{Orlov}}, in
  \emph{\bibinfo{booktitle}{Time-dependent quantum molecular dynamics}}, edited
  by \bibinfo{editor}{\bibfnamefont{J.}~\bibnamefont{Broeckhove}}
  \bibnamefont{and}
  \bibinfo{editor}{\bibfnamefont{L.}~\bibnamefont{Lathouwers}}
  (\bibinfo{publisher}{Plenum}, \bibinfo{year}{1992}), pp.
  \bibinfo{pages}{347--360}.

\bibitem[{\citenamefont{Gross et~al.}(1992)\citenamefont{Gross, Neuhauser, and
  Rabitz}}]{GrossJCP92}
\bibinfo{author}{\bibfnamefont{P.}~\bibnamefont{Gross}},
  \bibinfo{author}{\bibfnamefont{D.}~\bibnamefont{Neuhauser}},
  \bibnamefont{and} \bibinfo{author}{\bibfnamefont{H.}~\bibnamefont{Rabitz}},
  \bibinfo{journal}{J. Chem. Phys.} \textbf{\bibinfo{volume}{96}},
  \bibinfo{pages}{2834} (\bibinfo{year}{1992}).

\bibitem[{\citenamefont{Soml\'{o}i et~al.}(1993)\citenamefont{Soml\'{o}i,
  Kazakovski, and Tannor}}]{SomloiCP93}
\bibinfo{author}{\bibfnamefont{J.}~\bibnamefont{Soml\'{o}i}},
  \bibinfo{author}{\bibfnamefont{V.~A.} \bibnamefont{Kazakovski}},
  \bibnamefont{and} \bibinfo{author}{\bibfnamefont{D.~J.}
  \bibnamefont{Tannor}}, \bibinfo{journal}{Chem. Phys.}
  \textbf{\bibinfo{volume}{172}}, \bibinfo{pages}{85} (\bibinfo{year}{1993}).

\bibitem[{\citenamefont{Assion et~al.}(1998)\citenamefont{Assion, Baumert,
  Bergt, Brixner, Kiefer, Seyfried, Strehle, and Gerber}}]{AssionSci98}
\bibinfo{author}{\bibfnamefont{A.}~\bibnamefont{Assion}},
  \bibinfo{author}{\bibfnamefont{T.}~\bibnamefont{Baumert}},
  \bibinfo{author}{\bibfnamefont{X.}~\bibnamefont{Bergt}},
  \bibinfo{author}{\bibfnamefont{T.}~\bibnamefont{Brixner}},
  \bibinfo{author}{\bibfnamefont{X.}~\bibnamefont{Kiefer}},
  \bibinfo{author}{\bibfnamefont{V.}~\bibnamefont{Seyfried}},
  \bibinfo{author}{\bibfnamefont{X.}~\bibnamefont{Strehle}}, \bibnamefont{and}
  \bibinfo{author}{\bibfnamefont{G.}~\bibnamefont{Gerber}},
  \bibinfo{journal}{Science} \textbf{\bibinfo{volume}{282}},
  \bibinfo{pages}{919} (\bibinfo{year}{1998}).

\bibitem[{\citenamefont{Judson and Rabitz}(1992)}]{JudsonPRL92}
\bibinfo{author}{\bibfnamefont{R.~S.} \bibnamefont{Judson}} \bibnamefont{and}
  \bibinfo{author}{\bibfnamefont{H.}~\bibnamefont{Rabitz}},
  \bibinfo{journal}{Phys. Rev. Lett.} \textbf{\bibinfo{volume}{68}},
  \bibinfo{pages}{1500} (\bibinfo{year}{1992}).

\bibitem[{\citenamefont{Sch\"afer-Bung
  et~al.}(2004)\citenamefont{Sch\"afer-Bung, Mitri\'{c},
  Bona\v{c}i\'{c}-Kouteck\'{y}, Bartelt, Lupulescu, Lindinger, Vajda, Weber,
  and W\"oste}}]{BorisJPC04}
\bibinfo{author}{\bibfnamefont{B.}~\bibnamefont{Sch\"afer-Bung}},
  \bibinfo{author}{\bibfnamefont{R.}~\bibnamefont{Mitri\'{c}}},
  \bibinfo{author}{\bibfnamefont{V.}~\bibnamefont{Bona\v{c}i\'{c}-Kouteck\'{y}}},
  \bibinfo{author}{\bibfnamefont{A.}~\bibnamefont{Bartelt}},
  \bibinfo{author}{\bibfnamefont{C.}~\bibnamefont{Lupulescu}},
  \bibinfo{author}{\bibfnamefont{A.}~\bibnamefont{Lindinger}},
  \bibinfo{author}{\bibfnamefont{v.}~\bibnamefont{Vajda}},
  \bibinfo{author}{\bibfnamefont{S.~M.} \bibnamefont{Weber}}, \bibnamefont{and}
  \bibinfo{author}{\bibfnamefont{L.}~\bibnamefont{W\"oste}},
  \bibinfo{journal}{J. Phys. Chem. A} \textbf{\bibinfo{volume}{108}},
  \bibinfo{pages}{4175} (\bibinfo{year}{2004}).

\bibitem[{\citenamefont{Zhu et~al.}(1998)\citenamefont{Zhu, Botina, and
  Rabitz}}]{ZhuJCP98}
\bibinfo{author}{\bibfnamefont{W.}~\bibnamefont{Zhu}},
  \bibinfo{author}{\bibfnamefont{J.}~\bibnamefont{Botina}}, \bibnamefont{and}
  \bibinfo{author}{\bibfnamefont{H.}~\bibnamefont{Rabitz}},
  \bibinfo{journal}{J. Chem. Phys.} \textbf{\bibinfo{volume}{108}},
  \bibinfo{pages}{1953} (\bibinfo{year}{1998}).

\bibitem[{\citenamefont{Weiner}(2000)}]{WeinerRSI00}
\bibinfo{author}{\bibfnamefont{A.~M.} \bibnamefont{Weiner}},
  \bibinfo{journal}{Rev. Sci. Instrum.} \textbf{\bibinfo{volume}{71}},
  \bibinfo{pages}{1929} (\bibinfo{year}{2000}).

\bibitem[{\citenamefont{Wright et~al.}(2004)\citenamefont{Wright, Gould, and
  Gensemer}}]{WrightRSI04}
\bibinfo{author}{\bibfnamefont{M.~J.} \bibnamefont{Wright}},
  \bibinfo{author}{\bibfnamefont{P.~L.} \bibnamefont{Gould}}, \bibnamefont{and}
  \bibinfo{author}{\bibfnamefont{S.~D.} \bibnamefont{Gensemer}},
  \bibinfo{journal}{Rev. Sci. Instrum.} \textbf{\bibinfo{volume}{75}},
  \bibinfo{pages}{4718} (\bibinfo{year}{2004}).

\bibitem[{\citenamefont{Motzoi et~al.}(2011)\citenamefont{Motzoi, Gambetta,
  Merkel, and Wilhelm}}]{MotzoiPRA11}
\bibinfo{author}{\bibfnamefont{F.}~\bibnamefont{Motzoi}},
  \bibinfo{author}{\bibfnamefont{J.~M.} \bibnamefont{Gambetta}},
  \bibinfo{author}{\bibfnamefont{S.~T.} \bibnamefont{Merkel}},
  \bibnamefont{and} \bibinfo{author}{\bibfnamefont{F.~K.}
  \bibnamefont{Wilhelm}}, \bibinfo{journal}{Phys. Rev. A}
  \textbf{\bibinfo{volume}{84}}, \bibinfo{pages}{022307}
  (\bibinfo{year}{2011}).

\bibitem[{\citenamefont{J\"ager and Hohenester}(2013)}]{JaegerPRA13}
\bibinfo{author}{\bibfnamefont{G.}~\bibnamefont{J\"ager}} \bibnamefont{and}
  \bibinfo{author}{\bibfnamefont{U.}~\bibnamefont{Hohenester}},
  \bibinfo{journal}{Phys. Rev. A} \textbf{\bibinfo{volume}{88}},
  \bibinfo{pages}{035601} (\bibinfo{year}{2013}).

\bibitem[{\citenamefont{Kobzar et~al.}(2005)\citenamefont{Kobzar, Luy, Khaneja,
  and Glaser}}]{KobzarJMR05}
\bibinfo{author}{\bibfnamefont{K.}~\bibnamefont{Kobzar}},
  \bibinfo{author}{\bibfnamefont{B.}~\bibnamefont{Luy}},
  \bibinfo{author}{\bibfnamefont{N.}~\bibnamefont{Khaneja}}, \bibnamefont{and}
  \bibinfo{author}{\bibfnamefont{S.~J.} \bibnamefont{Glaser}},
  \bibinfo{journal}{J. Magn. Reson.} \textbf{\bibinfo{volume}{173}},
  \bibinfo{pages}{229 } (\bibinfo{year}{2005}).

\bibitem[{\citenamefont{Skinner et~al.}(2006)\citenamefont{Skinner, Kobzar,
  Luy, Bendall, Bermel, Khaneja, and Glaser}}]{SkinnerJMR06}
\bibinfo{author}{\bibfnamefont{T.~E.} \bibnamefont{Skinner}},
  \bibinfo{author}{\bibfnamefont{K.}~\bibnamefont{Kobzar}},
  \bibinfo{author}{\bibfnamefont{B.}~\bibnamefont{Luy}},
  \bibinfo{author}{\bibfnamefont{M.~R.} \bibnamefont{Bendall}},
  \bibinfo{author}{\bibfnamefont{W.}~\bibnamefont{Bermel}},
  \bibinfo{author}{\bibfnamefont{N.}~\bibnamefont{Khaneja}}, \bibnamefont{and}
  \bibinfo{author}{\bibfnamefont{S.~J.} \bibnamefont{Glaser}},
  \bibinfo{journal}{J. Magn. Reson.} \textbf{\bibinfo{volume}{179}},
  \bibinfo{pages}{241 } (\bibinfo{year}{2006}).

\bibitem[{\citenamefont{Nimbalkar et~al.}(2013)\citenamefont{Nimbalkar, Luy,
  Skinner, Neves, Gershenzon, Kobzar, Bermel, and Glaser}}]{NimbalkarJMR13}
\bibinfo{author}{\bibfnamefont{M.}~\bibnamefont{Nimbalkar}},
  \bibinfo{author}{\bibfnamefont{B.}~\bibnamefont{Luy}},
  \bibinfo{author}{\bibfnamefont{T.~E.} \bibnamefont{Skinner}},
  \bibinfo{author}{\bibfnamefont{J.~L.} \bibnamefont{Neves}},
  \bibinfo{author}{\bibfnamefont{N.~I.} \bibnamefont{Gershenzon}},
  \bibinfo{author}{\bibfnamefont{K.}~\bibnamefont{Kobzar}},
  \bibinfo{author}{\bibfnamefont{W.}~\bibnamefont{Bermel}}, \bibnamefont{and}
  \bibinfo{author}{\bibfnamefont{S.~J.} \bibnamefont{Glaser}},
  \bibinfo{journal}{J. Magn. Reson.} \textbf{\bibinfo{volume}{228}},
  \bibinfo{pages}{16 } (\bibinfo{year}{2013}).

\bibitem[{\citenamefont{Palao and Kosloff}(2002)}]{PalaoPRL02}
\bibinfo{author}{\bibfnamefont{J.~P.} \bibnamefont{Palao}} \bibnamefont{and}
  \bibinfo{author}{\bibfnamefont{R.}~\bibnamefont{Kosloff}},
  \bibinfo{journal}{Phys. Rev. Lett.} \textbf{\bibinfo{volume}{89}},
  \bibinfo{pages}{188301} (\bibinfo{year}{2002}).

\bibitem[{\citenamefont{Tesch and de~Vivie-Riedle}(2002)}]{TeschPRL02}
\bibinfo{author}{\bibfnamefont{C.}~\bibnamefont{Tesch}} \bibnamefont{and}
  \bibinfo{author}{\bibfnamefont{R.}~\bibnamefont{de~Vivie-Riedle}},
  \bibinfo{journal}{Phys. Rev. Lett.} \textbf{\bibinfo{volume}{89}},
  \bibinfo{pages}{157901} (\bibinfo{year}{2002}).

\bibitem[{\citenamefont{Palao and Kosloff}(2003)}]{PalaoPRA03}
\bibinfo{author}{\bibfnamefont{J.~P.} \bibnamefont{Palao}} \bibnamefont{and}
  \bibinfo{author}{\bibfnamefont{R.}~\bibnamefont{Kosloff}},
  \bibinfo{journal}{Phys. Rev. A} \textbf{\bibinfo{volume}{68}},
  \bibinfo{pages}{062308} (\bibinfo{year}{2003}).

\bibitem[{\citenamefont{Platzer et~al.}(2010)\citenamefont{Platzer, Mintert,
  and Buchleitner}}]{PlatzerPRL10}
\bibinfo{author}{\bibfnamefont{F.}~\bibnamefont{Platzer}},
  \bibinfo{author}{\bibfnamefont{F.}~\bibnamefont{Mintert}}, \bibnamefont{and}
  \bibinfo{author}{\bibfnamefont{A.}~\bibnamefont{Buchleitner}},
  \bibinfo{journal}{Phys. Rev. Lett.} \textbf{\bibinfo{volume}{105}},
  \bibinfo{pages}{020501} (\bibinfo{year}{2010}).

\bibitem[{\citenamefont{M\"uller et~al.}(2011)\citenamefont{M\"uller, Reich,
  Murphy, Yuan, Vala, Whaley, Calarco, and Koch}}]{MuellerPRA11}
\bibinfo{author}{\bibfnamefont{M.~M.} \bibnamefont{M\"uller}},
  \bibinfo{author}{\bibfnamefont{D.~M.} \bibnamefont{Reich}},
  \bibinfo{author}{\bibfnamefont{M.}~\bibnamefont{Murphy}},
  \bibinfo{author}{\bibfnamefont{H.}~\bibnamefont{Yuan}},
  \bibinfo{author}{\bibfnamefont{J.}~\bibnamefont{Vala}},
  \bibinfo{author}{\bibfnamefont{K.~B.} \bibnamefont{Whaley}},
  \bibinfo{author}{\bibfnamefont{T.}~\bibnamefont{Calarco}}, \bibnamefont{and}
  \bibinfo{author}{\bibfnamefont{C.~P.} \bibnamefont{Koch}},
  \bibinfo{journal}{Phys. Rev. A} \textbf{\bibinfo{volume}{84}},
  \bibinfo{pages}{042315} (\bibinfo{year}{2011}).

\bibitem[{\citenamefont{Caneva et~al.}(2012)\citenamefont{Caneva, Calarco, and
  Montangero}}]{CanevaNJP12}
\bibinfo{author}{\bibfnamefont{T.}~\bibnamefont{Caneva}},
  \bibinfo{author}{\bibfnamefont{T.}~\bibnamefont{Calarco}}, \bibnamefont{and}
  \bibinfo{author}{\bibfnamefont{S.}~\bibnamefont{Montangero}},
  \bibinfo{journal}{New J. Phys.} \textbf{\bibinfo{volume}{14}},
  \bibinfo{pages}{093041} (\bibinfo{year}{2012}).

\bibitem[{\citenamefont{Egger and Wilhelm}(2014)}]{EggerPRA14}
\bibinfo{author}{\bibfnamefont{D.~J.} \bibnamefont{Egger}} \bibnamefont{and}
  \bibinfo{author}{\bibfnamefont{F.~K.} \bibnamefont{Wilhelm}},
  \bibinfo{journal}{Phys. Rev. A} \textbf{\bibinfo{volume}{90}},
  \bibinfo{pages}{052331} (\bibinfo{year}{2014}).

\bibitem[{\citenamefont{Poulsen et~al.}(2010)\citenamefont{Poulsen, Sklarz,
  Tannor, and Calarco}}]{PoulsenPRA10}
\bibinfo{author}{\bibfnamefont{U.~V.} \bibnamefont{Poulsen}},
  \bibinfo{author}{\bibfnamefont{S.}~\bibnamefont{Sklarz}},
  \bibinfo{author}{\bibfnamefont{D.}~\bibnamefont{Tannor}}, \bibnamefont{and}
  \bibinfo{author}{\bibfnamefont{T.}~\bibnamefont{Calarco}},
  \bibinfo{journal}{Phys. Rev. A} \textbf{\bibinfo{volume}{82}},
  \bibinfo{pages}{012339} (\bibinfo{year}{2010}).

\bibitem[{\citenamefont{Treutlein et~al.}(2006)\citenamefont{Treutlein,
  H\"ansch, Reichel, Negretti, Cirone, and Calarco}}]{TreutleinPRA06}
\bibinfo{author}{\bibfnamefont{P.}~\bibnamefont{Treutlein}},
  \bibinfo{author}{\bibfnamefont{T.~W.} \bibnamefont{H\"ansch}},
  \bibinfo{author}{\bibfnamefont{J.}~\bibnamefont{Reichel}},
  \bibinfo{author}{\bibfnamefont{A.}~\bibnamefont{Negretti}},
  \bibinfo{author}{\bibfnamefont{M.~A.} \bibnamefont{Cirone}},
  \bibnamefont{and} \bibinfo{author}{\bibfnamefont{T.}~\bibnamefont{Calarco}},
  \bibinfo{journal}{Phys. Rev. A} \textbf{\bibinfo{volume}{74}},
  \bibinfo{pages}{022312} (\bibinfo{year}{2006}).

\bibitem[{\citenamefont{Goerz et~al.}(2011)\citenamefont{Goerz, Calarco, and
  Koch}}]{GoerzJPB11}
\bibinfo{author}{\bibfnamefont{M.~H.} \bibnamefont{Goerz}},
  \bibinfo{author}{\bibfnamefont{T.}~\bibnamefont{Calarco}}, \bibnamefont{and}
  \bibinfo{author}{\bibfnamefont{C.~P.} \bibnamefont{Koch}},
  \bibinfo{journal}{J. Phys. B} \textbf{\bibinfo{volume}{44}},
  \bibinfo{pages}{154011} (\bibinfo{year}{2011}).

\bibitem[{\citenamefont{Goerz et~al.}(2014{\natexlab{a}})\citenamefont{Goerz,
  Halperin, Aytac, Koch, and Whaley}}]{GoerzPRA14}
\bibinfo{author}{\bibfnamefont{M.~H.} \bibnamefont{Goerz}},
  \bibinfo{author}{\bibfnamefont{E.~J.} \bibnamefont{Halperin}},
  \bibinfo{author}{\bibfnamefont{J.~M.} \bibnamefont{Aytac}},
  \bibinfo{author}{\bibfnamefont{C.~P.} \bibnamefont{Koch}}, \bibnamefont{and}
  \bibinfo{author}{\bibfnamefont{K.~B.} \bibnamefont{Whaley}},
  \bibinfo{journal}{Phys. Rev. A} \textbf{\bibinfo{volume}{90}},
  \bibinfo{pages}{032329} (\bibinfo{year}{2014}{\natexlab{a}}).

\bibitem[{\citenamefont{Scheuer et~al.}(2014)\citenamefont{Scheuer, Kong, Said,
  Chen, Kurz, Marseglia, Du, Hemmer, Montangero, Calarco
  et~al.}}]{ScheuerNJP14}
\bibinfo{author}{\bibfnamefont{J.}~\bibnamefont{Scheuer}},
  \bibinfo{author}{\bibfnamefont{X.}~\bibnamefont{Kong}},
  \bibinfo{author}{\bibfnamefont{R.~S.} \bibnamefont{Said}},
  \bibinfo{author}{\bibfnamefont{J.}~\bibnamefont{Chen}},
  \bibinfo{author}{\bibfnamefont{A.}~\bibnamefont{Kurz}},
  \bibinfo{author}{\bibfnamefont{L.}~\bibnamefont{Marseglia}},
  \bibinfo{author}{\bibfnamefont{J.}~\bibnamefont{Du}},
  \bibinfo{author}{\bibfnamefont{P.~R.} \bibnamefont{Hemmer}},
  \bibinfo{author}{\bibfnamefont{S.}~\bibnamefont{Montangero}},
  \bibinfo{author}{\bibfnamefont{T.}~\bibnamefont{Calarco}},
  \bibnamefont{et~al.}, \bibinfo{journal}{New J. Phys.}
  \textbf{\bibinfo{volume}{16}}, \bibinfo{pages}{093022}
  (\bibinfo{year}{2014}).

\bibitem[{\citenamefont{Waldherr et~al.}(2014)\citenamefont{Waldherr, Wang,
  Zaiser, Jamali, Schulte-Herbr\"uggen, Abe, Ohshima, Isoya, Du, Neumann
  et~al.}}]{WaldherrNat14}
\bibinfo{author}{\bibfnamefont{G.}~\bibnamefont{Waldherr}},
  \bibinfo{author}{\bibfnamefont{Y.}~\bibnamefont{Wang}},
  \bibinfo{author}{\bibfnamefont{S.}~\bibnamefont{Zaiser}},
  \bibinfo{author}{\bibfnamefont{M.}~\bibnamefont{Jamali}},
  \bibinfo{author}{\bibfnamefont{T.}~\bibnamefont{Schulte-Herbr\"uggen}},
  \bibinfo{author}{\bibfnamefont{H.}~\bibnamefont{Abe}},
  \bibinfo{author}{\bibfnamefont{T.}~\bibnamefont{Ohshima}},
  \bibinfo{author}{\bibfnamefont{J.}~\bibnamefont{Isoya}},
  \bibinfo{author}{\bibfnamefont{J.~F.} \bibnamefont{Du}},
  \bibinfo{author}{\bibfnamefont{P.}~\bibnamefont{Neumann}},
  \bibnamefont{et~al.}, \bibinfo{journal}{Nature}
  \textbf{\bibinfo{volume}{506}}, \bibinfo{pages}{204} (\bibinfo{year}{2014}).

\bibitem[{\citenamefont{Rebentrost et~al.}(2009)\citenamefont{Rebentrost,
  Serban, Schulte-Herbr\"uggen, and Wilhelm}}]{RebentrostPRL09}
\bibinfo{author}{\bibfnamefont{P.}~\bibnamefont{Rebentrost}},
  \bibinfo{author}{\bibfnamefont{I.}~\bibnamefont{Serban}},
  \bibinfo{author}{\bibfnamefont{T.}~\bibnamefont{Schulte-Herbr\"uggen}},
  \bibnamefont{and} \bibinfo{author}{\bibfnamefont{F.~K.}
  \bibnamefont{Wilhelm}}, \bibinfo{journal}{Phys. Rev. Lett.}
  \textbf{\bibinfo{volume}{102}}, \bibinfo{pages}{090401}
  (\bibinfo{year}{2009}).

\bibitem[{\citenamefont{Goerz et~al.}(2014{\natexlab{b}})\citenamefont{Goerz,
  Reich, and Koch.}}]{GoerzNJP14}
\bibinfo{author}{\bibfnamefont{M.~H.} \bibnamefont{Goerz}},
  \bibinfo{author}{\bibfnamefont{D.~M.} \bibnamefont{Reich}}, \bibnamefont{and}
  \bibinfo{author}{\bibfnamefont{C.~P.} \bibnamefont{Koch.}},
  \bibinfo{journal}{New J. Phys.} \textbf{\bibinfo{volume}{16}},
  \bibinfo{pages}{055012} (\bibinfo{year}{2014}{\natexlab{b}}).

\bibitem[{\citenamefont{Grond et~al.}(2009)\citenamefont{Grond, Schmiedmayer,
  and Hohenester}}]{GrondPRA09}
\bibinfo{author}{\bibfnamefont{J.}~\bibnamefont{Grond}},
  \bibinfo{author}{\bibfnamefont{J.}~\bibnamefont{Schmiedmayer}},
  \bibnamefont{and}
  \bibinfo{author}{\bibfnamefont{U.}~\bibnamefont{Hohenester}},
  \bibinfo{journal}{Phys. Rev. A} \textbf{\bibinfo{volume}{79}},
  \bibinfo{pages}{021603} (\bibinfo{year}{2009}).

\bibitem[{\citenamefont{Wang et~al.}(2010)\citenamefont{Wang, Bayat, Bose, and
  Schirmer}}]{WangPRA10}
\bibinfo{author}{\bibfnamefont{X.}~\bibnamefont{Wang}},
  \bibinfo{author}{\bibfnamefont{A.}~\bibnamefont{Bayat}},
  \bibinfo{author}{\bibfnamefont{S.}~\bibnamefont{Bose}}, \bibnamefont{and}
  \bibinfo{author}{\bibfnamefont{S.~G.} \bibnamefont{Schirmer}},
  \bibinfo{journal}{Phys. Rev. A} \textbf{\bibinfo{volume}{82}},
  \bibinfo{pages}{012330} (\bibinfo{year}{2010}).

\bibitem[{\citenamefont{Machnes et~al.}(2011)\citenamefont{Machnes, Sander,
  Glaser, de~Fouqui\`eres, Gruslys, Schirmer, and
  Schulte-Herbr\"uggen}}]{MachnesPRA11}
\bibinfo{author}{\bibfnamefont{S.}~\bibnamefont{Machnes}},
  \bibinfo{author}{\bibfnamefont{U.}~\bibnamefont{Sander}},
  \bibinfo{author}{\bibfnamefont{S.~J.} \bibnamefont{Glaser}},
  \bibinfo{author}{\bibfnamefont{P.}~\bibnamefont{de~Fouqui\`eres}},
  \bibinfo{author}{\bibfnamefont{A.}~\bibnamefont{Gruslys}},
  \bibinfo{author}{\bibfnamefont{S.}~\bibnamefont{Schirmer}}, \bibnamefont{and}
  \bibinfo{author}{\bibfnamefont{T.}~\bibnamefont{Schulte-Herbr\"uggen}},
  \bibinfo{journal}{Phys. Rev. A} \textbf{\bibinfo{volume}{84}},
  \bibinfo{pages}{022305} (\bibinfo{year}{2011}).

\bibitem[{\citenamefont{Doria et~al.}(2011)\citenamefont{Doria, Calarco, and
  Montangero}}]{DoriaPRL11}
\bibinfo{author}{\bibfnamefont{P.}~\bibnamefont{Doria}},
  \bibinfo{author}{\bibfnamefont{T.}~\bibnamefont{Calarco}}, \bibnamefont{and}
  \bibinfo{author}{\bibfnamefont{S.}~\bibnamefont{Montangero}},
  \bibinfo{journal}{Phys. Rev. Lett.} \textbf{\bibinfo{volume}{106}},
  \bibinfo{pages}{190501} (\bibinfo{year}{2011}).

\bibitem[{\citenamefont{Rojan et~al.}(2014)\citenamefont{Rojan, Reich,
  Dotsenko, Raimond, Koch, and Morigi}}]{RojanPRA14}
\bibinfo{author}{\bibfnamefont{K.}~\bibnamefont{Rojan}},
  \bibinfo{author}{\bibfnamefont{D.~M.} \bibnamefont{Reich}},
  \bibinfo{author}{\bibfnamefont{I.}~\bibnamefont{Dotsenko}},
  \bibinfo{author}{\bibfnamefont{J.-M.} \bibnamefont{Raimond}},
  \bibinfo{author}{\bibfnamefont{C.~P.} \bibnamefont{Koch}}, \bibnamefont{and}
  \bibinfo{author}{\bibfnamefont{G.}~\bibnamefont{Morigi}},
  \bibinfo{journal}{Phys. Rev. A} \textbf{\bibinfo{volume}{90}},
  \bibinfo{pages}{023824} (\bibinfo{year}{2014}).

\bibitem[{\citenamefont{De~Chiara et~al.}(2008)\citenamefont{De~Chiara,
  Calarco, Anderlini, Montangero, Lee, Brown, Phillips, and
  Porto}}]{DeChiaraPRA08}
\bibinfo{author}{\bibfnamefont{G.}~\bibnamefont{De~Chiara}},
  \bibinfo{author}{\bibfnamefont{T.}~\bibnamefont{Calarco}},
  \bibinfo{author}{\bibfnamefont{M.}~\bibnamefont{Anderlini}},
  \bibinfo{author}{\bibfnamefont{S.}~\bibnamefont{Montangero}},
  \bibinfo{author}{\bibfnamefont{P.~J.} \bibnamefont{Lee}},
  \bibinfo{author}{\bibfnamefont{B.~L.} \bibnamefont{Brown}},
  \bibinfo{author}{\bibfnamefont{W.~D.} \bibnamefont{Phillips}},
  \bibnamefont{and} \bibinfo{author}{\bibfnamefont{J.~V.} \bibnamefont{Porto}},
  \bibinfo{journal}{Phys. Rev. A} \textbf{\bibinfo{volume}{77}},
  \bibinfo{pages}{052333} (\bibinfo{year}{2008}).

\bibitem[{\citenamefont{Caneva et~al.}(2009)\citenamefont{Caneva, Murphy,
  Calarco, Fazio, Montangero, Giovannetti, and Santoro}}]{CanevaPRL09}
\bibinfo{author}{\bibfnamefont{T.}~\bibnamefont{Caneva}},
  \bibinfo{author}{\bibfnamefont{M.}~\bibnamefont{Murphy}},
  \bibinfo{author}{\bibfnamefont{T.}~\bibnamefont{Calarco}},
  \bibinfo{author}{\bibfnamefont{R.}~\bibnamefont{Fazio}},
  \bibinfo{author}{\bibfnamefont{S.}~\bibnamefont{Montangero}},
  \bibinfo{author}{\bibfnamefont{V.}~\bibnamefont{Giovannetti}},
  \bibnamefont{and} \bibinfo{author}{\bibfnamefont{G.~E.}
  \bibnamefont{Santoro}}, \bibinfo{journal}{Phys. Rev. Lett.}
  \textbf{\bibinfo{volume}{103}}, \bibinfo{pages}{240501}
  (\bibinfo{year}{2009}).

\bibitem[{\citenamefont{F\"urst et~al.}(2014)\citenamefont{F\"urst, Goerz,
  Poschinger, Murphy, Montangero, Calarco, Schmidt-Kaler, Singer, and
  Koch.}}]{FuerstNJP14}
\bibinfo{author}{\bibfnamefont{H.~A.} \bibnamefont{F\"urst}},
  \bibinfo{author}{\bibfnamefont{M.~H.} \bibnamefont{Goerz}},
  \bibinfo{author}{\bibfnamefont{U.~G.} \bibnamefont{Poschinger}},
  \bibinfo{author}{\bibfnamefont{M.}~\bibnamefont{Murphy}},
  \bibinfo{author}{\bibfnamefont{S.}~\bibnamefont{Montangero}},
  \bibinfo{author}{\bibfnamefont{T.}~\bibnamefont{Calarco}},
  \bibinfo{author}{\bibfnamefont{F.}~\bibnamefont{Schmidt-Kaler}},
  \bibinfo{author}{\bibfnamefont{K.}~\bibnamefont{Singer}}, \bibnamefont{and}
  \bibinfo{author}{\bibfnamefont{C.~P.} \bibnamefont{Koch.}},
  \bibinfo{journal}{New J. Phys.} \textbf{\bibinfo{volume}{16}},
  \bibinfo{pages}{075007} (\bibinfo{year}{2014}).

\bibitem[{\citenamefont{Gorshkov et~al.}(2008)\citenamefont{Gorshkov, Calarco,
  Lukin, and S\o{}rensen}}]{GorshkovPRA08}
\bibinfo{author}{\bibfnamefont{A.~V.} \bibnamefont{Gorshkov}},
  \bibinfo{author}{\bibfnamefont{T.}~\bibnamefont{Calarco}},
  \bibinfo{author}{\bibfnamefont{M.~D.} \bibnamefont{Lukin}}, \bibnamefont{and}
  \bibinfo{author}{\bibfnamefont{A.~S.} \bibnamefont{S\o{}rensen}},
  \bibinfo{journal}{Phys. Rev. A} \textbf{\bibinfo{volume}{77}},
  \bibinfo{pages}{043806} (\bibinfo{year}{2008}).

\bibitem[{\citenamefont{Rosi et~al.}(2013)\citenamefont{Rosi, Bernard, Fabbri,
  Fallani, Fort, Inguscio, Calarco, and Montangero}}]{RosiPRA13}
\bibinfo{author}{\bibfnamefont{S.}~\bibnamefont{Rosi}},
  \bibinfo{author}{\bibfnamefont{A.}~\bibnamefont{Bernard}},
  \bibinfo{author}{\bibfnamefont{N.}~\bibnamefont{Fabbri}},
  \bibinfo{author}{\bibfnamefont{L.}~\bibnamefont{Fallani}},
  \bibinfo{author}{\bibfnamefont{C.}~\bibnamefont{Fort}},
  \bibinfo{author}{\bibfnamefont{M.}~\bibnamefont{Inguscio}},
  \bibinfo{author}{\bibfnamefont{T.}~\bibnamefont{Calarco}}, \bibnamefont{and}
  \bibinfo{author}{\bibfnamefont{S.}~\bibnamefont{Montangero}},
  \bibinfo{journal}{Phys. Rev. A} \textbf{\bibinfo{volume}{88}},
  \bibinfo{pages}{021601} (\bibinfo{year}{2013}).

\bibitem[{\citenamefont{van Frank et~al.}(2015)\citenamefont{van Frank,
  Bonneau, Schmiedmayer, Hild, Gross, Cheneau, Bloch, Pichler, Negretti,
  Calarco et~al.}}]{vanFrank15}
\bibinfo{author}{\bibfnamefont{S.}~\bibnamefont{van Frank}},
  \bibinfo{author}{\bibfnamefont{M.}~\bibnamefont{Bonneau}},
  \bibinfo{author}{\bibfnamefont{J.}~\bibnamefont{Schmiedmayer}},
  \bibinfo{author}{\bibfnamefont{S.}~\bibnamefont{Hild}},
  \bibinfo{author}{\bibfnamefont{C.}~\bibnamefont{Gross}},
  \bibinfo{author}{\bibfnamefont{M.}~\bibnamefont{Cheneau}},
  \bibinfo{author}{\bibfnamefont{I.}~\bibnamefont{Bloch}},
  \bibinfo{author}{\bibfnamefont{T.}~\bibnamefont{Pichler}},
  \bibinfo{author}{\bibfnamefont{A.}~\bibnamefont{Negretti}},
  \bibinfo{author}{\bibfnamefont{T.}~\bibnamefont{Calarco}},
  \bibnamefont{et~al.}, \bibinfo{journal}{arXiv:1511.02247}
  (\bibinfo{year}{2015}).

\bibitem[{\citenamefont{H\"aberle et~al.}(2013)\citenamefont{H\"aberle,
  Schmid-Lorch, Karrai, Reinhard, and Wrachtrup}}]{HaeberlePRL13}
\bibinfo{author}{\bibfnamefont{T.}~\bibnamefont{H\"aberle}},
  \bibinfo{author}{\bibfnamefont{D.}~\bibnamefont{Schmid-Lorch}},
  \bibinfo{author}{\bibfnamefont{K.}~\bibnamefont{Karrai}},
  \bibinfo{author}{\bibfnamefont{F.}~\bibnamefont{Reinhard}}, \bibnamefont{and}
  \bibinfo{author}{\bibfnamefont{J.}~\bibnamefont{Wrachtrup}},
  \bibinfo{journal}{Phys. Rev. Lett.} \textbf{\bibinfo{volume}{111}},
  \bibinfo{pages}{170801} (\bibinfo{year}{2013}).

\bibitem[{\citenamefont{Dolde et~al.}(2014)\citenamefont{Dolde, Bergholm, Wang,
  Jakobi, Naydenov, Pezzagna, Meijer, Jelezko, Neumann, Schulte-Herbr\"uggen
  et~al.}}]{DoldeNatComm14}
\bibinfo{author}{\bibfnamefont{F.}~\bibnamefont{Dolde}},
  \bibinfo{author}{\bibfnamefont{V.}~\bibnamefont{Bergholm}},
  \bibinfo{author}{\bibfnamefont{Y.}~\bibnamefont{Wang}},
  \bibinfo{author}{\bibfnamefont{I.}~\bibnamefont{Jakobi}},
  \bibinfo{author}{\bibfnamefont{B.}~\bibnamefont{Naydenov}},
  \bibinfo{author}{\bibfnamefont{S.}~\bibnamefont{Pezzagna}},
  \bibinfo{author}{\bibfnamefont{J.}~\bibnamefont{Meijer}},
  \bibinfo{author}{\bibfnamefont{F.}~\bibnamefont{Jelezko}},
  \bibinfo{author}{\bibfnamefont{P.}~\bibnamefont{Neumann}},
  \bibinfo{author}{\bibfnamefont{T.}~\bibnamefont{Schulte-Herbr\"uggen}},
  \bibnamefont{et~al.}, \bibinfo{journal}{Nat. Commun.}
  \textbf{\bibinfo{volume}{5}}, \bibinfo{pages}{3371} (\bibinfo{year}{2014}).

\bibitem[{\citenamefont{van Frank et~al.}(2014)\citenamefont{van Frank,
  Negretti, Berrada, B\"ucker, Montangero, Schaff, Schumm, Calarco, and
  Schmiedmayer}}]{vanFrankNatComm14}
\bibinfo{author}{\bibfnamefont{S.}~\bibnamefont{van Frank}},
  \bibinfo{author}{\bibfnamefont{A.}~\bibnamefont{Negretti}},
  \bibinfo{author}{\bibfnamefont{T.}~\bibnamefont{Berrada}},
  \bibinfo{author}{\bibfnamefont{R.}~\bibnamefont{B\"ucker}},
  \bibinfo{author}{\bibfnamefont{S.}~\bibnamefont{Montangero}},
  \bibinfo{author}{\bibfnamefont{J.-F.} \bibnamefont{Schaff}},
  \bibinfo{author}{\bibfnamefont{T.}~\bibnamefont{Schumm}},
  \bibinfo{author}{\bibfnamefont{T.}~\bibnamefont{Calarco}}, \bibnamefont{and}
  \bibinfo{author}{\bibfnamefont{J.}~\bibnamefont{Schmiedmayer}},
  \bibinfo{journal}{Nat. Commun.} \textbf{\bibinfo{volume}{5}},
  \bibinfo{pages}{4009} (\bibinfo{year}{2014}).

\bibitem[{\citenamefont{Weiss}(2012)}]{WeissBook}
\bibinfo{author}{\bibfnamefont{U.}~\bibnamefont{Weiss}},
  \emph{\bibinfo{title}{Quantum dissipative systems}}
  (\bibinfo{publisher}{World Scientific}, \bibinfo{address}{Singapore},
  \bibinfo{year}{2012}), \bibinfo{edition}{4th} ed.

\bibitem[{\citenamefont{Breuer and Petruccione}(2007)}]{BreuerBook}
\bibinfo{author}{\bibfnamefont{H.-P.} \bibnamefont{Breuer}} \bibnamefont{and}
  \bibinfo{author}{\bibfnamefont{F.}~\bibnamefont{Petruccione}},
  \emph{\bibinfo{title}{The theory of open quantum systems}}
  (\bibinfo{publisher}{Oxford University Press}, \bibinfo{year}{2007}).

\bibitem[{\citenamefont{Breuer et~al.}(2015)\citenamefont{Breuer, Laine, Piilo,
  and Vacchini}}]{BreuerReview}
\bibinfo{author}{\bibfnamefont{H.-P.} \bibnamefont{Breuer}},
  \bibinfo{author}{\bibfnamefont{E.-M.} \bibnamefont{Laine}},
  \bibinfo{author}{\bibfnamefont{J.}~\bibnamefont{Piilo}}, \bibnamefont{and}
  \bibinfo{author}{\bibfnamefont{B.}~\bibnamefont{Vacchini}},
  \bibinfo{journal}{arXiv:1505.01385}  (\bibinfo{year}{2015}).

\bibitem[{\citenamefont{Rivas et~al.}(2014)\citenamefont{Rivas, Huelga, and
  Plenio}}]{RivasRPP14}
\bibinfo{author}{\bibfnamefont{A.}~\bibnamefont{Rivas}},
  \bibinfo{author}{\bibfnamefont{S.~F.} \bibnamefont{Huelga}},
  \bibnamefont{and} \bibinfo{author}{\bibfnamefont{M.~B.}
  \bibnamefont{Plenio}}, \bibinfo{journal}{Rep. Prog. Phys.}
  \textbf{\bibinfo{volume}{77}}, \bibinfo{pages}{094001}
  (\bibinfo{year}{2014}).

\bibitem[{\citenamefont{Kosloff}(1994)}]{KosloffARPC94}
\bibinfo{author}{\bibfnamefont{R.}~\bibnamefont{Kosloff}},
  \bibinfo{journal}{Annu. Rev. Phys. Chem.} \textbf{\bibinfo{volume}{45}},
  \bibinfo{pages}{145} (\bibinfo{year}{1994}).

\bibitem[{\citenamefont{Lucas and Hornberger}(2014)}]{LucasPRA14}
\bibinfo{author}{\bibfnamefont{F.}~\bibnamefont{Lucas}} \bibnamefont{and}
  \bibinfo{author}{\bibfnamefont{K.}~\bibnamefont{Hornberger}},
  \bibinfo{journal}{Phys. Rev. A} \textbf{\bibinfo{volume}{89}},
  \bibinfo{pages}{012112} (\bibinfo{year}{2014}).

\bibitem[{\citenamefont{Sweke et~al.}(2015)\citenamefont{Sweke, Sinayskiy,
  Bernard, and Petruccione}}]{SwekePRA15}
\bibinfo{author}{\bibfnamefont{R.}~\bibnamefont{Sweke}},
  \bibinfo{author}{\bibfnamefont{I.}~\bibnamefont{Sinayskiy}},
  \bibinfo{author}{\bibfnamefont{D.}~\bibnamefont{Bernard}}, \bibnamefont{and}
  \bibinfo{author}{\bibfnamefont{F.}~\bibnamefont{Petruccione}},
  \bibinfo{journal}{Phys. Rev. A} \textbf{\bibinfo{volume}{91}},
  \bibinfo{pages}{062308} (\bibinfo{year}{2015}).

\bibitem[{\citenamefont{de~Vega and Alonso}(2015)}]{deVegaRMP16}
\bibinfo{author}{\bibfnamefont{I.}~\bibnamefont{de~Vega}} \bibnamefont{and}
  \bibinfo{author}{\bibfnamefont{D.}~\bibnamefont{Alonso}},
  \bibinfo{journal}{arXiv:1511.06994}  (\bibinfo{year}{2015}).

\bibitem[{\citenamefont{Piilo et~al.}(2008)\citenamefont{Piilo, Maniscalco,
  H\"ark\"onen, and Suominen}}]{PiiloPRL08}
\bibinfo{author}{\bibfnamefont{J.}~\bibnamefont{Piilo}},
  \bibinfo{author}{\bibfnamefont{S.}~\bibnamefont{Maniscalco}},
  \bibinfo{author}{\bibfnamefont{K.}~\bibnamefont{H\"ark\"onen}},
  \bibnamefont{and} \bibinfo{author}{\bibfnamefont{K.-A.}
  \bibnamefont{Suominen}}, \bibinfo{journal}{Phys. Rev. Lett.}
  \textbf{\bibinfo{volume}{100}}, \bibinfo{pages}{180402}
  (\bibinfo{year}{2008}).

\bibitem[{\citenamefont{Koch et~al.}(2008)\citenamefont{Koch, Gro\ss{}mann,
  Stockburger, and Ankerhold}}]{KochPRL08}
\bibinfo{author}{\bibfnamefont{W.}~\bibnamefont{Koch}},
  \bibinfo{author}{\bibfnamefont{F.}~\bibnamefont{Gro\ss{}mann}},
  \bibinfo{author}{\bibfnamefont{J.~T.} \bibnamefont{Stockburger}},
  \bibnamefont{and}
  \bibinfo{author}{\bibfnamefont{J.}~\bibnamefont{Ankerhold}},
  \bibinfo{journal}{Phys. Rev. Lett.} \textbf{\bibinfo{volume}{100}},
  \bibinfo{pages}{230402} (\bibinfo{year}{2008}).

\bibitem[{\citenamefont{Di\'osi and Ferialdi}(2014)}]{DiosiPRL14}
\bibinfo{author}{\bibfnamefont{L.}~\bibnamefont{Di\'osi}} \bibnamefont{and}
  \bibinfo{author}{\bibfnamefont{L.}~\bibnamefont{Ferialdi}},
  \bibinfo{journal}{Phys. Rev. Lett.} \textbf{\bibinfo{volume}{113}},
  \bibinfo{pages}{200403} (\bibinfo{year}{2014}).

\bibitem[{\citenamefont{Meier and Tannor}(1999)}]{MeierJCP99}
\bibinfo{author}{\bibfnamefont{C.}~\bibnamefont{Meier}} \bibnamefont{and}
  \bibinfo{author}{\bibfnamefont{D.~J.} \bibnamefont{Tannor}},
  \bibinfo{journal}{J. Chem. Phys.} \textbf{\bibinfo{volume}{111}},
  \bibinfo{pages}{3365} (\bibinfo{year}{1999}).

\bibitem[{\citenamefont{Koch et~al.}(2003)\citenamefont{Koch, Kl\"uner, Freund,
  and Kosloff}}]{KochPRL03}
\bibinfo{author}{\bibfnamefont{C.~P.} \bibnamefont{Koch}},
  \bibinfo{author}{\bibfnamefont{T.}~\bibnamefont{Kl\"uner}},
  \bibinfo{author}{\bibfnamefont{H.-J.} \bibnamefont{Freund}},
  \bibnamefont{and} \bibinfo{author}{\bibfnamefont{R.}~\bibnamefont{Kosloff}},
  \bibinfo{journal}{Phys. Rev. Lett.} \textbf{\bibinfo{volume}{90}},
  \bibinfo{pages}{117601} (\bibinfo{year}{2003}).

\bibitem[{\citenamefont{Hughes et~al.}(2009)\citenamefont{Hughes, Christ, and
  Burghardt}}]{HughesJCP09}
\bibinfo{author}{\bibfnamefont{K.~H.} \bibnamefont{Hughes}},
  \bibinfo{author}{\bibfnamefont{C.~D.} \bibnamefont{Christ}},
  \bibnamefont{and}
  \bibinfo{author}{\bibfnamefont{I.}~\bibnamefont{Burghardt}},
  \bibinfo{journal}{J. Chem. Phys.} \textbf{\bibinfo{volume}{131}},
  \bibinfo{pages}{024109} (\bibinfo{year}{2009}).

\bibitem[{\citenamefont{Georgescu et~al.}(2014)\citenamefont{Georgescu, Ashhab,
  and Nori}}]{GeorgescuRMP14}
\bibinfo{author}{\bibfnamefont{I.~M.} \bibnamefont{Georgescu}},
  \bibinfo{author}{\bibfnamefont{S.}~\bibnamefont{Ashhab}}, \bibnamefont{and}
  \bibinfo{author}{\bibfnamefont{F.}~\bibnamefont{Nori}},
  \bibinfo{journal}{Rev. Mod. Phys.} \textbf{\bibinfo{volume}{86}},
  \bibinfo{pages}{153} (\bibinfo{year}{2014}).

\bibitem[{\citenamefont{Gualdi and Koch}(2013)}]{GualdiPRA13}
\bibinfo{author}{\bibfnamefont{G.}~\bibnamefont{Gualdi}} \bibnamefont{and}
  \bibinfo{author}{\bibfnamefont{C.~P.} \bibnamefont{Koch}},
  \bibinfo{journal}{Phys. Rev. A} \textbf{\bibinfo{volume}{88}},
  \bibinfo{pages}{022122} (\bibinfo{year}{2013}).

\bibitem[{\citenamefont{Reich et~al.}(2015)\citenamefont{Reich, Katz, and
  Koch}}]{ReichSciRep15}
\bibinfo{author}{\bibfnamefont{D.~M.} \bibnamefont{Reich}},
  \bibinfo{author}{\bibfnamefont{N.}~\bibnamefont{Katz}}, \bibnamefont{and}
  \bibinfo{author}{\bibfnamefont{C.~P.} \bibnamefont{Koch}},
  \bibinfo{journal}{Sci. Rep.} \textbf{\bibinfo{volume}{5}},
  \bibinfo{pages}{12430} (\bibinfo{year}{2015}).

\bibitem[{\citenamefont{Katz et~al.}(2008)\citenamefont{Katz, Gelman, Ratner,
  and Kosloff}}]{KatzJCP08}
\bibinfo{author}{\bibfnamefont{G.}~\bibnamefont{Katz}},
  \bibinfo{author}{\bibfnamefont{D.}~\bibnamefont{Gelman}},
  \bibinfo{author}{\bibfnamefont{M.~A.} \bibnamefont{Ratner}},
  \bibnamefont{and} \bibinfo{author}{\bibfnamefont{R.}~\bibnamefont{Kosloff}},
  \bibinfo{journal}{J. Chem. Phys.} \textbf{\bibinfo{volume}{129}},
  \bibinfo{pages}{034108} (\bibinfo{year}{2008}).

\bibitem[{\citenamefont{Wolf et~al.}(2008)\citenamefont{Wolf, Eisert, Cubitt,
  and Cirac}}]{WolfPRL08}
\bibinfo{author}{\bibfnamefont{M.~M.} \bibnamefont{Wolf}},
  \bibinfo{author}{\bibfnamefont{J.}~\bibnamefont{Eisert}},
  \bibinfo{author}{\bibfnamefont{T.~S.} \bibnamefont{Cubitt}},
  \bibnamefont{and} \bibinfo{author}{\bibfnamefont{J.~I.} \bibnamefont{Cirac}},
  \bibinfo{journal}{Phys. Rev. Lett.} \textbf{\bibinfo{volume}{101}},
  \bibinfo{pages}{150402} (\bibinfo{year}{2008}).

\bibitem[{\citenamefont{Rivas et~al.}(2010)\citenamefont{Rivas, Huelga, and
  Plenio}}]{RivasPRL10}
\bibinfo{author}{\bibfnamefont{A.}~\bibnamefont{Rivas}},
  \bibinfo{author}{\bibfnamefont{S.~F.} \bibnamefont{Huelga}},
  \bibnamefont{and} \bibinfo{author}{\bibfnamefont{M.~B.}
  \bibnamefont{Plenio}}, \bibinfo{journal}{Phys. Rev. Lett.}
  \textbf{\bibinfo{volume}{105}}, \bibinfo{pages}{050403}
  (\bibinfo{year}{2010}).

\bibitem[{\citenamefont{Dhar et~al.}(2015)\citenamefont{Dhar, Bera, and
  Adesso}}]{DharPRA15}
\bibinfo{author}{\bibfnamefont{H.~S.} \bibnamefont{Dhar}},
  \bibinfo{author}{\bibfnamefont{M.~N.} \bibnamefont{Bera}}, \bibnamefont{and}
  \bibinfo{author}{\bibfnamefont{G.}~\bibnamefont{Adesso}},
  \bibinfo{journal}{Phys. Rev. A} \textbf{\bibinfo{volume}{91}},
  \bibinfo{pages}{032115} (\bibinfo{year}{2015}).

\bibitem[{\citenamefont{Breuer et~al.}(2009)\citenamefont{Breuer, Laine, and
  Piilo}}]{BreuerPRL09}
\bibinfo{author}{\bibfnamefont{H.-P.} \bibnamefont{Breuer}},
  \bibinfo{author}{\bibfnamefont{E.-M.} \bibnamefont{Laine}}, \bibnamefont{and}
  \bibinfo{author}{\bibfnamefont{J.}~\bibnamefont{Piilo}},
  \bibinfo{journal}{Phys. Rev. Lett.} \textbf{\bibinfo{volume}{103}},
  \bibinfo{pages}{210401} (\bibinfo{year}{2009}).

\bibitem[{\citenamefont{Laine et~al.}(2010)\citenamefont{Laine, Piilo, and
  Breuer}}]{LainePRA10}
\bibinfo{author}{\bibfnamefont{E.-M.} \bibnamefont{Laine}},
  \bibinfo{author}{\bibfnamefont{J.}~\bibnamefont{Piilo}}, \bibnamefont{and}
  \bibinfo{author}{\bibfnamefont{H.-P.} \bibnamefont{Breuer}},
  \bibinfo{journal}{Phys. Rev. A} \textbf{\bibinfo{volume}{81}},
  \bibinfo{pages}{062115} (\bibinfo{year}{2010}).

\bibitem[{\citenamefont{Vasile et~al.}(2011)\citenamefont{Vasile, Maniscalco,
  Paris, Breuer, and Piilo}}]{VasilePRA11}
\bibinfo{author}{\bibfnamefont{R.}~\bibnamefont{Vasile}},
  \bibinfo{author}{\bibfnamefont{S.}~\bibnamefont{Maniscalco}},
  \bibinfo{author}{\bibfnamefont{M.~G.~A.} \bibnamefont{Paris}},
  \bibinfo{author}{\bibfnamefont{H.-P.} \bibnamefont{Breuer}},
  \bibnamefont{and} \bibinfo{author}{\bibfnamefont{J.}~\bibnamefont{Piilo}},
  \bibinfo{journal}{Phys. Rev. A} \textbf{\bibinfo{volume}{84}},
  \bibinfo{pages}{052118} (\bibinfo{year}{2011}).

\bibitem[{\citenamefont{Lorenzo et~al.}(2013)\citenamefont{Lorenzo, Plastina,
  and Paternostro}}]{LorenzoPRA13}
\bibinfo{author}{\bibfnamefont{S.}~\bibnamefont{Lorenzo}},
  \bibinfo{author}{\bibfnamefont{F.}~\bibnamefont{Plastina}}, \bibnamefont{and}
  \bibinfo{author}{\bibfnamefont{M.}~\bibnamefont{Paternostro}},
  \bibinfo{journal}{Phys. Rev. A} \textbf{\bibinfo{volume}{88}},
  \bibinfo{pages}{020102} (\bibinfo{year}{2013}).

\bibitem[{\citenamefont{Bylicka et~al.}(2014)\citenamefont{Bylicka,
  Chru\'sci\'nski, and Maniscalco}}]{BylickaSciRep14}
\bibinfo{author}{\bibfnamefont{B.}~\bibnamefont{Bylicka}},
  \bibinfo{author}{\bibfnamefont{D.}~\bibnamefont{Chru\'sci\'nski}},
  \bibnamefont{and}
  \bibinfo{author}{\bibfnamefont{S.}~\bibnamefont{Maniscalco}},
  \bibinfo{journal}{Sci. Rep.} \textbf{\bibinfo{volume}{4}},
  \bibinfo{pages}{5720} (\bibinfo{year}{2014}).

\bibitem[{\citenamefont{Addis et~al.}(2014)\citenamefont{Addis, Bylicka,
  Chru\'{s}ci\'{n}ski, and Maniscalco}}]{AddisPRA14}
\bibinfo{author}{\bibfnamefont{C.}~\bibnamefont{Addis}},
  \bibinfo{author}{\bibfnamefont{B.}~\bibnamefont{Bylicka}},
  \bibinfo{author}{\bibfnamefont{D.}~\bibnamefont{Chru\'{s}ci\'{n}ski}},
  \bibnamefont{and}
  \bibinfo{author}{\bibfnamefont{S.}~\bibnamefont{Maniscalco}},
  \bibinfo{journal}{Phys. Rev. A} \textbf{\bibinfo{volume}{90}},
  \bibinfo{pages}{052103} (\bibinfo{year}{2014}).

\bibitem[{\citenamefont{Chen et~al.}(2015)\citenamefont{Chen, Lien, Chen, and
  Chen}}]{ChenPRA15}
\bibinfo{author}{\bibfnamefont{H.-B.} \bibnamefont{Chen}},
  \bibinfo{author}{\bibfnamefont{J.-Y.} \bibnamefont{Lien}},
  \bibinfo{author}{\bibfnamefont{G.-Y.} \bibnamefont{Chen}}, \bibnamefont{and}
  \bibinfo{author}{\bibfnamefont{Y.-N.} \bibnamefont{Chen}},
  \bibinfo{journal}{Phys. Rev. A} \textbf{\bibinfo{volume}{92}},
  \bibinfo{pages}{042105} (\bibinfo{year}{2015}).

\bibitem[{\citenamefont{Bae and Chruscinski}(2016)}]{Bae16}
\bibinfo{author}{\bibfnamefont{J.}~\bibnamefont{Bae}} \bibnamefont{and}
  \bibinfo{author}{\bibfnamefont{D.}~\bibnamefont{Chruscinski}},
  \bibinfo{journal}{arXiv:1601.05522}  (\bibinfo{year}{2016}).

\bibitem[{\citenamefont{Liu et~al.}(2011)\citenamefont{Liu, Li, Huang, Li, Guo,
  Laine, Breuer, and Piilo}}]{LuiNatPhys11}
\bibinfo{author}{\bibfnamefont{B.-H.} \bibnamefont{Liu}},
  \bibinfo{author}{\bibfnamefont{L.}~\bibnamefont{Li}},
  \bibinfo{author}{\bibfnamefont{Y.-F.} \bibnamefont{Huang}},
  \bibinfo{author}{\bibfnamefont{C.-F.} \bibnamefont{Li}},
  \bibinfo{author}{\bibfnamefont{G.-C.} \bibnamefont{Guo}},
  \bibinfo{author}{\bibfnamefont{E.-M.} \bibnamefont{Laine}},
  \bibinfo{author}{\bibfnamefont{H.-P.} \bibnamefont{Breuer}},
  \bibnamefont{and} \bibinfo{author}{\bibfnamefont{J.}~\bibnamefont{Piilo}},
  \bibinfo{journal}{Nature Phys.} \textbf{\bibinfo{volume}{7}},
  \bibinfo{pages}{931} (\bibinfo{year}{2011}).

\bibitem[{\citenamefont{Gessner et~al.}(2014)\citenamefont{Gessner, Ramm,
  Pruttivarasin, Buchleitner, Breuer, and H\"affner}}]{GessnerNatPhys14}
\bibinfo{author}{\bibfnamefont{M.}~\bibnamefont{Gessner}},
  \bibinfo{author}{\bibfnamefont{M.}~\bibnamefont{Ramm}},
  \bibinfo{author}{\bibfnamefont{T.}~\bibnamefont{Pruttivarasin}},
  \bibinfo{author}{\bibfnamefont{A.}~\bibnamefont{Buchleitner}},
  \bibinfo{author}{\bibfnamefont{H.-P.} \bibnamefont{Breuer}},
  \bibnamefont{and}
  \bibinfo{author}{\bibfnamefont{H.}~\bibnamefont{H\"affner}},
  \bibinfo{journal}{Nature Phys.} \textbf{\bibinfo{volume}{10}},
  \bibinfo{pages}{105} (\bibinfo{year}{2014}).

\bibitem[{\citenamefont{Bartana et~al.}(1997)\citenamefont{Bartana, Kosloff,
  and Tannor}}]{BartanaJCP97}
\bibinfo{author}{\bibfnamefont{A.}~\bibnamefont{Bartana}},
  \bibinfo{author}{\bibfnamefont{R.}~\bibnamefont{Kosloff}}, \bibnamefont{and}
  \bibinfo{author}{\bibfnamefont{D.~J.} \bibnamefont{Tannor}},
  \bibinfo{journal}{J. Chem. Phys.} \textbf{\bibinfo{volume}{106}},
  \bibinfo{pages}{1435} (\bibinfo{year}{1997}).

\bibitem[{\citenamefont{Schmidt et~al.}(2011)\citenamefont{Schmidt, Negretti,
  Ankerhold, Calarco, and Stockburger}}]{SchmidtPRL11}
\bibinfo{author}{\bibfnamefont{R.}~\bibnamefont{Schmidt}},
  \bibinfo{author}{\bibfnamefont{A.}~\bibnamefont{Negretti}},
  \bibinfo{author}{\bibfnamefont{J.}~\bibnamefont{Ankerhold}},
  \bibinfo{author}{\bibfnamefont{T.}~\bibnamefont{Calarco}}, \bibnamefont{and}
  \bibinfo{author}{\bibfnamefont{J.~T.} \bibnamefont{Stockburger}},
  \bibinfo{journal}{Phys. Rev. Lett.} \textbf{\bibinfo{volume}{107}},
  \bibinfo{pages}{130404} (\bibinfo{year}{2011}).

\bibitem[{\citenamefont{Viteau et~al.}(2008)\citenamefont{Viteau, Chotia,
  Allegrini, Bouloufa, Dulieu, Comparat, and Pillet}}]{ViteauSci08}
\bibinfo{author}{\bibfnamefont{M.}~\bibnamefont{Viteau}},
  \bibinfo{author}{\bibfnamefont{A.}~\bibnamefont{Chotia}},
  \bibinfo{author}{\bibfnamefont{M.}~\bibnamefont{Allegrini}},
  \bibinfo{author}{\bibfnamefont{N.}~\bibnamefont{Bouloufa}},
  \bibinfo{author}{\bibfnamefont{O.}~\bibnamefont{Dulieu}},
  \bibinfo{author}{\bibfnamefont{D.}~\bibnamefont{Comparat}}, \bibnamefont{and}
  \bibinfo{author}{\bibfnamefont{P.}~\bibnamefont{Pillet}},
  \bibinfo{journal}{Science} \textbf{\bibinfo{volume}{321}},
  \bibinfo{pages}{232} (\bibinfo{year}{2008}).

\bibitem[{\citenamefont{Reich and Koch}(2013)}]{ReichNJP13}
\bibinfo{author}{\bibfnamefont{D.~M.} \bibnamefont{Reich}} \bibnamefont{and}
  \bibinfo{author}{\bibfnamefont{C.~P.} \bibnamefont{Koch}},
  \bibinfo{journal}{New J. Phys.} \textbf{\bibinfo{volume}{15}},
  \bibinfo{pages}{125028} (\bibinfo{year}{2013}).

\bibitem[{\citenamefont{Kallush and Kosloff}(2006)}]{KallushPRA06}
\bibinfo{author}{\bibfnamefont{S.}~\bibnamefont{Kallush}} \bibnamefont{and}
  \bibinfo{author}{\bibfnamefont{R.}~\bibnamefont{Kosloff}},
  \bibinfo{journal}{Phys. Rev. A} \textbf{\bibinfo{volume}{73}},
  \bibinfo{pages}{032324} (\bibinfo{year}{2006}).

\bibitem[{\citenamefont{Schulte-Herbr\"uggen
  et~al.}(2011)\citenamefont{Schulte-Herbr\"uggen, Sp\"orl, Khaneja, and
  Glaser}}]{ToSHJPB11}
\bibinfo{author}{\bibfnamefont{T.}~\bibnamefont{Schulte-Herbr\"uggen}},
  \bibinfo{author}{\bibfnamefont{A.}~\bibnamefont{Sp\"orl}},
  \bibinfo{author}{\bibfnamefont{N.}~\bibnamefont{Khaneja}}, \bibnamefont{and}
  \bibinfo{author}{\bibfnamefont{S.~J.} \bibnamefont{Glaser}},
  \bibinfo{journal}{J. Phys. B} \textbf{\bibinfo{volume}{44}},
  \bibinfo{pages}{154013} (\bibinfo{year}{2011}).

\bibitem[{\citenamefont{Bendersky et~al.}(2008)\citenamefont{Bendersky,
  Pastawski, and Paz}}]{BenderskyPRL08}
\bibinfo{author}{\bibfnamefont{A.}~\bibnamefont{Bendersky}},
  \bibinfo{author}{\bibfnamefont{F.}~\bibnamefont{Pastawski}},
  \bibnamefont{and} \bibinfo{author}{\bibfnamefont{J.~P.} \bibnamefont{Paz}},
  \bibinfo{journal}{Phys. Rev. Lett.} \textbf{\bibinfo{volume}{100}},
  \bibinfo{pages}{190403} (\bibinfo{year}{2008}).

\bibitem[{\citenamefont{da~Silva et~al.}(2011)\citenamefont{da~Silva,
  Landon-Cardinal, and Poulin}}]{daSilvaPRL11}
\bibinfo{author}{\bibfnamefont{M.~P.} \bibnamefont{da~Silva}},
  \bibinfo{author}{\bibfnamefont{O.}~\bibnamefont{Landon-Cardinal}},
  \bibnamefont{and} \bibinfo{author}{\bibfnamefont{D.}~\bibnamefont{Poulin}},
  \bibinfo{journal}{Phys. Rev. Lett.} \textbf{\bibinfo{volume}{107}},
  \bibinfo{pages}{210404} (\bibinfo{year}{2011}).

\bibitem[{\citenamefont{Flammia and Liu}(2011)}]{FlammiaPRL11}
\bibinfo{author}{\bibfnamefont{S.~T.} \bibnamefont{Flammia}} \bibnamefont{and}
  \bibinfo{author}{\bibfnamefont{Y.-K.} \bibnamefont{Liu}},
  \bibinfo{journal}{Phys. Rev. Lett.} \textbf{\bibinfo{volume}{106}},
  \bibinfo{pages}{230501} (\bibinfo{year}{2011}).

\bibitem[{\citenamefont{Magesan et~al.}(2011)\citenamefont{Magesan, Gambetta,
  and Emerson}}]{MagesanPRL11}
\bibinfo{author}{\bibfnamefont{E.}~\bibnamefont{Magesan}},
  \bibinfo{author}{\bibfnamefont{J.~M.} \bibnamefont{Gambetta}},
  \bibnamefont{and} \bibinfo{author}{\bibfnamefont{J.}~\bibnamefont{Emerson}},
  \bibinfo{journal}{Phys. Rev. Lett.} \textbf{\bibinfo{volume}{106}},
  \bibinfo{pages}{180504} (\bibinfo{year}{2011}).

\bibitem[{\citenamefont{Reich et~al.}(2013{\natexlab{a}})\citenamefont{Reich,
  Gualdi, and Koch}}]{ReichPRL13}
\bibinfo{author}{\bibfnamefont{D.~M.} \bibnamefont{Reich}},
  \bibinfo{author}{\bibfnamefont{G.}~\bibnamefont{Gualdi}}, \bibnamefont{and}
  \bibinfo{author}{\bibfnamefont{C.~P.} \bibnamefont{Koch}},
  \bibinfo{journal}{Phys. Rev. Lett.} \textbf{\bibinfo{volume}{111}},
  \bibinfo{pages}{200401} (\bibinfo{year}{2013}{\natexlab{a}}).

\bibitem[{\citenamefont{Reich et~al.}(2013{\natexlab{b}})\citenamefont{Reich,
  Gualdi, and Koch}}]{ReichPRA13}
\bibinfo{author}{\bibfnamefont{D.~M.} \bibnamefont{Reich}},
  \bibinfo{author}{\bibfnamefont{G.}~\bibnamefont{Gualdi}}, \bibnamefont{and}
  \bibinfo{author}{\bibfnamefont{C.~P.} \bibnamefont{Koch}},
  \bibinfo{journal}{Phys. Rev. A} \textbf{\bibinfo{volume}{88}},
  \bibinfo{pages}{042309} (\bibinfo{year}{2013}{\natexlab{b}}).

\bibitem[{\citenamefont{Reich et~al.}(2012)\citenamefont{Reich, Ndong, and
  Koch}}]{ReichJCP12}
\bibinfo{author}{\bibfnamefont{D.~M.} \bibnamefont{Reich}},
  \bibinfo{author}{\bibfnamefont{M.}~\bibnamefont{Ndong}}, \bibnamefont{and}
  \bibinfo{author}{\bibfnamefont{C.~P.} \bibnamefont{Koch}},
  \bibinfo{journal}{J. Chem. Phys.} \textbf{\bibinfo{volume}{136}},
  \bibinfo{pages}{104103} (\bibinfo{year}{2012}).

\bibitem[{\citenamefont{Watts et~al.}(2015)\citenamefont{Watts, Vala, M\"uller,
  Calarco, Whaley, Reich, Goerz, and Koch}}]{WattsPRA15}
\bibinfo{author}{\bibfnamefont{P.}~\bibnamefont{Watts}},
  \bibinfo{author}{\bibfnamefont{J.}~\bibnamefont{Vala}},
  \bibinfo{author}{\bibfnamefont{M.~M.} \bibnamefont{M\"uller}},
  \bibinfo{author}{\bibfnamefont{T.}~\bibnamefont{Calarco}},
  \bibinfo{author}{\bibfnamefont{K.~B.} \bibnamefont{Whaley}},
  \bibinfo{author}{\bibfnamefont{D.~M.} \bibnamefont{Reich}},
  \bibinfo{author}{\bibfnamefont{M.~H.} \bibnamefont{Goerz}}, \bibnamefont{and}
  \bibinfo{author}{\bibfnamefont{C.~P.} \bibnamefont{Koch}},
  \bibinfo{journal}{Phys. Rev. A} \textbf{\bibinfo{volume}{91}},
  \bibinfo{pages}{062306} (\bibinfo{year}{2015}).

\bibitem[{\citenamefont{Goerz et~al.}(2015{\natexlab{a}})\citenamefont{Goerz,
  Gualdi, Reich, Koch, Motzoi, Whaley, Vala, M\"uller, Montangero, and
  Calarco}}]{GoerzPRA15}
\bibinfo{author}{\bibfnamefont{M.~H.} \bibnamefont{Goerz}},
  \bibinfo{author}{\bibfnamefont{G.}~\bibnamefont{Gualdi}},
  \bibinfo{author}{\bibfnamefont{D.~M.} \bibnamefont{Reich}},
  \bibinfo{author}{\bibfnamefont{C.~P.} \bibnamefont{Koch}},
  \bibinfo{author}{\bibfnamefont{F.}~\bibnamefont{Motzoi}},
  \bibinfo{author}{\bibfnamefont{K.~B.} \bibnamefont{Whaley}},
  \bibinfo{author}{\bibfnamefont{J.}~\bibnamefont{Vala}},
  \bibinfo{author}{\bibfnamefont{M.~M.} \bibnamefont{M\"uller}},
  \bibinfo{author}{\bibfnamefont{S.}~\bibnamefont{Montangero}},
  \bibnamefont{and} \bibinfo{author}{\bibfnamefont{T.}~\bibnamefont{Calarco}},
  \bibinfo{journal}{Phys. Rev. A} \textbf{\bibinfo{volume}{91}},
  \bibinfo{pages}{062307} (\bibinfo{year}{2015}{\natexlab{a}}).

\bibitem[{\citenamefont{Reich}(2015)}]{ReichPhD}
\bibinfo{author}{\bibfnamefont{D.~M.} \bibnamefont{Reich}}, Ph.D. thesis,
  \bibinfo{school}{Universit\"at Kassel} (\bibinfo{year}{2015}),
  \urlprefix\url{http://nbn-resolving.de/urn:nbn:de:hebis:34-2015061948576}.

\bibitem[{\citenamefont{Jurdjevic}(1997)}]{JurdjevicBook}
\bibinfo{author}{\bibfnamefont{V.}~\bibnamefont{Jurdjevic}},
  \emph{\bibinfo{title}{Geometric Control Theory}}, Cambridge Studies in
  Advanced Mathematics (\bibinfo{publisher}{Cambridge Univ. Press},
  \bibinfo{address}{Cambridge}, \bibinfo{year}{1997}).

\bibitem[{\citenamefont{Jurdjevic and Sussmann}(1972)}]{Jurdjevic72}
\bibinfo{author}{\bibfnamefont{V.}~\bibnamefont{Jurdjevic}} \bibnamefont{and}
  \bibinfo{author}{\bibfnamefont{H.~J.} \bibnamefont{Sussmann}},
  \bibinfo{journal}{J. Diff. Eqns.} \textbf{\bibinfo{volume}{12}},
  \bibinfo{pages}{313} (\bibinfo{year}{1972}), ISSN \bibinfo{issn}{0022-0396}.

\bibitem[{\citenamefont{Huang et~al.}(1983)\citenamefont{Huang, Tarn, and
  Clark}}]{HuangJMP83}
\bibinfo{author}{\bibfnamefont{G.~M.} \bibnamefont{Huang}},
  \bibinfo{author}{\bibfnamefont{T.~J.} \bibnamefont{Tarn}}, \bibnamefont{and}
  \bibinfo{author}{\bibfnamefont{C.~W.} \bibnamefont{Clark}},
  \bibinfo{journal}{J. Math. Phys.} \textbf{\bibinfo{volume}{24}},
  \bibinfo{pages}{2608} (\bibinfo{year}{1983}).

\bibitem[{\citenamefont{Polack et~al.}(2009)\citenamefont{Polack, Suchowski,
  and Tannor}}]{PolackPRA09}
\bibinfo{author}{\bibfnamefont{T.}~\bibnamefont{Polack}},
  \bibinfo{author}{\bibfnamefont{H.}~\bibnamefont{Suchowski}},
  \bibnamefont{and} \bibinfo{author}{\bibfnamefont{D.~J.}
  \bibnamefont{Tannor}}, \bibinfo{journal}{Phys. Rev. A}
  \textbf{\bibinfo{volume}{79}}, \bibinfo{pages}{053403}
  (\bibinfo{year}{2009}).

\bibitem[{\citenamefont{Arenz et~al.}(2014)\citenamefont{Arenz, Gualdi, and
  Burgarth}}]{ArenzNJP14}
\bibinfo{author}{\bibfnamefont{C.}~\bibnamefont{Arenz}},
  \bibinfo{author}{\bibfnamefont{G.}~\bibnamefont{Gualdi}}, \bibnamefont{and}
  \bibinfo{author}{\bibfnamefont{D.}~\bibnamefont{Burgarth}},
  \bibinfo{journal}{New J. Phys.} \textbf{\bibinfo{volume}{16}},
  \bibinfo{pages}{065023} (\bibinfo{year}{2014}).

\bibitem[{\citenamefont{Wu et~al.}(2007)\citenamefont{Wu, Pechen, Brif, and
  Rabitz}}]{WuJPhysA07}
\bibinfo{author}{\bibfnamefont{R.}~\bibnamefont{Wu}},
  \bibinfo{author}{\bibfnamefont{A.}~\bibnamefont{Pechen}},
  \bibinfo{author}{\bibfnamefont{C.}~\bibnamefont{Brif}}, \bibnamefont{and}
  \bibinfo{author}{\bibfnamefont{H.}~\bibnamefont{Rabitz}},
  \bibinfo{journal}{J. Phys. A} \textbf{\bibinfo{volume}{40}},
  \bibinfo{pages}{5681} (\bibinfo{year}{2007}).

\bibitem[{\citenamefont{Ticozzi and Viola}(2014)}]{TicozziSciRep14}
\bibinfo{author}{\bibfnamefont{F.}~\bibnamefont{Ticozzi}} \bibnamefont{and}
  \bibinfo{author}{\bibfnamefont{L.}~\bibnamefont{Viola}},
  \bibinfo{journal}{Sci. Rep.} \textbf{\bibinfo{volume}{4}},
  \bibinfo{pages}{5192} (\bibinfo{year}{2014}).

\bibitem[{\citenamefont{Chambrion et~al.}(2009)\citenamefont{Chambrion, Mason,
  Sigalotti, and Boscain}}]{ChambrionAnnHP09}
\bibinfo{author}{\bibfnamefont{T.}~\bibnamefont{Chambrion}},
  \bibinfo{author}{\bibfnamefont{P.}~\bibnamefont{Mason}},
  \bibinfo{author}{\bibfnamefont{M.}~\bibnamefont{Sigalotti}},
  \bibnamefont{and} \bibinfo{author}{\bibfnamefont{U.}~\bibnamefont{Boscain}},
  \bibinfo{journal}{Ann. Inst. Henri Poincar\'e} \textbf{\bibinfo{volume}{26}},
  \bibinfo{pages}{329} (\bibinfo{year}{2009}).

\bibitem[{\citenamefont{Beauchard et~al.}(2010)\citenamefont{Beauchard, Coron,
  and Rouchon}}]{BeauchardCMP10}
\bibinfo{author}{\bibfnamefont{K.}~\bibnamefont{Beauchard}},
  \bibinfo{author}{\bibfnamefont{J.-M.} \bibnamefont{Coron}}, \bibnamefont{and}
  \bibinfo{author}{\bibfnamefont{P.}~\bibnamefont{Rouchon}},
  \bibinfo{journal}{Commun. Math. Phys.} \textbf{\bibinfo{volume}{296}},
  \bibinfo{pages}{525} (\bibinfo{year}{2010}).

\bibitem[{\citenamefont{Boussaid et~al.}(2013)\citenamefont{Boussaid,
  Caponigro, and Chambrion}}]{BoussaidIEEETrans13}
\bibinfo{author}{\bibfnamefont{N.}~\bibnamefont{Boussaid}},
  \bibinfo{author}{\bibfnamefont{M.}~\bibnamefont{Caponigro}},
  \bibnamefont{and}
  \bibinfo{author}{\bibfnamefont{T.}~\bibnamefont{Chambrion}},
  \bibinfo{journal}{IEEE Trans. Automat. Control}
  \textbf{\bibinfo{volume}{58}}, \bibinfo{pages}{2205} (\bibinfo{year}{2013}).

\bibitem[{\citenamefont{Boscain et~al.}(2015)\citenamefont{Boscain, Gauthier,
  Rossi, and Sigalotti}}]{BoscainCMP15}
\bibinfo{author}{\bibfnamefont{U.}~\bibnamefont{Boscain}},
  \bibinfo{author}{\bibfnamefont{J.-P.} \bibnamefont{Gauthier}},
  \bibinfo{author}{\bibfnamefont{F.}~\bibnamefont{Rossi}}, \bibnamefont{and}
  \bibinfo{author}{\bibfnamefont{M.}~\bibnamefont{Sigalotti}},
  \bibinfo{journal}{Commun. Math. Phys.} \textbf{\bibinfo{volume}{333}},
  \bibinfo{pages}{1225} (\bibinfo{year}{2015}), ISSN \bibinfo{issn}{0010-3616}.

\bibitem[{\citenamefont{Altafini}(2003)}]{AltafiniJMP03}
\bibinfo{author}{\bibfnamefont{C.}~\bibnamefont{Altafini}},
  \bibinfo{journal}{J. Math. Phys.} \textbf{\bibinfo{volume}{44}},
  \bibinfo{pages}{2357} (\bibinfo{year}{2003}).

\bibitem[{\citenamefont{Lloyd and Viola}(2001)}]{LorenzaPRA01}
\bibinfo{author}{\bibfnamefont{S.}~\bibnamefont{Lloyd}} \bibnamefont{and}
  \bibinfo{author}{\bibfnamefont{L.}~\bibnamefont{Viola}},
  \bibinfo{journal}{Phys. Rev. A} \textbf{\bibinfo{volume}{65}},
  \bibinfo{pages}{010101} (\bibinfo{year}{2001}).

\bibitem[{\citenamefont{Dirr et~al.}(2009)\citenamefont{Dirr, Helmke,
  Kurniawan, and Schulte-Herbr\"uggen}}]{ToSHRMP09}
\bibinfo{author}{\bibfnamefont{G.}~\bibnamefont{Dirr}},
  \bibinfo{author}{\bibfnamefont{U.}~\bibnamefont{Helmke}},
  \bibinfo{author}{\bibfnamefont{I.}~\bibnamefont{Kurniawan}},
  \bibnamefont{and}
  \bibinfo{author}{\bibfnamefont{T.}~\bibnamefont{Schulte-Herbr\"uggen}},
  \bibinfo{journal}{Rep. Math. Phys.} \textbf{\bibinfo{volume}{64}},
  \bibinfo{pages}{93 } (\bibinfo{year}{2009}).

\bibitem[{\citenamefont{O'Meara et~al.}(2012)\citenamefont{O'Meara, Dirr, and
  Schulte-Herbruggen}}]{MearaIEEE12}
\bibinfo{author}{\bibfnamefont{C.}~\bibnamefont{O'Meara}},
  \bibinfo{author}{\bibfnamefont{G.}~\bibnamefont{Dirr}}, \bibnamefont{and}
  \bibinfo{author}{\bibfnamefont{T.}~\bibnamefont{Schulte-Herbruggen}},
  \bibinfo{journal}{IEEE Trans. Automatic Control}
  \textbf{\bibinfo{volume}{57}}, \bibinfo{pages}{2050} (\bibinfo{year}{2012}),
  ISSN \bibinfo{issn}{0018-9286}.

\bibitem[{\citenamefont{Altafini and Ticozzi}(2012)}]{AltafiniAC12}
\bibinfo{author}{\bibfnamefont{C.}~\bibnamefont{Altafini}} \bibnamefont{and}
  \bibinfo{author}{\bibfnamefont{F.}~\bibnamefont{Ticozzi}},
  \bibinfo{journal}{IEEE Trans. Automatic Control}
  \textbf{\bibinfo{volume}{57}}, \bibinfo{pages}{1898} (\bibinfo{year}{2012}),
  ISSN \bibinfo{issn}{0018-9286}.

\bibitem[{\citenamefont{Dive et~al.}(2015)\citenamefont{Dive, Burgarth, and
  Mintert}}]{Dive16}
\bibinfo{author}{\bibfnamefont{B.}~\bibnamefont{Dive}},
  \bibinfo{author}{\bibfnamefont{D.}~\bibnamefont{Burgarth}}, \bibnamefont{and}
  \bibinfo{author}{\bibfnamefont{F.}~\bibnamefont{Mintert}},
  \bibinfo{journal}{arXiv:1509.07163}  (\bibinfo{year}{2015}).

\bibitem[{\citenamefont{Lidar et~al.}(1998)\citenamefont{Lidar, Chuang, and
  Whaley}}]{LidarPRL98}
\bibinfo{author}{\bibfnamefont{D.~A.} \bibnamefont{Lidar}},
  \bibinfo{author}{\bibfnamefont{I.~L.} \bibnamefont{Chuang}},
  \bibnamefont{and} \bibinfo{author}{\bibfnamefont{K.~B.}
  \bibnamefont{Whaley}}, \bibinfo{journal}{Phys. Rev. Lett.}
  \textbf{\bibinfo{volume}{81}}, \bibinfo{pages}{2594} (\bibinfo{year}{1998}).

\bibitem[{\citenamefont{Lidar and Whaley}(2003)}]{LidarReviewDFS03}
\bibinfo{author}{\bibfnamefont{D.~A.} \bibnamefont{Lidar}} \bibnamefont{and}
  \bibinfo{author}{\bibfnamefont{K.~B.} \bibnamefont{Whaley}}, in
  \emph{\bibinfo{booktitle}{Irreversible Quantum Dynamics}}, edited by
  \bibinfo{editor}{\bibfnamefont{F.}~\bibnamefont{Benatti}} \bibnamefont{and}
  \bibinfo{editor}{\bibfnamefont{R.}~\bibnamefont{Floreani}}
  (\bibinfo{address}{Berlin}, \bibinfo{year}{2003}), vol. \bibinfo{volume}{622}
  of \emph{\bibinfo{series}{Springer Lecture Notes in Physics}}, pp.
  \bibinfo{pages}{83--120}.

\bibitem[{\citenamefont{Duan and Guo}(1997)}]{LMDuanPRL97}
\bibinfo{author}{\bibfnamefont{L.-M.} \bibnamefont{Duan}} \bibnamefont{and}
  \bibinfo{author}{\bibfnamefont{G.-C.} \bibnamefont{Guo}},
  \bibinfo{journal}{Phys. Rev. Lett.} \textbf{\bibinfo{volume}{79}},
  \bibinfo{pages}{1953} (\bibinfo{year}{1997}).

\bibitem[{\citenamefont{Fortunato et~al.}(2002)\citenamefont{Fortunato, Viola,
  Hodges, Teklemariam, and Cory}}]{FortunatoNJP02}
\bibinfo{author}{\bibfnamefont{E.~M.} \bibnamefont{Fortunato}},
  \bibinfo{author}{\bibfnamefont{L.}~\bibnamefont{Viola}},
  \bibinfo{author}{\bibfnamefont{J.}~\bibnamefont{Hodges}},
  \bibinfo{author}{\bibfnamefont{G.}~\bibnamefont{Teklemariam}},
  \bibnamefont{and} \bibinfo{author}{\bibfnamefont{D.~G.} \bibnamefont{Cory}},
  \bibinfo{journal}{New J. Phys.} \textbf{\bibinfo{volume}{4}},
  \bibinfo{pages}{5} (\bibinfo{year}{2002}).

\bibitem[{\citenamefont{Monz et~al.}(2009)\citenamefont{Monz, Kim, Villar,
  Schindler, Chwalla, Riebe, Roos, H\"{a}ffner, H\"{a}nsel, Hennrich
  et~al.}}]{MonzPRL09}
\bibinfo{author}{\bibfnamefont{T.}~\bibnamefont{Monz}},
  \bibinfo{author}{\bibfnamefont{K.}~\bibnamefont{Kim}},
  \bibinfo{author}{\bibfnamefont{A.~S.} \bibnamefont{Villar}},
  \bibinfo{author}{\bibfnamefont{P.}~\bibnamefont{Schindler}},
  \bibinfo{author}{\bibfnamefont{M.}~\bibnamefont{Chwalla}},
  \bibinfo{author}{\bibfnamefont{M.}~\bibnamefont{Riebe}},
  \bibinfo{author}{\bibfnamefont{C.~F.} \bibnamefont{Roos}},
  \bibinfo{author}{\bibfnamefont{H.}~\bibnamefont{H\"{a}ffner}},
  \bibinfo{author}{\bibfnamefont{W.}~\bibnamefont{H\"{a}nsel}},
  \bibinfo{author}{\bibfnamefont{M.}~\bibnamefont{Hennrich}},
  \bibnamefont{et~al.}, \bibinfo{journal}{Phys. Rev. Lett.}
  \textbf{\bibinfo{volume}{103}}, \bibinfo{pages}{200503}
  (\bibinfo{year}{2009}).

\bibitem[{\citenamefont{Wang et~al.}(2013)\citenamefont{Wang, Byrd, and
  Jacobs}}]{WangJacobsPRA13}
\bibinfo{author}{\bibfnamefont{X.}~\bibnamefont{Wang}},
  \bibinfo{author}{\bibfnamefont{M.}~\bibnamefont{Byrd}}, \bibnamefont{and}
  \bibinfo{author}{\bibfnamefont{K.}~\bibnamefont{Jacobs}},
  \bibinfo{journal}{Phys. Rev. A} \textbf{\bibinfo{volume}{87}},
  \bibinfo{pages}{012338} (\bibinfo{year}{2013}).

\bibitem[{\citenamefont{Yi et~al.}(2009)\citenamefont{Yi, Huang, Wu, and
  Oh}}]{YiPRA09}
\bibinfo{author}{\bibfnamefont{X.~X.} \bibnamefont{Yi}},
  \bibinfo{author}{\bibfnamefont{X.~L.} \bibnamefont{Huang}},
  \bibinfo{author}{\bibfnamefont{C.}~\bibnamefont{Wu}}, \bibnamefont{and}
  \bibinfo{author}{\bibfnamefont{C.~H.} \bibnamefont{Oh}},
  \bibinfo{journal}{Phys. Rev. A} \textbf{\bibinfo{volume}{80}},
  \bibinfo{pages}{052316} (\bibinfo{year}{2009}).

\bibitem[{\citenamefont{Knill et~al.}(2000)\citenamefont{Knill, Laflamme, and
  Viola}}]{LorenzaPRL00}
\bibinfo{author}{\bibfnamefont{E.}~\bibnamefont{Knill}},
  \bibinfo{author}{\bibfnamefont{R.}~\bibnamefont{Laflamme}}, \bibnamefont{and}
  \bibinfo{author}{\bibfnamefont{L.}~\bibnamefont{Viola}},
  \bibinfo{journal}{Phys. Rev. Lett.} \textbf{\bibinfo{volume}{84}},
  \bibinfo{pages}{2525} (\bibinfo{year}{2000}).

\bibitem[{\citenamefont{Viola et~al.}(2001)\citenamefont{Viola, Fortunato,
  Pravia, Knill, Laflamme, and Cory}}]{LorenzaSci01}
\bibinfo{author}{\bibfnamefont{L.}~\bibnamefont{Viola}},
  \bibinfo{author}{\bibfnamefont{E.~M.} \bibnamefont{Fortunato}},
  \bibinfo{author}{\bibfnamefont{M.~A.} \bibnamefont{Pravia}},
  \bibinfo{author}{\bibfnamefont{E.}~\bibnamefont{Knill}},
  \bibinfo{author}{\bibfnamefont{R.}~\bibnamefont{Laflamme}}, \bibnamefont{and}
  \bibinfo{author}{\bibfnamefont{D.~G.} \bibnamefont{Cory}},
  \bibinfo{journal}{Science} \textbf{\bibinfo{volume}{293}},
  \bibinfo{pages}{2059} (\bibinfo{year}{2001}).

\bibitem[{\citenamefont{Waugh}(2007)}]{AHT}
\bibinfo{author}{\bibfnamefont{J.~S.} \bibnamefont{Waugh}},
  \emph{\bibinfo{title}{Average Hamiltonian Theory}} (\bibinfo{publisher}{John
  Wiley \& Sons, Ltd.}, \bibinfo{year}{2007}).

\bibitem[{\citenamefont{Viola et~al.}(1999)\citenamefont{Viola, Knill, and
  Lloyd}}]{LorenzaPRL99}
\bibinfo{author}{\bibfnamefont{L.}~\bibnamefont{Viola}},
  \bibinfo{author}{\bibfnamefont{E.}~\bibnamefont{Knill}}, \bibnamefont{and}
  \bibinfo{author}{\bibfnamefont{S.}~\bibnamefont{Lloyd}},
  \bibinfo{journal}{Phys. Rev. Lett.} \textbf{\bibinfo{volume}{82}},
  \bibinfo{pages}{2417} (\bibinfo{year}{1999}).

\bibitem[{\citenamefont{Viola}(2004)}]{LorenzaJMO04}
\bibinfo{author}{\bibfnamefont{L.}~\bibnamefont{Viola}}, \bibinfo{journal}{J.
  Mod. Opt.} \textbf{\bibinfo{volume}{51}}, \bibinfo{pages}{2357}
  (\bibinfo{year}{2004}).

\bibitem[{\citenamefont{Souza et~al.}(2012)\citenamefont{Souza, {\'A}lvarez,
  and Suter}}]{SouzaPTRSA12}
\bibinfo{author}{\bibfnamefont{A.~M.} \bibnamefont{Souza}},
  \bibinfo{author}{\bibfnamefont{G.~A.} \bibnamefont{{\'A}lvarez}},
  \bibnamefont{and} \bibinfo{author}{\bibfnamefont{D.}~\bibnamefont{Suter}},
  \bibinfo{journal}{Phil. Trans. R. Soc. A} \textbf{\bibinfo{volume}{370}},
  \bibinfo{pages}{4748} (\bibinfo{year}{2012}), ISSN \bibinfo{issn}{1364-503X}.

\bibitem[{\citenamefont{Suter and \'Alvarez}(2016)}]{SuterRMP15}
\bibinfo{author}{\bibfnamefont{D.}~\bibnamefont{Suter}} \bibnamefont{and}
  \bibinfo{author}{\bibfnamefont{G.~A.} \bibnamefont{\'Alvarez}},
  \bibinfo{journal}{Rev. Mod. Phys.} p. \bibinfo{pages}{in press}
  (\bibinfo{year}{2016}).

\bibitem[{\citenamefont{Khodjasteh and Lidar}(2005)}]{KhodjastehPRL05}
\bibinfo{author}{\bibfnamefont{K.}~\bibnamefont{Khodjasteh}} \bibnamefont{and}
  \bibinfo{author}{\bibfnamefont{D.~A.} \bibnamefont{Lidar}},
  \bibinfo{journal}{Phys. Rev. Lett.} \textbf{\bibinfo{volume}{95}},
  \bibinfo{pages}{180501} (\bibinfo{year}{2005}).

\bibitem[{\citenamefont{Biercuk et~al.}(2009)\citenamefont{Biercuk, Uys,
  VanDevender, Shiga, Itano, and Bollinger}}]{BiercukNature09}
\bibinfo{author}{\bibfnamefont{M.~J.} \bibnamefont{Biercuk}},
  \bibinfo{author}{\bibfnamefont{H.}~\bibnamefont{Uys}},
  \bibinfo{author}{\bibfnamefont{A.~P.} \bibnamefont{VanDevender}},
  \bibinfo{author}{\bibfnamefont{N.}~\bibnamefont{Shiga}},
  \bibinfo{author}{\bibfnamefont{W.~M.} \bibnamefont{Itano}}, \bibnamefont{and}
  \bibinfo{author}{\bibfnamefont{J.~J.} \bibnamefont{Bollinger}},
  \bibinfo{journal}{Nature} \textbf{\bibinfo{volume}{458}},
  \bibinfo{pages}{996} (\bibinfo{year}{2009}).

\bibitem[{\citenamefont{Bylander et~al.}(2011)\citenamefont{Bylander,
  Gustavsson, Yan, Yoshihara, Harrabi, Fitch, Cory, Nakamura, Tsai, and
  Oliver}}]{BylanderNatPhys11}
\bibinfo{author}{\bibfnamefont{J.}~\bibnamefont{Bylander}},
  \bibinfo{author}{\bibfnamefont{S.}~\bibnamefont{Gustavsson}},
  \bibinfo{author}{\bibfnamefont{F.}~\bibnamefont{Yan}},
  \bibinfo{author}{\bibfnamefont{F.}~\bibnamefont{Yoshihara}},
  \bibinfo{author}{\bibfnamefont{K.}~\bibnamefont{Harrabi}},
  \bibinfo{author}{\bibfnamefont{G.}~\bibnamefont{Fitch}},
  \bibinfo{author}{\bibfnamefont{D.~G.} \bibnamefont{Cory}},
  \bibinfo{author}{\bibfnamefont{Y.}~\bibnamefont{Nakamura}},
  \bibinfo{author}{\bibfnamefont{J.-S.} \bibnamefont{Tsai}}, \bibnamefont{and}
  \bibinfo{author}{\bibfnamefont{W.~D.} \bibnamefont{Oliver}},
  \bibinfo{journal}{Nature Phys.} \textbf{\bibinfo{volume}{7}},
  \bibinfo{pages}{565–570} (\bibinfo{year}{2011}).

\bibitem[{\citenamefont{Green et~al.}(2012)\citenamefont{Green, Uys, and
  Biercuk}}]{GreenPRL12}
\bibinfo{author}{\bibfnamefont{T.}~\bibnamefont{Green}},
  \bibinfo{author}{\bibfnamefont{H.}~\bibnamefont{Uys}}, \bibnamefont{and}
  \bibinfo{author}{\bibfnamefont{M.~J.} \bibnamefont{Biercuk}},
  \bibinfo{journal}{Phys. Rev. Lett.} \textbf{\bibinfo{volume}{109}},
  \bibinfo{pages}{020501} (\bibinfo{year}{2012}).

\bibitem[{\citenamefont{Paz-Silva and Viola}(2014)}]{Paz-SilvaPRL14}
\bibinfo{author}{\bibfnamefont{G.~A.} \bibnamefont{Paz-Silva}}
  \bibnamefont{and} \bibinfo{author}{\bibfnamefont{L.}~\bibnamefont{Viola}},
  \bibinfo{journal}{Phys. Rev. Lett.} \textbf{\bibinfo{volume}{113}},
  \bibinfo{pages}{250501} (\bibinfo{year}{2014}).

\bibitem[{\citenamefont{Soare et~al.}(2014)\citenamefont{Soare, Ball, Hayes,
  Sastrawan, Jarratt, McLoughlin, Zhen, Green, and Biercuk}}]{SoareNatPhys14}
\bibinfo{author}{\bibfnamefont{A.}~\bibnamefont{Soare}},
  \bibinfo{author}{\bibfnamefont{H.}~\bibnamefont{Ball}},
  \bibinfo{author}{\bibfnamefont{D.}~\bibnamefont{Hayes}},
  \bibinfo{author}{\bibfnamefont{J.}~\bibnamefont{Sastrawan}},
  \bibinfo{author}{\bibfnamefont{M.~C.} \bibnamefont{Jarratt}},
  \bibinfo{author}{\bibfnamefont{J.~J.} \bibnamefont{McLoughlin}},
  \bibinfo{author}{\bibfnamefont{X.}~\bibnamefont{Zhen}},
  \bibinfo{author}{\bibfnamefont{T.~J.} \bibnamefont{Green}}, \bibnamefont{and}
  \bibinfo{author}{\bibfnamefont{M.~J.} \bibnamefont{Biercuk}},
  \bibinfo{journal}{Nature Phys.} \textbf{\bibinfo{volume}{10}},
  \bibinfo{pages}{825} (\bibinfo{year}{2014}).

\bibitem[{\citenamefont{Clausen et~al.}(2010)\citenamefont{Clausen, Bensky, and
  Kurizki}}]{ClausenPRL10}
\bibinfo{author}{\bibfnamefont{J.}~\bibnamefont{Clausen}},
  \bibinfo{author}{\bibfnamefont{G.}~\bibnamefont{Bensky}}, \bibnamefont{and}
  \bibinfo{author}{\bibfnamefont{G.}~\bibnamefont{Kurizki}},
  \bibinfo{journal}{Phys. Rev. Lett.} \textbf{\bibinfo{volume}{104}},
  \bibinfo{pages}{040401} (\bibinfo{year}{2010}).

\bibitem[{\citenamefont{Viyuela et~al.}(2012)\citenamefont{Viyuela, Rivas, and
  Martin-Delgado}}]{ViyuelaPRB12}
\bibinfo{author}{\bibfnamefont{O.}~\bibnamefont{Viyuela}},
  \bibinfo{author}{\bibfnamefont{A.}~\bibnamefont{Rivas}}, \bibnamefont{and}
  \bibinfo{author}{\bibfnamefont{M.~A.} \bibnamefont{Martin-Delgado}},
  \bibinfo{journal}{Phys. Rev. B} \textbf{\bibinfo{volume}{86}},
  \bibinfo{pages}{155140} (\bibinfo{year}{2012}).

\bibitem[{\citenamefont{Rivas et~al.}(2013)\citenamefont{Rivas, Viyuela, and
  Martin-Delgado}}]{RivasPRB13}
\bibinfo{author}{\bibfnamefont{A.}~\bibnamefont{Rivas}},
  \bibinfo{author}{\bibfnamefont{O.}~\bibnamefont{Viyuela}}, \bibnamefont{and}
  \bibinfo{author}{\bibfnamefont{M.~A.} \bibnamefont{Martin-Delgado}},
  \bibinfo{journal}{Phys. Rev. B} \textbf{\bibinfo{volume}{88}},
  \bibinfo{pages}{155141} (\bibinfo{year}{2013}).

\bibitem[{\citenamefont{Cywi\ifmmode~\acute{n}\else \'{n}\fi{}ski
  et~al.}(2008)\citenamefont{Cywi\ifmmode~\acute{n}\else \'{n}\fi{}ski,
  Lutchyn, Nave, and Das~Sarma}}]{CywinskiPRB08}
\bibinfo{author}{\bibfnamefont{L.}~\bibnamefont{Cywi\ifmmode~\acute{n}\else
  \'{n}\fi{}ski}}, \bibinfo{author}{\bibfnamefont{R.~M.}
  \bibnamefont{Lutchyn}}, \bibinfo{author}{\bibfnamefont{C.~P.}
  \bibnamefont{Nave}}, \bibnamefont{and}
  \bibinfo{author}{\bibfnamefont{S.}~\bibnamefont{Das~Sarma}},
  \bibinfo{journal}{Phys. Rev. B} \textbf{\bibinfo{volume}{77}},
  \bibinfo{pages}{174509} (\bibinfo{year}{2008}).

\bibitem[{\citenamefont{Kotler et~al.}(2011)\citenamefont{Kotler, Akerman,
  Glickman, Keselman, and Ozeri}}]{KotlerNature11}
\bibinfo{author}{\bibfnamefont{S.}~\bibnamefont{Kotler}},
  \bibinfo{author}{\bibfnamefont{N.}~\bibnamefont{Akerman}},
  \bibinfo{author}{\bibfnamefont{Y.}~\bibnamefont{Glickman}},
  \bibinfo{author}{\bibfnamefont{A.}~\bibnamefont{Keselman}}, \bibnamefont{and}
  \bibinfo{author}{\bibfnamefont{R.}~\bibnamefont{Ozeri}},
  \bibinfo{journal}{Nature} \textbf{\bibinfo{volume}{473}}, \bibinfo{pages}{61}
  (\bibinfo{year}{2011}).

\bibitem[{\citenamefont{Kotler et~al.}(2013)\citenamefont{Kotler, Akerman,
  Glickman, and Ozeri}}]{KotlerPRL13}
\bibinfo{author}{\bibfnamefont{S.}~\bibnamefont{Kotler}},
  \bibinfo{author}{\bibfnamefont{N.}~\bibnamefont{Akerman}},
  \bibinfo{author}{\bibfnamefont{Y.}~\bibnamefont{Glickman}}, \bibnamefont{and}
  \bibinfo{author}{\bibfnamefont{R.}~\bibnamefont{Ozeri}},
  \bibinfo{journal}{Phys. Rev. Lett.} \textbf{\bibinfo{volume}{110}},
  \bibinfo{pages}{110503} (\bibinfo{year}{2013}).

\bibitem[{\citenamefont{Khaneja
  et~al.}(2003{\natexlab{a}})\citenamefont{Khaneja, Luy, and
  Glaser}}]{KhanejaPNAS03}
\bibinfo{author}{\bibfnamefont{N.}~\bibnamefont{Khaneja}},
  \bibinfo{author}{\bibfnamefont{B.}~\bibnamefont{Luy}}, \bibnamefont{and}
  \bibinfo{author}{\bibfnamefont{S.~J.} \bibnamefont{Glaser}},
  \bibinfo{journal}{Proc. Natl. Acad. Sci. USA} \textbf{\bibinfo{volume}{100}},
  \bibinfo{pages}{13162} (\bibinfo{year}{2003}{\natexlab{a}}).

\bibitem[{\citenamefont{Khaneja
  et~al.}(2003{\natexlab{b}})\citenamefont{Khaneja, Reiss, Luy, and
  Glaser}}]{KhanejaJMR03}
\bibinfo{author}{\bibfnamefont{N.}~\bibnamefont{Khaneja}},
  \bibinfo{author}{\bibfnamefont{T.}~\bibnamefont{Reiss}},
  \bibinfo{author}{\bibfnamefont{B.}~\bibnamefont{Luy}}, \bibnamefont{and}
  \bibinfo{author}{\bibfnamefont{S.~J.} \bibnamefont{Glaser}},
  \bibinfo{journal}{J. Magn. Reson.} \textbf{\bibinfo{volume}{162}},
  \bibinfo{pages}{311 } (\bibinfo{year}{2003}{\natexlab{b}}).

\bibitem[{\citenamefont{Lapert et~al.}(2010)\citenamefont{Lapert, Zhang, Braun,
  Glaser, and Sugny}}]{LapertPRL10}
\bibinfo{author}{\bibfnamefont{M.}~\bibnamefont{Lapert}},
  \bibinfo{author}{\bibfnamefont{Y.}~\bibnamefont{Zhang}},
  \bibinfo{author}{\bibfnamefont{M.}~\bibnamefont{Braun}},
  \bibinfo{author}{\bibfnamefont{S.~J.} \bibnamefont{Glaser}},
  \bibnamefont{and} \bibinfo{author}{\bibfnamefont{D.}~\bibnamefont{Sugny}},
  \bibinfo{journal}{Phys. Rev. Lett.} \textbf{\bibinfo{volume}{104}},
  \bibinfo{pages}{083001} (\bibinfo{year}{2010}).

\bibitem[{\citenamefont{Lapert et~al.}(2013)\citenamefont{Lapert, Ass\'emat,
  Glaser, and Sugny}}]{LapertPRA13}
\bibinfo{author}{\bibfnamefont{M.}~\bibnamefont{Lapert}},
  \bibinfo{author}{\bibfnamefont{E.}~\bibnamefont{Ass\'emat}},
  \bibinfo{author}{\bibfnamefont{S.~J.} \bibnamefont{Glaser}},
  \bibnamefont{and} \bibinfo{author}{\bibfnamefont{D.}~\bibnamefont{Sugny}},
  \bibinfo{journal}{Phys. Rev. A} \textbf{\bibinfo{volume}{88}},
  \bibinfo{pages}{033407} (\bibinfo{year}{2013}).

\bibitem[{\citenamefont{Mukherjee et~al.}(2013)\citenamefont{Mukherjee,
  Carlini, Mari, Caneva, Montangero, Calarco, Fazio, and
  Giovannetti}}]{MukherjeePRA13}
\bibinfo{author}{\bibfnamefont{V.}~\bibnamefont{Mukherjee}},
  \bibinfo{author}{\bibfnamefont{A.}~\bibnamefont{Carlini}},
  \bibinfo{author}{\bibfnamefont{A.}~\bibnamefont{Mari}},
  \bibinfo{author}{\bibfnamefont{T.}~\bibnamefont{Caneva}},
  \bibinfo{author}{\bibfnamefont{S.}~\bibnamefont{Montangero}},
  \bibinfo{author}{\bibfnamefont{T.}~\bibnamefont{Calarco}},
  \bibinfo{author}{\bibfnamefont{R.}~\bibnamefont{Fazio}}, \bibnamefont{and}
  \bibinfo{author}{\bibfnamefont{V.}~\bibnamefont{Giovannetti}},
  \bibinfo{journal}{Phys. Rev. A} \textbf{\bibinfo{volume}{88}},
  \bibinfo{pages}{062326} (\bibinfo{year}{2013}).

\bibitem[{\citenamefont{Mukherjee et~al.}(2015)\citenamefont{Mukherjee,
  Giovannetti, Fazio, Huelga, Calarco, and Montangero}}]{MukherjeeNJP15}
\bibinfo{author}{\bibfnamefont{V.}~\bibnamefont{Mukherjee}},
  \bibinfo{author}{\bibfnamefont{V.}~\bibnamefont{Giovannetti}},
  \bibinfo{author}{\bibfnamefont{R.}~\bibnamefont{Fazio}},
  \bibinfo{author}{\bibfnamefont{S.~F.} \bibnamefont{Huelga}},
  \bibinfo{author}{\bibfnamefont{T.}~\bibnamefont{Calarco}}, \bibnamefont{and}
  \bibinfo{author}{\bibfnamefont{S.}~\bibnamefont{Montangero}},
  \bibinfo{journal}{New J. Phys.} \textbf{\bibinfo{volume}{17}},
  \bibinfo{pages}{063031} (\bibinfo{year}{2015}).

\bibitem[{\citenamefont{Rezek et~al.}(2009)\citenamefont{Rezek, Salamon,
  Hoffmann, and Kosloff}}]{RezekEPL09}
\bibinfo{author}{\bibfnamefont{Y.}~\bibnamefont{Rezek}},
  \bibinfo{author}{\bibfnamefont{P.}~\bibnamefont{Salamon}},
  \bibinfo{author}{\bibfnamefont{K.~H.} \bibnamefont{Hoffmann}},
  \bibnamefont{and} \bibinfo{author}{\bibfnamefont{R.}~\bibnamefont{Kosloff}},
  \bibinfo{journal}{Europhys. Lett.} \textbf{\bibinfo{volume}{85}},
  \bibinfo{pages}{30008} (\bibinfo{year}{2009}).

\bibitem[{\citenamefont{Hoffmann et~al.}(2011)\citenamefont{Hoffmann, Salamon,
  Rezek, and Kosloff}}]{HoffmannEPL11}
\bibinfo{author}{\bibfnamefont{K.~H.} \bibnamefont{Hoffmann}},
  \bibinfo{author}{\bibfnamefont{P.}~\bibnamefont{Salamon}},
  \bibinfo{author}{\bibfnamefont{Y.}~\bibnamefont{Rezek}}, \bibnamefont{and}
  \bibinfo{author}{\bibfnamefont{R.}~\bibnamefont{Kosloff}},
  \bibinfo{journal}{Europhys. Lett.} \textbf{\bibinfo{volume}{96}},
  \bibinfo{pages}{60015} (\bibinfo{year}{2011}).

\bibitem[{\citenamefont{Sklarz et~al.}(2004)\citenamefont{Sklarz, Tannor, and
  Khaneja}}]{SklarzPRA04}
\bibinfo{author}{\bibfnamefont{S.~E.} \bibnamefont{Sklarz}},
  \bibinfo{author}{\bibfnamefont{D.~J.} \bibnamefont{Tannor}},
  \bibnamefont{and} \bibinfo{author}{\bibfnamefont{N.}~\bibnamefont{Khaneja}},
  \bibinfo{journal}{Phys. Rev. A} \textbf{\bibinfo{volume}{69}},
  \bibinfo{pages}{053408} (\bibinfo{year}{2004}).

\bibitem[{\citenamefont{Yuan et~al.}(2012)\citenamefont{Yuan, Koch, Salamon,
  and Tannor}}]{HaidongPRA12}
\bibinfo{author}{\bibfnamefont{H.}~\bibnamefont{Yuan}},
  \bibinfo{author}{\bibfnamefont{C.~P.} \bibnamefont{Koch}},
  \bibinfo{author}{\bibfnamefont{P.}~\bibnamefont{Salamon}}, \bibnamefont{and}
  \bibinfo{author}{\bibfnamefont{D.~J.} \bibnamefont{Tannor}},
  \bibinfo{journal}{Phys. Rev. A} \textbf{\bibinfo{volume}{85}},
  \bibinfo{pages}{033417} (\bibinfo{year}{2012}).

\bibitem[{\citenamefont{Bartana et~al.}(1993)\citenamefont{Bartana, Kosloff,
  and Tannor}}]{BartanaJCP93}
\bibinfo{author}{\bibfnamefont{A.}~\bibnamefont{Bartana}},
  \bibinfo{author}{\bibfnamefont{R.}~\bibnamefont{Kosloff}}, \bibnamefont{and}
  \bibinfo{author}{\bibfnamefont{D.~J.} \bibnamefont{Tannor}},
  \bibinfo{journal}{J. Chem. Phys.} \textbf{\bibinfo{volume}{99}},
  \bibinfo{pages}{196} (\bibinfo{year}{1993}).

\bibitem[{\citenamefont{Tannor and Bartana}(1999)}]{TannorJPCA99}
\bibinfo{author}{\bibfnamefont{D.~J.} \bibnamefont{Tannor}} \bibnamefont{and}
  \bibinfo{author}{\bibfnamefont{A.}~\bibnamefont{Bartana}},
  \bibinfo{journal}{J. Phys. Chem. A} \textbf{\bibinfo{volume}{103}},
  \bibinfo{pages}{10359} (\bibinfo{year}{1999}).

\bibitem[{\citenamefont{Schmidt et~al.}(2012)\citenamefont{Schmidt, Rohrer,
  Ankerhold, and Stockburger}}]{SchmidtPS12}
\bibinfo{author}{\bibfnamefont{R.}~\bibnamefont{Schmidt}},
  \bibinfo{author}{\bibfnamefont{S.}~\bibnamefont{Rohrer}},
  \bibinfo{author}{\bibfnamefont{J.}~\bibnamefont{Ankerhold}},
  \bibnamefont{and} \bibinfo{author}{\bibfnamefont{J.~T.}
  \bibnamefont{Stockburger}}, \bibinfo{journal}{Physica Scripta}
  \textbf{\bibinfo{volume}{2012}}, \bibinfo{pages}{014034}
  (\bibinfo{year}{2012}).

\bibitem[{\citenamefont{Ohtsuki et~al.}(1999)\citenamefont{Ohtsuki, Zhu, and
  Rabitz}}]{OhtsukiJCP99}
\bibinfo{author}{\bibfnamefont{Y.}~\bibnamefont{Ohtsuki}},
  \bibinfo{author}{\bibfnamefont{W.}~\bibnamefont{Zhu}}, \bibnamefont{and}
  \bibinfo{author}{\bibfnamefont{H.}~\bibnamefont{Rabitz}},
  \bibinfo{journal}{J. Chem. Phys.} \textbf{\bibinfo{volume}{110}},
  \bibinfo{pages}{9825} (\bibinfo{year}{1999}).

\bibitem[{\citenamefont{Khaneja et~al.}(2005)\citenamefont{Khaneja, Reiss,
  Kehlet, Schulte-Herbr\"uggen, and Glaser}}]{KhanejaJMR05}
\bibinfo{author}{\bibfnamefont{N.}~\bibnamefont{Khaneja}},
  \bibinfo{author}{\bibfnamefont{T.}~\bibnamefont{Reiss}},
  \bibinfo{author}{\bibfnamefont{C.}~\bibnamefont{Kehlet}},
  \bibinfo{author}{\bibfnamefont{T.}~\bibnamefont{Schulte-Herbr\"uggen}},
  \bibnamefont{and} \bibinfo{author}{\bibfnamefont{S.~J.}
  \bibnamefont{Glaser}}, \bibinfo{journal}{J. Magn. Reson.}
  \textbf{\bibinfo{volume}{172}}, \bibinfo{pages}{296 } (\bibinfo{year}{2005}).

\bibitem[{\citenamefont{Sundermann and Vivie-Riedle}(1999)}]{SundermannJCP99}
\bibinfo{author}{\bibfnamefont{K.}~\bibnamefont{Sundermann}} \bibnamefont{and}
  \bibinfo{author}{\bibfnamefont{R.~d.} \bibnamefont{Vivie-Riedle}},
  \bibinfo{journal}{J. Chem. Phys.} \textbf{\bibinfo{volume}{110}},
  \bibinfo{pages}{1896} (\bibinfo{year}{1999}).

\bibitem[{\citenamefont{Eitan et~al.}(2011)\citenamefont{Eitan, Mundt, and
  Tannor}}]{EitanPRA11}
\bibinfo{author}{\bibfnamefont{R.}~\bibnamefont{Eitan}},
  \bibinfo{author}{\bibfnamefont{M.}~\bibnamefont{Mundt}}, \bibnamefont{and}
  \bibinfo{author}{\bibfnamefont{D.~J.} \bibnamefont{Tannor}},
  \bibinfo{journal}{Phys. Rev. A} \textbf{\bibinfo{volume}{83}},
  \bibinfo{pages}{053426} (\bibinfo{year}{2011}).

\bibitem[{\citenamefont{Goerz et~al.}(2015{\natexlab{b}})\citenamefont{Goerz,
  Whaley, and Koch}}]{GoerzEPJQT15}
\bibinfo{author}{\bibfnamefont{M.~H.} \bibnamefont{Goerz}},
  \bibinfo{author}{\bibfnamefont{K.~B.} \bibnamefont{Whaley}},
  \bibnamefont{and} \bibinfo{author}{\bibfnamefont{C.~P.} \bibnamefont{Koch}},
  \bibinfo{journal}{EPJ Quantum Technology} \textbf{\bibinfo{volume}{2}},
  \bibinfo{pages}{21} (\bibinfo{year}{2015}{\natexlab{b}}), ISSN
  \bibinfo{issn}{2196-0763}.

\bibitem[{\citenamefont{Palao et~al.}(2013)\citenamefont{Palao, Reich, and
  Koch}}]{PalaoPRA13}
\bibinfo{author}{\bibfnamefont{J.~P.} \bibnamefont{Palao}},
  \bibinfo{author}{\bibfnamefont{D.~M.} \bibnamefont{Reich}}, \bibnamefont{and}
  \bibinfo{author}{\bibfnamefont{C.~P.} \bibnamefont{Koch}},
  \bibinfo{journal}{Phys. Rev. A} \textbf{\bibinfo{volume}{88}},
  \bibinfo{pages}{053409} (\bibinfo{year}{2013}).

\bibitem[{\citenamefont{Reich et~al.}(2014{\natexlab{a}})\citenamefont{Reich,
  Palao, and Koch}}]{ReichJMO14}
\bibinfo{author}{\bibfnamefont{D.~M.} \bibnamefont{Reich}},
  \bibinfo{author}{\bibfnamefont{J.~P.} \bibnamefont{Palao}}, \bibnamefont{and}
  \bibinfo{author}{\bibfnamefont{C.~P.} \bibnamefont{Koch}},
  \bibinfo{journal}{J. Mod. Opt.} \textbf{\bibinfo{volume}{61}},
  \bibinfo{pages}{822} (\bibinfo{year}{2014}{\natexlab{a}}).

\bibitem[{\citenamefont{Goetz et~al.}(2016)\citenamefont{Goetz, Karamatskou,
  Santra, and Koch}}]{GoetzPRA16}
\bibinfo{author}{\bibfnamefont{R.~E.} \bibnamefont{Goetz}},
  \bibinfo{author}{\bibfnamefont{A.}~\bibnamefont{Karamatskou}},
  \bibinfo{author}{\bibfnamefont{R.}~\bibnamefont{Santra}}, \bibnamefont{and}
  \bibinfo{author}{\bibfnamefont{C.~P.} \bibnamefont{Koch}},
  \bibinfo{journal}{Phys. Rev. A} \textbf{\bibinfo{volume}{93}},
  \bibinfo{pages}{013413} (\bibinfo{year}{2016}).

\bibitem[{\citenamefont{Moore and Rabitz}(2012)}]{MooreJCP12}
\bibinfo{author}{\bibfnamefont{K.~W.} \bibnamefont{Moore}} \bibnamefont{and}
  \bibinfo{author}{\bibfnamefont{H.}~\bibnamefont{Rabitz}},
  \bibinfo{journal}{J. Chem. Phys.} \textbf{\bibinfo{volume}{137}},
  \bibinfo{pages}{134113} (\bibinfo{year}{2012}).

\bibitem[{\citenamefont{Riviello et~al.}(2015)\citenamefont{Riviello,
  Moore~Tibbetts, Brif, Long, Wu, Ho, and Rabitz}}]{RivielloPRA15}
\bibinfo{author}{\bibfnamefont{G.}~\bibnamefont{Riviello}},
  \bibinfo{author}{\bibfnamefont{K.}~\bibnamefont{Moore~Tibbetts}},
  \bibinfo{author}{\bibfnamefont{C.}~\bibnamefont{Brif}},
  \bibinfo{author}{\bibfnamefont{R.}~\bibnamefont{Long}},
  \bibinfo{author}{\bibfnamefont{R.-B.} \bibnamefont{Wu}},
  \bibinfo{author}{\bibfnamefont{T.-S.} \bibnamefont{Ho}}, \bibnamefont{and}
  \bibinfo{author}{\bibfnamefont{H.}~\bibnamefont{Rabitz}},
  \bibinfo{journal}{Phys. Rev. A} \textbf{\bibinfo{volume}{91}},
  \bibinfo{pages}{043401} (\bibinfo{year}{2015}).

\bibitem[{\citenamefont{Gorman et~al.}(2012)\citenamefont{Gorman, Young, and
  Whaley}}]{GormanPRA12}
\bibinfo{author}{\bibfnamefont{D.~J.} \bibnamefont{Gorman}},
  \bibinfo{author}{\bibfnamefont{K.~C.} \bibnamefont{Young}}, \bibnamefont{and}
  \bibinfo{author}{\bibfnamefont{K.~B.} \bibnamefont{Whaley}},
  \bibinfo{journal}{Phys. Rev. A} \textbf{\bibinfo{volume}{86}},
  \bibinfo{pages}{012317} (\bibinfo{year}{2012}).

\bibitem[{\citenamefont{Quiroz and Lidar}(2013)}]{QuirozPRA13}
\bibinfo{author}{\bibfnamefont{G.}~\bibnamefont{Quiroz}} \bibnamefont{and}
  \bibinfo{author}{\bibfnamefont{D.~A.} \bibnamefont{Lidar}},
  \bibinfo{journal}{Phys. Rev. A} \textbf{\bibinfo{volume}{88}},
  \bibinfo{pages}{052306} (\bibinfo{year}{2013}).

\bibitem[{\citenamefont{Paz-Silva and Lidar}(2013)}]{Paz-SilvaSciRep13}
\bibinfo{author}{\bibfnamefont{G.~A.} \bibnamefont{Paz-Silva}}
  \bibnamefont{and} \bibinfo{author}{\bibfnamefont{D.~A.} \bibnamefont{Lidar}},
  \bibinfo{journal}{Sci. Rep.} \textbf{\bibinfo{volume}{3}},
  \bibinfo{pages}{1530} (\bibinfo{year}{2013}).

\bibitem[{\citenamefont{Katz et~al.}(2007)\citenamefont{Katz, Ratner, and
  Kosloff}}]{KatzPRL07}
\bibinfo{author}{\bibfnamefont{G.}~\bibnamefont{Katz}},
  \bibinfo{author}{\bibfnamefont{M.~A.} \bibnamefont{Ratner}},
  \bibnamefont{and} \bibinfo{author}{\bibfnamefont{R.}~\bibnamefont{Kosloff}},
  \bibinfo{journal}{Phys. Rev. Lett.} \textbf{\bibinfo{volume}{98}},
  \bibinfo{pages}{203006} (\bibinfo{year}{2007}).

\bibitem[{\citenamefont{Stefanatos et~al.}(2004)\citenamefont{Stefanatos,
  Khaneja, and Glaser}}]{StefanatosPRA04}
\bibinfo{author}{\bibfnamefont{D.}~\bibnamefont{Stefanatos}},
  \bibinfo{author}{\bibfnamefont{N.}~\bibnamefont{Khaneja}}, \bibnamefont{and}
  \bibinfo{author}{\bibfnamefont{S.~J.} \bibnamefont{Glaser}},
  \bibinfo{journal}{Phys. Rev. A} \textbf{\bibinfo{volume}{69}},
  \bibinfo{pages}{022319} (\bibinfo{year}{2004}).

\bibitem[{\citenamefont{Gershenzon et~al.}(2007)\citenamefont{Gershenzon,
  Kobzar, Luy, Glaser, and Skinner}}]{GershenzonJMR08}
\bibinfo{author}{\bibfnamefont{N.~I.} \bibnamefont{Gershenzon}},
  \bibinfo{author}{\bibfnamefont{K.}~\bibnamefont{Kobzar}},
  \bibinfo{author}{\bibfnamefont{B.}~\bibnamefont{Luy}},
  \bibinfo{author}{\bibfnamefont{S.~J.} \bibnamefont{Glaser}},
  \bibnamefont{and} \bibinfo{author}{\bibfnamefont{T.~E.}
  \bibnamefont{Skinner}}, \bibinfo{journal}{J. Magn. Reson.}
  \textbf{\bibinfo{volume}{188}}, \bibinfo{pages}{330 } (\bibinfo{year}{2007}).

\bibitem[{\citenamefont{Giovannetti et~al.}(2003)\citenamefont{Giovannetti,
  Lloyd, and Maccone}}]{GiovannettiPRA03}
\bibinfo{author}{\bibfnamefont{V.}~\bibnamefont{Giovannetti}},
  \bibinfo{author}{\bibfnamefont{S.}~\bibnamefont{Lloyd}}, \bibnamefont{and}
  \bibinfo{author}{\bibfnamefont{L.}~\bibnamefont{Maccone}},
  \bibinfo{journal}{Phys. Rev. A} \textbf{\bibinfo{volume}{67}},
  \bibinfo{pages}{052109} (\bibinfo{year}{2003}).

\bibitem[{\citenamefont{Levitin and Toffoli}(2009)}]{LevitinPRL09}
\bibinfo{author}{\bibfnamefont{L.~B.} \bibnamefont{Levitin}} \bibnamefont{and}
  \bibinfo{author}{\bibfnamefont{T.}~\bibnamefont{Toffoli}},
  \bibinfo{journal}{Phys. Rev. Lett.} \textbf{\bibinfo{volume}{103}},
  \bibinfo{pages}{160502} (\bibinfo{year}{2009}).

\bibitem[{\citenamefont{Hegerfeldt}(2013)}]{HegerfeldtPRL13}
\bibinfo{author}{\bibfnamefont{G.~C.} \bibnamefont{Hegerfeldt}},
  \bibinfo{journal}{Phys. Rev. Lett.} \textbf{\bibinfo{volume}{111}},
  \bibinfo{pages}{260501} (\bibinfo{year}{2013}).

\bibitem[{\citenamefont{Goerz}(2015)}]{GoerzPhD}
\bibinfo{author}{\bibfnamefont{M.~H.} \bibnamefont{Goerz}}, Ph.D. thesis,
  \bibinfo{school}{Universit\"at Kassel} (\bibinfo{year}{2015}),
  \urlprefix\url{http://nbn-resolving.de/urn:nbn:de:hebis:34-2015052748381}.

\bibitem[{\citenamefont{Hohenester and Stadler}(2004)}]{HohenesterPRL04}
\bibinfo{author}{\bibfnamefont{U.}~\bibnamefont{Hohenester}} \bibnamefont{and}
  \bibinfo{author}{\bibfnamefont{G.}~\bibnamefont{Stadler}},
  \bibinfo{journal}{Phys. Rev. Lett.} \textbf{\bibinfo{volume}{92}},
  \bibinfo{pages}{196801} (\bibinfo{year}{2004}).

\bibitem[{\citenamefont{Hohenester}(2007)}]{HohenesterJPB07}
\bibinfo{author}{\bibfnamefont{U.}~\bibnamefont{Hohenester}},
  \bibinfo{journal}{J. Phys. B} \textbf{\bibinfo{volume}{40}},
  \bibinfo{pages}{S315} (\bibinfo{year}{2007}).

\bibitem[{\citenamefont{Deffner and Lutz}(2013)}]{DeffnerPRL13}
\bibinfo{author}{\bibfnamefont{S.}~\bibnamefont{Deffner}} \bibnamefont{and}
  \bibinfo{author}{\bibfnamefont{E.}~\bibnamefont{Lutz}},
  \bibinfo{journal}{Phys. Rev. Lett.} \textbf{\bibinfo{volume}{111}},
  \bibinfo{pages}{010402} (\bibinfo{year}{2013}).

\bibitem[{\citenamefont{Cimmarusti et~al.}(2015)\citenamefont{Cimmarusti, Yan,
  Patterson, Corcos, Orozco, and Deffner}}]{CimmarustiPRL15}
\bibinfo{author}{\bibfnamefont{A.~D.} \bibnamefont{Cimmarusti}},
  \bibinfo{author}{\bibfnamefont{Z.}~\bibnamefont{Yan}},
  \bibinfo{author}{\bibfnamefont{B.~D.} \bibnamefont{Patterson}},
  \bibinfo{author}{\bibfnamefont{L.~P.} \bibnamefont{Corcos}},
  \bibinfo{author}{\bibfnamefont{L.~A.} \bibnamefont{Orozco}},
  \bibnamefont{and} \bibinfo{author}{\bibfnamefont{S.}~\bibnamefont{Deffner}},
  \bibinfo{journal}{Phys. Rev. Lett.} \textbf{\bibinfo{volume}{114}},
  \bibinfo{pages}{233602} (\bibinfo{year}{2015}).

\bibitem[{\citenamefont{Machnes et~al.}(2012)\citenamefont{Machnes, Cerrillo,
  Aspelmeyer, Wieczorek, Plenio, and Retzker}}]{MachnesPRL12}
\bibinfo{author}{\bibfnamefont{S.}~\bibnamefont{Machnes}},
  \bibinfo{author}{\bibfnamefont{J.}~\bibnamefont{Cerrillo}},
  \bibinfo{author}{\bibfnamefont{M.}~\bibnamefont{Aspelmeyer}},
  \bibinfo{author}{\bibfnamefont{W.}~\bibnamefont{Wieczorek}},
  \bibinfo{author}{\bibfnamefont{M.~B.} \bibnamefont{Plenio}},
  \bibnamefont{and} \bibinfo{author}{\bibfnamefont{A.}~\bibnamefont{Retzker}},
  \bibinfo{journal}{Phys. Rev. Lett.} \textbf{\bibinfo{volume}{108}},
  \bibinfo{pages}{153601} (\bibinfo{year}{2012}).

\bibitem[{\citenamefont{Rahmani et~al.}(2013)\citenamefont{Rahmani, Kitagawa,
  Demler, and Chamon}}]{RahmaniPRA13}
\bibinfo{author}{\bibfnamefont{A.}~\bibnamefont{Rahmani}},
  \bibinfo{author}{\bibfnamefont{T.}~\bibnamefont{Kitagawa}},
  \bibinfo{author}{\bibfnamefont{E.}~\bibnamefont{Demler}}, \bibnamefont{and}
  \bibinfo{author}{\bibfnamefont{C.}~\bibnamefont{Chamon}},
  \bibinfo{journal}{Phys. Rev. A} \textbf{\bibinfo{volume}{87}},
  \bibinfo{pages}{043607} (\bibinfo{year}{2013}).

\bibitem[{\citenamefont{Poyatos et~al.}(1996)\citenamefont{Poyatos, Cirac, and
  Zoller}}]{PoyatosPRL96}
\bibinfo{author}{\bibfnamefont{J.~F.} \bibnamefont{Poyatos}},
  \bibinfo{author}{\bibfnamefont{J.~I.} \bibnamefont{Cirac}}, \bibnamefont{and}
  \bibinfo{author}{\bibfnamefont{P.}~\bibnamefont{Zoller}},
  \bibinfo{journal}{Phys. Rev. Lett.} \textbf{\bibinfo{volume}{77}},
  \bibinfo{pages}{4728} (\bibinfo{year}{1996}).

\bibitem[{\citenamefont{Pielawa et~al.}(2007)\citenamefont{Pielawa, Morigi,
  Vitali, and Davidovich}}]{PielawaPRL07}
\bibinfo{author}{\bibfnamefont{S.}~\bibnamefont{Pielawa}},
  \bibinfo{author}{\bibfnamefont{G.}~\bibnamefont{Morigi}},
  \bibinfo{author}{\bibfnamefont{D.}~\bibnamefont{Vitali}}, \bibnamefont{and}
  \bibinfo{author}{\bibfnamefont{L.}~\bibnamefont{Davidovich}},
  \bibinfo{journal}{Phys. Rev. Lett.} \textbf{\bibinfo{volume}{98}},
  \bibinfo{pages}{240401} (\bibinfo{year}{2007}).

\bibitem[{\citenamefont{Diehl et~al.}(2008)\citenamefont{Diehl, Micheli,
  Kantian, Kraus, B\"uchler, and Zoller}}]{DiehlNatPhys08}
\bibinfo{author}{\bibfnamefont{S.}~\bibnamefont{Diehl}},
  \bibinfo{author}{\bibfnamefont{A.}~\bibnamefont{Micheli}},
  \bibinfo{author}{\bibfnamefont{A.}~\bibnamefont{Kantian}},
  \bibinfo{author}{\bibfnamefont{B.}~\bibnamefont{Kraus}},
  \bibinfo{author}{\bibfnamefont{H.-P.} \bibnamefont{B\"uchler}},
  \bibnamefont{and} \bibinfo{author}{\bibfnamefont{P.}~\bibnamefont{Zoller}},
  \bibinfo{journal}{Nature Phys.} \textbf{\bibinfo{volume}{4}},
  \bibinfo{pages}{878} (\bibinfo{year}{2008}).

\bibitem[{\citenamefont{Schirmer and Wang}(2010)}]{SchirmerPRA10}
\bibinfo{author}{\bibfnamefont{S.~G.} \bibnamefont{Schirmer}} \bibnamefont{and}
  \bibinfo{author}{\bibfnamefont{X.}~\bibnamefont{Wang}},
  \bibinfo{journal}{Phys. Rev. A} \textbf{\bibinfo{volume}{81}},
  \bibinfo{pages}{062306} (\bibinfo{year}{2010}).

\bibitem[{\citenamefont{Motzoi et~al.}(2015)\citenamefont{Motzoi, Halperin,
  Wang, Whaley, and Schirmer}}]{Motzoi16}
\bibinfo{author}{\bibfnamefont{F.}~\bibnamefont{Motzoi}},
  \bibinfo{author}{\bibfnamefont{E.}~\bibnamefont{Halperin}},
  \bibinfo{author}{\bibfnamefont{X.}~\bibnamefont{Wang}},
  \bibinfo{author}{\bibfnamefont{K.~B.} \bibnamefont{Whaley}},
  \bibnamefont{and} \bibinfo{author}{\bibfnamefont{S.}~\bibnamefont{Schirmer}},
  \bibinfo{journal}{arXiv:1512.03415}  (\bibinfo{year}{2015}).

\bibitem[{\citenamefont{Pechen and Rabitz}(2006)}]{PechenPRA06}
\bibinfo{author}{\bibfnamefont{A.}~\bibnamefont{Pechen}} \bibnamefont{and}
  \bibinfo{author}{\bibfnamefont{H.}~\bibnamefont{Rabitz}},
  \bibinfo{journal}{Phys. Rev. A} \textbf{\bibinfo{volume}{73}},
  \bibinfo{pages}{062102} (\bibinfo{year}{2006}).

\bibitem[{\citenamefont{Pechen et~al.}(2006)\citenamefont{Pechen, Il'in,
  Shuang, and Rabitz}}]{PechenPRA06b}
\bibinfo{author}{\bibfnamefont{A.}~\bibnamefont{Pechen}},
  \bibinfo{author}{\bibfnamefont{N.}~\bibnamefont{Il'in}},
  \bibinfo{author}{\bibfnamefont{F.}~\bibnamefont{Shuang}}, \bibnamefont{and}
  \bibinfo{author}{\bibfnamefont{H.}~\bibnamefont{Rabitz}},
  \bibinfo{journal}{Phys. Rev. A} \textbf{\bibinfo{volume}{74}},
  \bibinfo{pages}{052102} (\bibinfo{year}{2006}).

\bibitem[{\citenamefont{Burgarth et~al.}(2014)\citenamefont{Burgarth, Facchi,
  Giovannetti, Nakazato, Pascazio, and Yuasa}}]{BurgarthNatComm14}
\bibinfo{author}{\bibfnamefont{D.~K.} \bibnamefont{Burgarth}},
  \bibinfo{author}{\bibfnamefont{P.}~\bibnamefont{Facchi}},
  \bibinfo{author}{\bibfnamefont{V.}~\bibnamefont{Giovannetti}},
  \bibinfo{author}{\bibfnamefont{H.}~\bibnamefont{Nakazato}},
  \bibinfo{author}{\bibfnamefont{S.}~\bibnamefont{Pascazio}}, \bibnamefont{and}
  \bibinfo{author}{\bibfnamefont{K.}~\bibnamefont{Yuasa}},
  \bibinfo{journal}{Nature Commun.} \textbf{\bibinfo{volume}{5}},
  \bibinfo{pages}{5173} (\bibinfo{year}{2014}).

\bibitem[{\citenamefont{Arenz et~al.}(2016)\citenamefont{Arenz, Burgarth,
  Facchi, Giovannetti, Nakazato, Pascazio, and Yuasa}}]{Arenz16}
\bibinfo{author}{\bibfnamefont{C.}~\bibnamefont{Arenz}},
  \bibinfo{author}{\bibfnamefont{D.}~\bibnamefont{Burgarth}},
  \bibinfo{author}{\bibfnamefont{P.}~\bibnamefont{Facchi}},
  \bibinfo{author}{\bibfnamefont{V.}~\bibnamefont{Giovannetti}},
  \bibinfo{author}{\bibfnamefont{H.}~\bibnamefont{Nakazato}},
  \bibinfo{author}{\bibfnamefont{S.}~\bibnamefont{Pascazio}}, \bibnamefont{and}
  \bibinfo{author}{\bibfnamefont{K.}~\bibnamefont{Yuasa}},
  \bibinfo{journal}{arXiv:1601.01212}  (\bibinfo{year}{2016}).

\bibitem[{\citenamefont{Shuang et~al.}(2007)\citenamefont{Shuang, Pechen, Ho,
  and Rabitz}}]{ShuangJCP07}
\bibinfo{author}{\bibfnamefont{F.}~\bibnamefont{Shuang}},
  \bibinfo{author}{\bibfnamefont{A.}~\bibnamefont{Pechen}},
  \bibinfo{author}{\bibfnamefont{T.-S.} \bibnamefont{Ho}}, \bibnamefont{and}
  \bibinfo{author}{\bibfnamefont{H.}~\bibnamefont{Rabitz}},
  \bibinfo{journal}{J. Chem. Phys.} \textbf{\bibinfo{volume}{126}},
  \bibinfo{pages}{134303} (\bibinfo{year}{2007}).

\bibitem[{\citenamefont{Floether et~al.}(2012)\citenamefont{Floether,
  de~Fouquieres, and Schirmer}}]{FloetherNJP12}
\bibinfo{author}{\bibfnamefont{F.~F.} \bibnamefont{Floether}},
  \bibinfo{author}{\bibfnamefont{P.}~\bibnamefont{de~Fouquieres}},
  \bibnamefont{and} \bibinfo{author}{\bibfnamefont{S.~G.}
  \bibnamefont{Schirmer}}, \bibinfo{journal}{New J. Phys.}
  \textbf{\bibinfo{volume}{14}}, \bibinfo{pages}{073023}
  (\bibinfo{year}{2012}).

\bibitem[{\citenamefont{Laine et~al.}(2014)\citenamefont{Laine, Breuer, and
  Piilo}}]{LaineSciRep14}
\bibinfo{author}{\bibfnamefont{E.-M.} \bibnamefont{Laine}},
  \bibinfo{author}{\bibfnamefont{H.-P.} \bibnamefont{Breuer}},
  \bibnamefont{and} \bibinfo{author}{\bibfnamefont{J.}~\bibnamefont{Piilo}},
  \bibinfo{journal}{Sci. Rep.} \textbf{\bibinfo{volume}{4}},
  \bibinfo{pages}{4620} (\bibinfo{year}{2014}).

\bibitem[{\citenamefont{Stefanatos}(2014)}]{StefanatosPRE14}
\bibinfo{author}{\bibfnamefont{D.}~\bibnamefont{Stefanatos}},
  \bibinfo{journal}{Phys. Rev. E} \textbf{\bibinfo{volume}{90}},
  \bibinfo{pages}{012119} (\bibinfo{year}{2014}).

\bibitem[{\citenamefont{Br\"uggemann et~al.}(2007)\citenamefont{Br\"uggemann,
  Pullerits, and May}}]{BrueggemannJPPA07}
\bibinfo{author}{\bibfnamefont{B.}~\bibnamefont{Br\"uggemann}},
  \bibinfo{author}{\bibfnamefont{T.}~\bibnamefont{Pullerits}},
  \bibnamefont{and} \bibinfo{author}{\bibfnamefont{V.}~\bibnamefont{May}},
  \bibinfo{journal}{J. Photochem. Photobiol. A} \textbf{\bibinfo{volume}{190}},
  \bibinfo{pages}{372 } (\bibinfo{year}{2007}).

\bibitem[{\citenamefont{Hoyer et~al.}(2014)\citenamefont{Hoyer, Caruso,
  Montangero, Sarovar, Calarco, Plenio, and Whaley}}]{HoyerNJP14}
\bibinfo{author}{\bibfnamefont{S.}~\bibnamefont{Hoyer}},
  \bibinfo{author}{\bibfnamefont{F.}~\bibnamefont{Caruso}},
  \bibinfo{author}{\bibfnamefont{S.}~\bibnamefont{Montangero}},
  \bibinfo{author}{\bibfnamefont{M.}~\bibnamefont{Sarovar}},
  \bibinfo{author}{\bibfnamefont{T.}~\bibnamefont{Calarco}},
  \bibinfo{author}{\bibfnamefont{M.~B.} \bibnamefont{Plenio}},
  \bibnamefont{and} \bibinfo{author}{\bibfnamefont{K.~B.}
  \bibnamefont{Whaley}}, \bibinfo{journal}{New J. Phys.}
  \textbf{\bibinfo{volume}{16}}, \bibinfo{pages}{045007}
  (\bibinfo{year}{2014}).

\bibitem[{\citenamefont{Pelzer et~al.}(2008)\citenamefont{Pelzer, Ramakrishna,
  and Seideman}}]{PelzerJCP08}
\bibinfo{author}{\bibfnamefont{A.}~\bibnamefont{Pelzer}},
  \bibinfo{author}{\bibfnamefont{S.}~\bibnamefont{Ramakrishna}},
  \bibnamefont{and} \bibinfo{author}{\bibfnamefont{T.}~\bibnamefont{Seideman}},
  \bibinfo{journal}{J. Chem. Phys.} \textbf{\bibinfo{volume}{129}},
  \bibinfo{pages}{134301} (\bibinfo{year}{2008}).

\bibitem[{\citenamefont{Zhdanov and Rabitz}(2011)}]{ZhdanovPRA11}
\bibinfo{author}{\bibfnamefont{D.}~\bibnamefont{Zhdanov}} \bibnamefont{and}
  \bibinfo{author}{\bibfnamefont{H.}~\bibnamefont{Rabitz}},
  \bibinfo{journal}{Phys. Rev. A} \textbf{\bibinfo{volume}{83}},
  \bibinfo{pages}{061402} (\bibinfo{year}{2011}).

\bibitem[{\citenamefont{Vieillard et~al.}(2013)\citenamefont{Vieillard,
  Chaussard, Billard, Sugny, Faucher, Ivanov, Hartmann, Boulet, and
  Lavorel}}]{VieillardPRA13}
\bibinfo{author}{\bibfnamefont{T.}~\bibnamefont{Vieillard}},
  \bibinfo{author}{\bibfnamefont{F.}~\bibnamefont{Chaussard}},
  \bibinfo{author}{\bibfnamefont{F.}~\bibnamefont{Billard}},
  \bibinfo{author}{\bibfnamefont{D.}~\bibnamefont{Sugny}},
  \bibinfo{author}{\bibfnamefont{O.}~\bibnamefont{Faucher}},
  \bibinfo{author}{\bibfnamefont{S.}~\bibnamefont{Ivanov}},
  \bibinfo{author}{\bibfnamefont{J.-M.} \bibnamefont{Hartmann}},
  \bibinfo{author}{\bibfnamefont{C.}~\bibnamefont{Boulet}}, \bibnamefont{and}
  \bibinfo{author}{\bibfnamefont{B.}~\bibnamefont{Lavorel}},
  \bibinfo{journal}{Phys. Rev. A} \textbf{\bibinfo{volume}{87}},
  \bibinfo{pages}{023409} (\bibinfo{year}{2013}).

\bibitem[{\citenamefont{Li and Kleinekath\"ofer}(2010)}]{KleinekathoeferEPJB10}
\bibinfo{author}{\bibfnamefont{G.-Q.} \bibnamefont{Li}} \bibnamefont{and}
  \bibinfo{author}{\bibfnamefont{U.}~\bibnamefont{Kleinekath\"ofer}},
  \bibinfo{journal}{Eur. Phys. J. B} \textbf{\bibinfo{volume}{76}},
  \bibinfo{pages}{309} (\bibinfo{year}{2010}).

\bibitem[{\citenamefont{Sugny et~al.}(2007)\citenamefont{Sugny, Ndong,
  Lauvergnat, Justum, and Desouter-Lecomte}}]{SugnyJPPA07}
\bibinfo{author}{\bibfnamefont{D.}~\bibnamefont{Sugny}},
  \bibinfo{author}{\bibfnamefont{M.}~\bibnamefont{Ndong}},
  \bibinfo{author}{\bibfnamefont{D.}~\bibnamefont{Lauvergnat}},
  \bibinfo{author}{\bibfnamefont{Y.}~\bibnamefont{Justum}}, \bibnamefont{and}
  \bibinfo{author}{\bibfnamefont{M.}~\bibnamefont{Desouter-Lecomte}},
  \bibinfo{journal}{J. Photochem. Photobiol. A} \textbf{\bibinfo{volume}{190}},
  \bibinfo{pages}{359 } (\bibinfo{year}{2007}), ISSN \bibinfo{issn}{1010-6030},
  \bibinfo{note}{theoretical Aspects of Photoinduced Processes in Complex
  Systems}.

\bibitem[{\citenamefont{Chenel et~al.}(2012)\citenamefont{Chenel, Dive, Meier,
  and Desouter-Lecomte}}]{ChenelJPCA12}
\bibinfo{author}{\bibfnamefont{A.}~\bibnamefont{Chenel}},
  \bibinfo{author}{\bibfnamefont{G.}~\bibnamefont{Dive}},
  \bibinfo{author}{\bibfnamefont{C.}~\bibnamefont{Meier}}, \bibnamefont{and}
  \bibinfo{author}{\bibfnamefont{M.}~\bibnamefont{Desouter-Lecomte}},
  \bibinfo{journal}{J. Phys. Chem. A} \textbf{\bibinfo{volume}{116}},
  \bibinfo{pages}{11273} (\bibinfo{year}{2012}).

\bibitem[{\citenamefont{Chenel et~al.}(2015)\citenamefont{Chenel, Meier, Dive,
  and Desouter-Lecomte}}]{ChenelJCP15}
\bibinfo{author}{\bibfnamefont{A.}~\bibnamefont{Chenel}},
  \bibinfo{author}{\bibfnamefont{C.}~\bibnamefont{Meier}},
  \bibinfo{author}{\bibfnamefont{G.}~\bibnamefont{Dive}}, \bibnamefont{and}
  \bibinfo{author}{\bibfnamefont{M.}~\bibnamefont{Desouter-Lecomte}},
  \bibinfo{journal}{J. Chem. Phys.} \textbf{\bibinfo{volume}{142}},
  \bibinfo{pages}{024307} (\bibinfo{year}{2015}).

\bibitem[{\citenamefont{Tremblay and Saalfrank}(2008)}]{TremblayPRA08}
\bibinfo{author}{\bibfnamefont{J.~C.} \bibnamefont{Tremblay}} \bibnamefont{and}
  \bibinfo{author}{\bibfnamefont{P.}~\bibnamefont{Saalfrank}},
  \bibinfo{journal}{Phys. Rev. A} \textbf{\bibinfo{volume}{78}},
  \bibinfo{pages}{063408} (\bibinfo{year}{2008}).

\bibitem[{\citenamefont{Mishima and Yamashita}(2009)}]{MishimaJCP09}
\bibinfo{author}{\bibfnamefont{K.}~\bibnamefont{Mishima}} \bibnamefont{and}
  \bibinfo{author}{\bibfnamefont{K.}~\bibnamefont{Yamashita}},
  \bibinfo{journal}{J. Chem. Phys.} \textbf{\bibinfo{volume}{131}},
  \bibinfo{pages}{014109} (\bibinfo{year}{2009}).

\bibitem[{\citenamefont{Asplund and Kl\"uner}(2011)}]{AsplundPRL11}
\bibinfo{author}{\bibfnamefont{E.}~\bibnamefont{Asplund}} \bibnamefont{and}
  \bibinfo{author}{\bibfnamefont{T.}~\bibnamefont{Kl\"uner}},
  \bibinfo{journal}{Phys. Rev. Lett.} \textbf{\bibinfo{volume}{106}},
  \bibinfo{pages}{140404} (\bibinfo{year}{2011}).

\bibitem[{\citenamefont{Tremblay et~al.}(2012)\citenamefont{Tremblay,
  F\"uchsel, and Saalfrank}}]{TremblayPRB12}
\bibinfo{author}{\bibfnamefont{J.~C.} \bibnamefont{Tremblay}},
  \bibinfo{author}{\bibfnamefont{G.}~\bibnamefont{F\"uchsel}},
  \bibnamefont{and}
  \bibinfo{author}{\bibfnamefont{P.}~\bibnamefont{Saalfrank}},
  \bibinfo{journal}{Phys. Rev. B} \textbf{\bibinfo{volume}{86}},
  \bibinfo{pages}{045438} (\bibinfo{year}{2012}).

\bibitem[{\citenamefont{Gualdi et~al.}(2014)\citenamefont{Gualdi, Licht, Reich,
  and Koch}}]{GualdiPRA14}
\bibinfo{author}{\bibfnamefont{G.}~\bibnamefont{Gualdi}},
  \bibinfo{author}{\bibfnamefont{D.}~\bibnamefont{Licht}},
  \bibinfo{author}{\bibfnamefont{D.~M.} \bibnamefont{Reich}}, \bibnamefont{and}
  \bibinfo{author}{\bibfnamefont{C.~P.} \bibnamefont{Koch}},
  \bibinfo{journal}{Phys. Rev. A} \textbf{\bibinfo{volume}{90}},
  \bibinfo{pages}{032317} (\bibinfo{year}{2014}).

\bibitem[{\citenamefont{Reich et~al.}(2014{\natexlab{b}})\citenamefont{Reich,
  Gualdi, and Koch}}]{ReichJPA14}
\bibinfo{author}{\bibfnamefont{D.~M.} \bibnamefont{Reich}},
  \bibinfo{author}{\bibfnamefont{G.}~\bibnamefont{Gualdi}}, \bibnamefont{and}
  \bibinfo{author}{\bibfnamefont{C.~P.} \bibnamefont{Koch}},
  \bibinfo{journal}{J. Phys. A: Math. Theor.} \textbf{\bibinfo{volume}{47}},
  \bibinfo{pages}{385305} (\bibinfo{year}{2014}{\natexlab{b}}).

\end{thebibliography}

\end{document}